\documentclass[aps,prl,showpacs,notitlepage,nofootinbib,superscriptaddress,floatfix,showkeys,twocolumn,preprintnumbers]{revtex4-1}

\usepackage{blindtext}
\usepackage{hyperref}
\usepackage{amsmath,amssymb}
\usepackage{float}
\usepackage{microtype}
\usepackage{graphicx}
\usepackage{bm}
\usepackage{latexsym}
\usepackage{epsfig}
\usepackage{psfrag}
\usepackage{ulem}
\usepackage{color}
\usepackage[dvipsnames]{xcolor}
\usepackage{subfigure}
\usepackage[section]{placeins}
\usepackage{comment}

\def\e1i{\epsilon_{1\mathrm{i}}}

\DeclareMathOperator{\s}{s}

\newcommand{\bq}{{\bf q}}
\newcommand{\bp}{{\bf p}}

\allowdisplaybreaks[1]

\begin{document}

\preprint{KCL-2025-02}
\title{In-flight positron annihilation as a probe of feebly interacting particles}

\author{Shyam Balaji}
\email{shyam.balaji@kcl.ac.uk}
\affiliation{Physics Department, King’s College London, Strand, London, WC2R 2LS, United Kingdom}

\author{Pierluca~Carenza}\email{pierluca.carenza@fysik.su.se}
\affiliation{The Oskar Klein Centre, Department of Physics, Stockholm University, Stockholm 106 91, Sweden}

\author{Pedro De la Torre Luque}\email{pedro.delatorre@uam.es}
\affiliation{Departamento de F\'{i}sica Te\'{o}rica, M-15, Universidad Aut\'{o}noma de Madrid, E-28049 Madrid, Spain}
\affiliation{Instituto de F\'{i}sica Te\'{o}rica UAM-CSIC, Universidad Aut\'{o}noma de Madrid, C/ Nicol\'{a}s Cabrera, 13-15, 28049 Madrid, Spain}
\affiliation{The Oskar Klein Centre, Department of Physics, Stockholm University, Stockholm 106 91, Sweden}

\author{Alessandro Lella}
\email{alessandro.lella@ba.infn.it}
\affiliation{Dipartimento Interateneo di Fisica “Michelangelo Merlin”, Via Amendola 173, 70126 Bari, Italy}
\affiliation{Istituto Nazionale di Fisica Nucleare - Sezione di Bari, Via Orabona 4, 70126 Bari, Italy}

\author{Leonardo Mastrototaro}
\email{lmastrototaro@unisa.it}
\affiliation{Dipartimento di Fisica ``E.R. Caianiello'', Università degli Studi di Salerno, Via Giovanni Paolo II, 132 - 84084 Fisciano (SA), Italy}
\affiliation{INFN - Gruppo Collegato di Salerno, Via Giovanni Paolo II, 132 - 84084 Fisciano (SA), Italy.}

\smallskip
\begin{abstract}
Core-collapse supernovae (SNe) provide a unique environment to study Feebly Interacting Particles (FIPs) such as Axion-Like Particles (ALPs), sterile neutrinos, and Dark Photons (DPs). This paper focuses on heavy FIPs produced in SNe, whose decay produces electrons and positrons, generating observable secondary signals during their propagation and annihilation. We focus on the In-flight Annihilation (IA) of positrons, which emerge as the most significant contribution to the resulting $\gamma$-ray spectrum. Using data from COMPTEL and EGRET we derive the most stringent bounds on the FIP-electron couplings for heavy ALPs, sterile neutrinos, and DPs. These results strengthen existing bounds by one to two orders of magnitude, depending on the FIP model.
\end{abstract}
\maketitle

\section{Introduction}

Core-collapse supernovae (SNe) are commonly recognized as nature's most powerful factories for light and weakly interacting particles. Therefore, they can be used to investigate neutrino physics~\cite{Raffelt:1996wa,Mirizzi:2015eza,Horiuchi:2018ofe} and the existence of exotic Feebly Interacting Particles (FIPs)~\cite{Raffelt:1996wa,Caputo:2024oqc}. FIPs include, among others, axions and Axion-Like Particles (ALPs)~\cite{Raffelt:1987yt,Keil:1996ju,Chang:2018rso,Carenza:2019pxu,Carenza:2020cis,Caputo:2022rca}, scalar bosons~\cite{Caputo:2021rux}, sterile neutrinos~\cite{Kolb:1996pa,Raffelt:2011nc,Mastrototaro:2019vug,Carenza:2023old, Akita:2023iwq}, Dark Photons (DPs)~\cite{Chang:2016ntp, Linden:2024fby}, light $CP$-even scalars~\cite{Dev:2020eam,Balaji:2022noj}, dark flavored particles~\cite{Camalich:2020wac} and unparticles~\cite{Hannestad:2007ys}. All of these FIPs are characterized by the common feature of having extremely suppressed interactions with the Standard Model (SM) particles, making their direct detection extremely challenging~\cite{Antel:2023hkf}. Therefore, SNe are often the most favorable environments to probe FIPs, lying at the frontier of low-energy and high-intensity physics.

Notably, heavy FIPs in a SN core might be able to escape the stellar volume decaying into SM particles along their path in the interstellar medium. Therefore, the investigation of observable astrophysical fluxes in coincidence with SN events is a promising opportunity to probe FIPs and their properties. An interesting research line is linked to the study of electrophilic FIPs more massive than $\sim1$~MeV and decaying into electron-positron pairs~\cite{DelaTorreLuque:2024zsr,Carenza:2023old,DelaTorreLuque:2023nhh,DelaTorreLuque:2023huu,Calore:2021lih,Calore:2021klc}. Considering Galactic SNe, the propagation and annihilation of the electrons and positrons injected by FIP decays, gives rise to a  diffuse signal spanning from X-ray to $\gamma$-ray energies. In particular, Inverse-Compton (IC) scattering of leptons on diffuse photon fluxes, generates a sub-MeV flux detectable in X-ray Multi-Mirror Mission (XMM-Newton)~\cite{Foster:2021ngm} and low-energy  Spectrometer on INTEGRAL (SPI) data~\cite{Bouchet:2008rp, Bouchet:2010dj, Siegert:2015knp, Berteaud:2022tws}.
At $511$~keV, the electron-positron annihilation line accurately measured by SPI, gives stringent information on the positron injection.
In the $\gamma$-ray range, bremsstrahlung emission produces a sizable signal that can be probed by SPI high-energy data, the Imaging Compton Telescope (COMPTEL)~\cite{Sreekumar:1997yg} and the Energetic Gamma Ray Experiment
Telescope (EGRET)~\cite{Strong:2004de}. Moreover, it was recently realized that In-flight positron Annihilation (IA)~\cite{Stecker} is the most important contribution to the $\gamma$-ray signal ~\cite{DelaTorreLuque:2024zsr}. Moreover,  since the primary lepton flux injected by FIP decays is measurable by Voyager-1~\cite{Stone:2013zlg}, it can be interpreted as a sensitive probe of new physics as well as photon observables.

In this work, we analyze several FIP models, from the production in SNe, to the induced secondary photon fluxes. We motivate that, in general, the IA signal is the most constraining probe of the FIP-electron coupling. Therefore, we evaluate the IA bound for ALPs, sterile neutrinos and DPs. We structure this paper as follows: 
in Sec.~\ref{sec:Snu_flux}, we describe the case of SNe as sources of FIPs injecting positrons in the interstellar medium~(ISM), in Sec.~\ref{sec:results} we present our results by comparing the secondary $\gamma$-ray production with data and finally in Sec.~\ref{sec:Conclusion} we conclude.

\section{Positron sources: the case of Feebly interacting particles}
\label{sec:Snu_flux}

In our previous paper~\cite{DelaTorreLuque:2024zsr} we made use of a very generic parametrization of the $e^\pm$ emission via FIP decay, $X\to e^{+}e^{-}$, that is given by~\cite{DelaTorreLuque:2023huu}
\begin{equation}
\begin{split}
    \frac{dN_{e}}{dE_{e}}&=N_{e}C_{0}\left(\frac{4E_{e}^{2}-m_{X}^{2}}{E_{0}^{2}}\right)^{\beta/2}e^{-(1+\beta)\frac{2E_{e}}{E_{0}}}\,,\\
      C_{0}&=\frac{2\sqrt{\pi}\left(\frac{1+\beta}{2m_{X}}\right)^{\frac{1+\beta}{2}}E_{0}^{\frac{\beta-1}{2}}}{K_{\frac{1+\beta}{2}}\left((1+\beta)\frac{m_{X}}{E_{0}}\right)\Gamma\left(1+\frac{\beta}{2}\right)}\,,
\end{split}
\label{eq:spectrum}
\end{equation}
where $E_{e}$ is the emitted lepton energy and we assume that FIPs are emitted with a modified blackbody spectrum, where $E_{0}$ is related to the FIP average energy, its mass is $m_{X}>2m_{e}$, $\beta$ is the spectral index, $K_{\frac{1+\beta}{2}}$ is the modified Bessel function of the second kind of order $(1+\beta)/2$, $\Gamma$ is the Euler-Gamma function and this flux is normalized such that 
\begin{equation}
    \int_{m_{X}/2}^{\infty} dE_{e}\frac{dN_{e}}{dE_{e}}=N_{e}\,.
\end{equation}
Here, we use $N_{e}$ to denote the number of electrons, which is equal to the number of positrons, produced in a SN explosion via FIP decays, i.e.~$N_{e}=N_{e^{+}}=N_{e^{-}}$. 
The simple prescription in Eq.~\eqref{eq:spectrum} does not depend on the type of FIP model.
This flux is obtained by assuming a FIP decaying into an electron-positron pair. Thus, it cannot be strictly valid for sterile neutrinos, which has more involved decay channels.
However, by changing the parameters in 
this simple prescription for the injected lepton flux enabled us to place constraints on the electron-coupling of different electrophilic FIPs in a model-independent way. 

For the sake of completeness, in this work we go beyond this simplification and we derive the full positron injection spectra for the cases of ALPs, DPs and sterile neutrinos (coupling to $\nu_{\tau}$ and $\nu_{\mu}$), evaluating SN FIP emission spectra for each specific case.

For definiteness, in the following all the FIP emission rates will be compute by employing the state-of-the-art 1D spherical-symmetric {\tt GARCHING} group's SN model SFHo-s18.8 provided in Ref.~\cite{SNarchive}, already used in Refs.~\cite{Bollig:2020xdr,Caputo:2021rux,Caputo:2022mah,Lella:2022uwi,Lella:2023bfb,Lella:2024hfk,Lella:2024dmx,Manzari:2024jns}. This model is developed from a stellar progenitor with mass $18.8~M_\odot$~\cite{Sukhbold:2017cnt} and  leads to a NS with baryonic mass $1.35~M_\odot$. The simulation is based on the neutrino-hydrodynamics code {\tt PROMETHEUS-VERTEX}~\cite{Rampp:2002bq}, taking into account all neutrino reactions relevant for core-collapse SNe~\cite{Buras:2005rp,Janka:2012wk,Bollig:2017lki,Fiorillo:2023frv}. Moreover, it also accounts for a 1D treatment of PNS convection via a mixing-length description of the convective fluxes~\cite{Mirizzi:2015eza} as well as muon physics~\cite{Bollig:2017lki}.
For the sake of clarity, we highlight that the choice of a given SN model in the analysis introduce unavoidable uncertainties on FIP fluxes, which are typically sensitive to the SN core temperatures and densities. However, we remark that the SN model adopted is characterized by the coldest PNS profile among those of the {\tt GARCHING} group's archive~\cite{SNarchive}~(see Ref.~\cite{Manzari:2024jns} e.g).  Thus, since a high SN core temperature significantly power FIP emission, our results have to be considered as conservative.
Finally, we point out that, since we work in the setup where FIPs couple weakly with ordinary matter, we neglect possible feedback of FIP emission on the SN profile employed~(see Refs.~\cite{Rembiasz:2018lok,Fischer:2021jfm,Betranhandy:2022bvr,Mori:2024vrf} for some works on possible impact of FIPs on the SN explosion mechanism.

\subsection{ALPs coupled to electrons}

Axion interactions with electrons are described by the following Lagrangian~\cite{Raffelt:1996wa}
\begin{equation}
 \mathcal{L}_{ae}=\frac{g_{ae}}{2\,m_e}\bar{\psi}_{e}\gamma^{\mu}\gamma^{5}\psi_{e}\,\partial_\mu a\,,
\label{eq:lagrangian}
\end{equation}
where $\psi_{e}$ and $a$ are, respectively, the electron and axion fields, $m_e$ is the electron mass, and $g_{ae}$ is the dimensionless axion-electron coupling. Starting from this Lagrangian, it is possible to realize that electron-ion bremsstrahlung is the dominant axion production process in core-collapse SNe for a vast range of ALP masses. The number of axions produced per unit volume and time can be evaluated as~\cite{Carenza:2021osu,Carenza:2021pcm}
\begin{equation}
\begin{split}
    \frac{d^{2}n_{a}}{dt\,d\omega_{a}}=&2\pi\int\frac{2d^{3}\bp_{i}}{(2\pi)^{3}2E_{i}}\frac{2d^{3}\bp_{f}}{(2\pi)^{3}2E_{f}}\frac{|\bp_{a}|}{(2\pi)^{3}}\\
    &(2\pi)\delta(E_{i}-E_{f}-\omega_{a})\,|\mathcal{M}|^{2}f_{i}(1-f_{f})=\\
    =&\frac{1}{64\pi^{6}}\int d\cos\theta_{ia}\,d\cos\theta_{if}\,d\delta\,dE_{f}\\
    &|{\bf p}_{i}||{\bf p}_{f}||{\bf p}_{a}||\mathcal{M}|^{2} f_{i}(1-f_{f})\,,
\end{split}
\label{eq:flux}
\end{equation}
where $\omega_{a}$, $E_{i}$ and $E_{f}$ are the energies of the axion, initial and final electrons respectively; $f_{i,f}$ are the electron distribution functions; $\theta_{ia}$, $\theta_{if}\in [0,\pi]$ are the angles between the initial electron and the axion and the final electron momenta respectively; $\delta\in[0,2\pi]$ is the angle between the two planes determined by the vectors ${\bf p}_i-{\bf p}_a$ and ${\bf p}_i-{\bf p}_f$ and the matrix element averaged over the electron spins and summed over all the target ions is~\cite{Carenza:2021osu}
\begin{equation}
\begin{split}
 &|\mathcal{M}|^{2}=\frac{1}{4}\sum_{j}n_{j}\sum_{\rm s}|\mathcal{M}_{j}|^{2}=\frac{g_{ae}^{2}e^{2}}{2}\frac{k_{S}^{2}T}{|\bq|^{2}(|\bq|^{2}+k_{S}^{2})}\\
 &\left[2\omega_{a}^{2}\frac{p_{i}\cdot p_{f}-m_{e}^{2}-K\cdot p_{a}}{(p_{i}\cdot p_{a})(p_{f}\cdot p_{a})}+2-\frac{p_{f}\cdot p_{a}}{p_{i}\cdot p_{a}}-\frac{p_{i}\cdot p_{a}}{p_{f}\cdot p_{a}}\right]\,,
  \end{split}
  \label{eq:matel2}
\end{equation}
where $K=p_{f}-p_{i}$, for massless axions. 
The complete result for massive axions is shown in Appendix~A of Ref.~\cite{Carenza:2021osu}. 

Nevertheless, electron-positron fusion $e^++e^-\rightarrow a$ could become dominant over bremsstrahlung in certain mass ranges. The electron fusion ALP emission spectrum per unit volume is given by~\cite{Carenza:2021pcm}
\begin{equation}
    \frac{d^2 n_a}{d\omega_a\,dt}=\frac{g_{ae}^2 m_a^2}{16\pi^3}\int_{E_{\rm min}}^{E_{\rm max}}dE_{+}f_+f_-\,,
\end{equation}
where where $E_{-,+}$ and $f_{-,+}$ are the electron and positron energies and distribution functions, respectively, while
\begin{equation}
    E_{\rm min,max}=\frac{\omega_a}{2}\pm\frac{\sqrt{\omega_a^2m_a^2+m_a^4-2|{\bf p}_a|^2m_e^2}}{2m_a}\,,
\end{equation}
in which ${\bf p}_a$ is the ALP momentum.

\begin{figure}[t!]
\includegraphics[width=1\columnwidth]{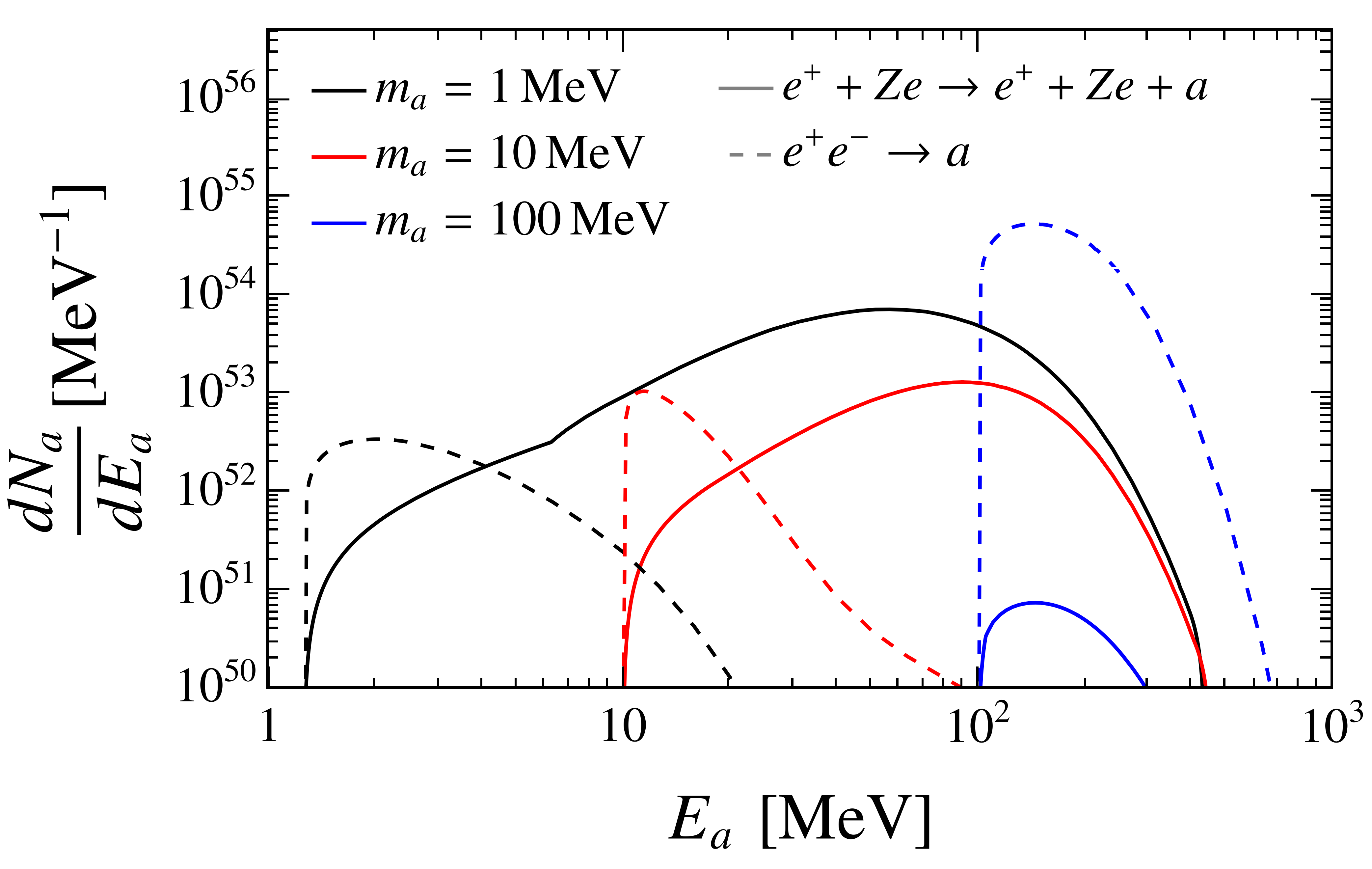}
\caption{SNe ALP production spectra via electron coupling $g_{ae}$ integrated over the SN volume and over the duration of the burst. Solid lines depict electron bremsstrahlung emission spectra, while electron fusion spectra are displayed a dashed lines. The different colors refer to different ALP masses. In this figure we set $g_{ae}=7.5\times10^{-10}$.}
\label{fig:ALPElectronProduction}
\end{figure}

Fig.~\ref{fig:ALPElectronProduction} displays SN emission spectra for ALPs coupled to electrons integrated over the whole SN volume and over the duration of the SN cooling phase $\Delta t\simeq10\,\s$. We observe that electron bremsstrahlung is characterized by quasi-thermal spectra peaking at energies $E\simeq m_a+\frac{3}{2}\,T$, where $T\simeq30\,$MeV is the inner SN core temperature at the beginning of the cooling phase. In particular, the reduction in bremsstrahlung emission for $m_a=100\,$MeV is related to the strong Boltzmann suppression observed for ALPs with masses $m_a\gtrsim 3T$. Conversely, as discussed in Ref.~\cite{Carenza:2021pcm}, the electron fusion contribution is suppressed for $m_a\lesssim 20\,$MeV while it becomes dominant only for higher masses. This trend is related to the fact that low-mass ALP never meet the threshold condition for this process $m_a\gtrsim2\,m_e^*$ for effective electron masses observed in the inner core during the first instants of the cooling phase $m_e^*\sim 10\,$MeV. Therefore, ALP emission through electron fusion plays a role only at $t_{\rm pb}\gtrsim3\,\s$, when the SN core is colder and ALP production dramatically decreases. On the other hand, electron fusion production is active for the whole duration of the SN cooling phase for ALPs with masses $m_a\gtrsim30\,$MeV, resulting in a larger contribution compared to electron bremsstrahlung.

\subsection{ALPs coupled to nucleons and electrons}
\label{subsec:NucProduction}

Building upon the analysis developed for ALPs coupled to electrons only, we now consider the physics case of ALPs interacting with both nucleons and electrons~\cite{Calore:2021klc}. This phenomenological scenario is of definite interest for our study, since a UV complete model accounting for ALP-lepton couplings naturally includes also ALP-quark couplings. Then, at energies below the confinement scale, interactions with the quark content of QCD induce ALP couplings to nuclear matter, which are typically described by the following Lagrangian~\cite{DiLuzio:2020wdo,Chang:1993gm}
\begin{equation}
    \begin{split}
        \mathcal{L}_{\rm{int}}&=g_a\frac{\partial_\mu a}{2m_N}\Bigg[C_{ap}\Bar{p}\gamma^\mu\gamma_5p+C_{an}\Bar{n}\gamma^\mu\gamma_5n+\\
        &+\frac{C_{a\pi N}}{f_\pi}(i\pi^+\Bar{p}\gamma^\mu n-i\pi^-\Bar{n}\gamma^\mu p)+\\
        &+C_{aN\Delta}\left(\Bar{p}\,\Delta^+_\mu+\overline{\Delta^+_\mu}\,p+\Bar{n}\,\Delta^0_\mu+\overline{\Delta^0_\mu}\,n\right)\Bigg]\,,
    \end{split}
\label{eq:NuclearInteractions}
\end{equation}
where $g_a=m_N/f_a$ is the dimensionless axion-nucleon coupling, $m_{N}=938$~MeV is the nucleon mass, $C_{aN}$ with $N=p,n$ are model-dependent $\mathcal{O}(1)$ coupling constants, $f_{\pi}=92.4~$MeV is the pion decay constant, $C_{a\pi N}=(C_{ap}-C_{an})/\sqrt{2}g_{A}$~\cite{Choi:2021ign} is the axion-pion-nucleon coupling and ${C_{aN\Delta}=-\sqrt{3}/2\,(C_{ap}-C_{an})}$ is the axion-nucleon-$\Delta$ baryon coupling, with $g_{A}\simeq1.28$~\cite{ParticleDataGroup:2022pth} the axial coupling. For convenience, we define the axion-proton and axion-neutron coupling as $g_{aN}=g_{a}C_{aN}$ for $N=p,n$. Inspired by the Kim-Shifman-Vainshtein-Zakharov (KSVZ) axion model~\cite{GrillidiCortona:2015jxo}, here we set $g_{an}=0$.\\ 
If ALPs are coupled to nuclear matter, then nuclear interactions are the main channel for ALP production in the hot and dense SN core, which is almost entirely constituted by neutrons and protons. In particular, the interaction Lagrangian in Eq.~\eqref{eq:NuclearInteractions} allows for two competing processes: nucleon-nucleon~($NN$) bremsstrahlung $NN\to NNa$, and pionic Compton-like scatterings (also called pion conversions) $\pi N\to N a$. The computation of the axion emissivity via $NN$ bremsstrahlung has been investigated in a number of works on the topic~\cite{Carena:1988kr,Brinkmann:1988vi,Raffelt:1993ix,Raffelt:1996wa} and the state-of-the-art calculation for the related axion emission rate has been introduced in Ref.~\cite{Carenza:2019pxu} accounting for a non-vanishing mass for the pion exchanged by the interacting nucleons~\cite{Stoica:2009zh}, the contribution from the two-pion exchange~\cite{Ericson:1988wr}, effective in-medium nucleon masses and multiple nucleon scattering effects~\cite{Raffelt:1991pw,Janka:1995ir}. Interestingly, Ref.~\cite{Springmann:2024mjp} has demonstrated that finite density effects on axion-nucleon interactions could magnify the bremsstrahlung emission rate. On the other hand, the impact of pion conversion processes~\cite{Turner:1991ax,Raffelt:1993ix,Keil:1996ju} has been recently reevaluated in Ref.~\cite{Carenza:2020cis}, since the authors of Ref.~\cite{Fore:2019wib} pointed out that strong interactions can magnify the abundance of negatively-charged pions. Under this assumption, the pion conversion emission rate could become comparable and even dominant over the bremsstrahlung contribution. Furthermore, Refs.~\cite{Choi:2021ign,Ho:2022oaw} have recently shown that the contribution from the contact interaction term and the $\Delta(1232)$ resonance significantly enhance axion emissivity via pionic Compton-like processes.

\begin{figure}[t!]
\includegraphics[width=1\columnwidth]{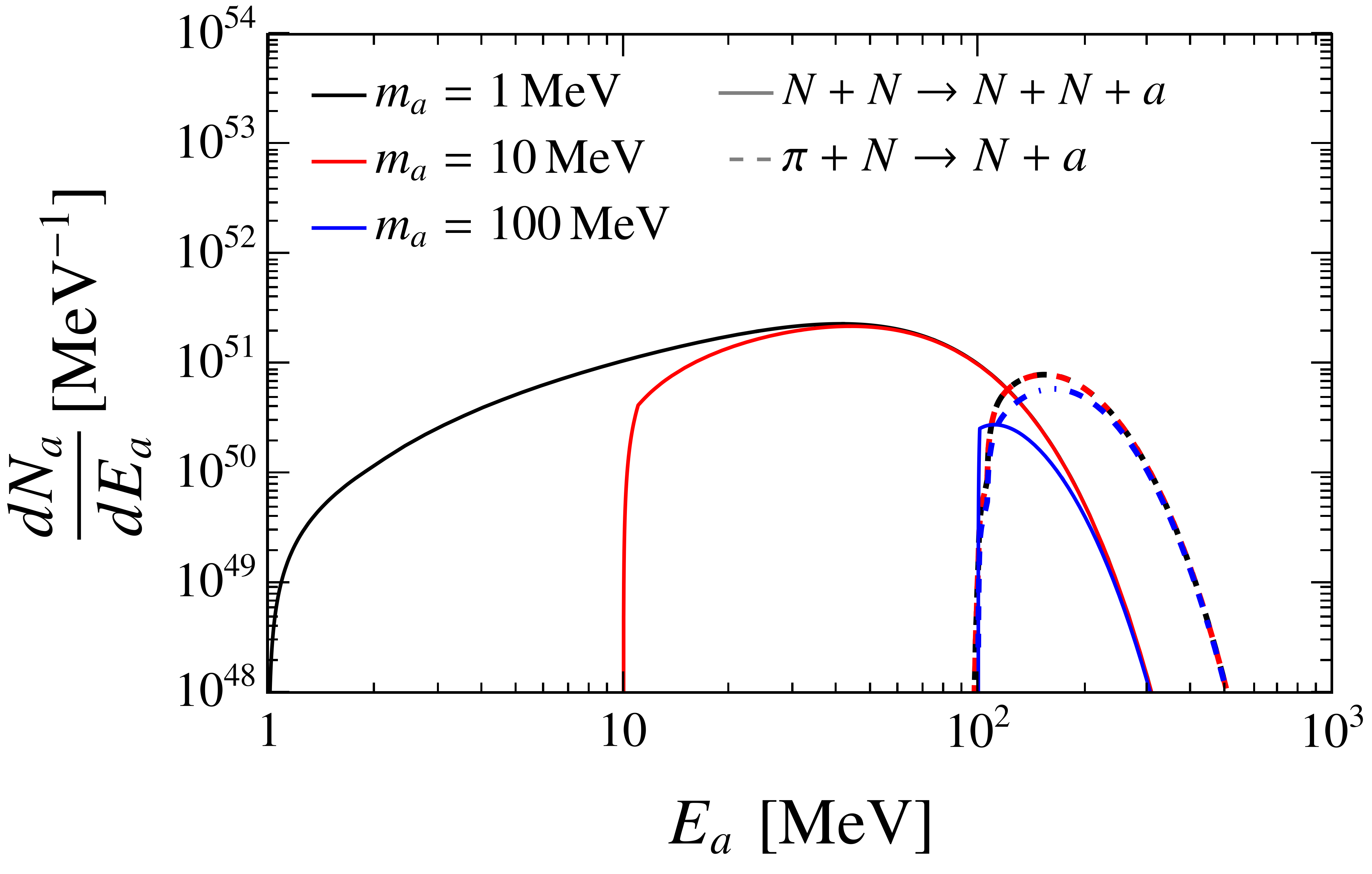}
\caption{ALP production spectra via nuclear couplings  $g_{aN}$ integrated over the SN volume and over the duration of the burst. Solid lines depict $NN$ bremsstrahlung emission spectra, while pion conversion spectra are displayed as dashed lines. The different colors refer to different ALP masses. In this figure we set the ALP-proton coupling $g_{ap}=2\times10^{-11}$ and $g_{an}=0$.}
\label{fig:ALPNuclearProduction}
\end{figure}

In this work, we follow the detailed calculation of Refs.~\cite{Lella:2022uwi, Carenza:2023lci} to implement the emission of MeV-scale ALPs via nuclear processes. As discussed in Refs.~\cite{Lella:2022uwi, Lella:2023bfb}, under the assumption of a sufficiently high pion fraction in the SN core, the ALP emission spectrum in the free-streaming regime $g_{aN}<10^{-8}$ is characterized by a bimodal shape. The ALP spectrum induced by nuclear couplings observed at large distance from the source is displayed in Fig.~\ref{fig:ALPNuclearProduction}, where the different colors refer to different ALP masses. In particular, all the spectra are integrated over the SN volume and over the duration of the SN cooling phase $\Delta t\sim10\,\s$. $NN$ bremsstrahlung emission spectrum~(solid lines) is characterized by a quasi-thermal spectral shape peaking at energies $E_a\sim m_a+50\,$MeV, while the pion conversion contribution~(dashed lines) shows a peak around $E_a\sim m_\pi+\frac{3}{2} T\sim200\,$MeV~\cite{Lella:2022uwi,Lella:2023bfb, Lella:2024hfk}. We can observe that for $m_a\gtrsim 3T\sim90\,$MeV bremsstrahlung is suppressed compared to pionic processes, which become dominant for ALP masses $m_a\gtrsim30\,$MeV.
The ALP production spectra can be plugged into Eq.~\eqref{eq:posspec} to determine the associated positron spectra. In this context, ALPs produced in the SN core via nuclear processes decay in electron-positron pairs $a\to e^{+}e^{-}$ as argued above, yielding a positron injection in the ISM. We highlight that, distinctly from the previous section, in this scenario ALP production and ALP decays depend on different unrelated couplings.

Nevertheless, Ref.~\cite{Lella:2024dmx} has pointed out that ALPs coupled to nuclear matter are naturally provided with a QCD-induced photon coupling, which can be estimated as
\begin{equation}
    \begin{split}
    g_{a\gamma}\simeq\frac{\alpha_{em}}{2\pi m_N}\times&\Big[-\frac{1.92}{1.59\,c_d-0.52}\,c_g\\
    &-\left(\frac{0.71}{1.59\,c_d-0.52}\,c_g+0.79\right)\frac{m_a^2}{m_\pi^2-m_a^2}\Big]
    \end{split}
\label{eq:Cgamma}
\end{equation}
where $m_\pi=135\,$MeV is the pion mass, $\alpha_{em}=1/137$, while $c_g$ and $c_d$ are model-dependent constants. As a benchmark case, here we set $c_g=1$, $c_d=0$. The presence of a non-vanishing photon coupling allows MeV-scale ALPs to efficiently decay into photon pairs, opening an additional decay channel which is competitive with electronic decays for ALP-electron couplings $g_{ae}\lesssim 10^{-16}$. Therefore, in this scenario, the total ALP decay length is given by
\begin{equation}
    \lambda_{\rm ALP}=\left (\lambda_{\mathrm{ALP},e}^{-1}+\lambda_{\mathrm{ALP},\gamma}^{-1}\right )^{-1}\,,
\end{equation}
where $\lambda_{\mathrm{ALP},e}$ and $\lambda_{\mathrm{ALP},\gamma}$ are the ALP decay lengths through electronic and photon decays, respectively~(see Refs.~\cite{Calore:2021lih,Altmann:1995bw} for their explicit expression). Thus, the induced positron spectra have to be rescaled by the branching ratio for ALP decays into electron-positron pairs
\begin{equation}
    \frac{dN_{\rm pos}}{dE_{\rm pos}}\rightarrow BR(a\rightarrow e^+e^-)\frac{dN_{\rm pos}}{dE_{\rm pos}}\,,
\end{equation}
in which $BR(a\rightarrow e^+e^-)=\lambda_{\rm ALP}/\lambda_{\mathrm{ALP},e}$. 
In this regard, we point out that in the scenario considered the induced photon coupling vanishes around $m_a\simeq147\,$MeV and electronic decays result in the only viable decay channel. On the other hand, around $m_a\simeq135\,$MeV the induced ALP-photon coupling shows a pole, so that the total number of emitted ALPs decay via photon pairs. We refer the reader to Ref.~\cite{Lella:2024dmx} for further discussions on these aspects. As an exemplary case, in the following we will set $g_{ap}=2\times10^{-11}$, which is the maximum value allowed by current astrophysical constraints introduced in Ref.~\cite{Lella:2024dmx}.

\subsection{Sterile neutrinos}

MeV-scale sterile neutrinos have interesting phenomenological implications. In the following, we will consider sterile neutrinos with a mass above $10$~MeV to safely avoid any possible resonant production and feedback on the SN, which usually occurs below this threshold~\cite{Raffelt:2011nc,Suliga:2020vpz,Arguelles:2016uwb}. 
In this mass range the mixing of a sterile neutrino with electron neutrino is very constrained~\cite{Alekhin:2015byh}, therefore we assume that the sterile neutrino is mixed dominantly with one active neutrino $\nu_\alpha$, with $\alpha=\mu,\tau$, such as
\begin{equation}
\begin{split}
\nu_\alpha &= U_{\alpha 1} \,\nu_\ell + U_{\alpha 4} \,\nu_4 \,\ ,   \\ 
\nu_s &= -U_{\alpha 4} \,\nu_\ell + U_{s4}\, \nu_4 \,\ ,
\end{split}
\end{equation}
where $\nu_\ell$ and $\nu_4$ are the light and the heavy mass eigenstates, respectively, $U$ is the unitary mixing matrix, linking mass and flavour states, and the most interesting parameter space corresponds to 
$|U_{\alpha 4}|^2\ll 1$, i.e.  $\nu_\ell$ is mostly active and $\nu_4$ is mostly sterile. The SN sterile neutrino production is discussed extensively in Ref.~\cite{Carenza:2023old}. In summary, the most important production channels are the neutral current scattering on nucleons, $\nu_{\alpha}N\to \nu_{s}N$, and the charged current one, $\mu^{-}N\to \nu_{s}N$. Subleading processes are elastic $2\to2$ neutrino scatterings.
A more detailed discussion on sterile neutrinos and their constraints arising from IA observations can be found in Ref.~\cite{DelaTorreLuque:2024zsr}, where the importance of IA as probe of new physics was originally realized.

\subsection{Dark photons}
The DP is a $U(1)^\prime$ gauge boson kinetically mixed with the SM photon~\cite{Okun:1982xi,Holdom:1985ag, Nguyen:2024kwy}. In this context, the relevant terms in the DP Lagrangian are~\cite{Holdom:1985ag,Foot:1991kb}
\begin{equation}
    \mathcal{L}=\frac{1}{2}m_{A^\prime}\,A^\prime_{\mu}\, A^{\prime\mu}-\frac{1}{4}\,F^\prime_{\mu\nu}\,F^{\prime\mu\nu}-\frac{\epsilon}{2}\,F^\prime_{\mu\nu}F^{\mu\nu}\,,
\end{equation}
where $A'$ is the DP field, $\epsilon$ the mixing parameter, $F_{\mu\nu}$ the electromagnetic field strength tensor and $F'_{\mu\nu}$ the analogue for the DP.
Being massive, DPs are provided with both transverse ($T$) and longitudinal ($L$) degrees of freedom. DP production in the SN core has been estimated in a series of papers~\cite{Kazanas:2014mca,Chang:2016ntp,Hardy:2016kme,Stetina:2017ozh,DeRocco:2019njg,Syvolap:2024hdh}. In particular, DPs are mainly emitted via proton bremsstrahlung $p\,N\to A\,p\,N$.

Following Ref.~\cite{DeRocco:2019njg,Calore:2021lih}, the number of DPs produced per unit volume and time in a SN is given by
\begin{equation}
\begin{split}
    \frac{dN^{0}_{A'}}{dV dt}&=
    \frac{dN^{0}_{A'}}{dV dt}\bigg|_L+\frac{dN^{0}_{A'}}{dV dt}\bigg|_T =
    \\
    &= \int \frac{d E E^2 v}{2\,\pi^2}\,e^{-E/T}(\Gamma'_{\rm abs,L}+2 \Gamma'_{\rm abs,T}) \,,
\end{split}
\end{equation}
where $v$ is the velocity, $\Gamma'_{\rm abs, L/T}$ is the absorptive width of the DP for the longitudinal and transverse modes. The main absorption process in the SN core is inverse bremsstrahlung (ibr), thus we can write
\begin{eqnarray}
    \Gamma'_{\rm ibr, L/T}& =&\frac{32}{3\pi}\frac{\alpha (\epsilon_m)^2_{\rm L/T} n_n n_p}{E^3} \left(\frac{\pi T}{m_N}\right)^{3/2}\nonumber \\ &\times &  \langle\sigma_{np}^{(2)}(T) \rangle\,\left(\frac{m_{A'}^{2}}{E^2}\right)_{\rm L}\,,
\end{eqnarray}
with $n_n$ and $n_p$ the neutron and the proton number density, $m_N=938$~MeV, $\langle\sigma_{np}^{(2)}(T) \rangle$ the averaged neutron-proton dipole scattering cross section from Ref.~\cite{Rrapaj:2015wgs}, $(\epsilon_m)^2_{\rm L/T}$ the in-medium mixing angle
\begin{equation}
    (\epsilon_m)^2_{\rm L/T}=\frac{\epsilon^2}{(1-{\rm Re}\Pi_{\rm L/T}/m_{A'}^{2})^2+({\rm Im}\,\Pi_{\rm L/T}/m_{A'}^{2})^2}\,,
\end{equation}
where $\Pi$ is the photon polarization tensor. The real part of the polarization tensor for the two modes is given by
\begin{equation}
\begin{split}
    {\rm Re}\,\Pi_{\rm L}&=\frac{3\omega^2_p}{v^2}\,(1-v^2)\left[\frac{1}{2v}\ln\left(\frac{1+v}{1-v}\right)-1\right]\,,\\
    {\rm Re}\,\Pi_{\rm T}&=\frac{3\omega^2_p}{2v^2}\,(1-v^2)\left[1-\frac{1-v^2}{2v}\ln\left(\frac{1+v}{1-v}\right)\right]\,,
\end{split}
\label{eq:pi}
\end{equation}
with $\omega_p$ the plasma frequency in the SN core. For typical SN conditions it can be estimated as
\begin{equation}
    \omega_p^2=\frac{4\pi \alpha_{\rm EM} n_e}{\sqrt{m_e^2+(3\pi^2\,n_e)^{2/3}}}\,,
\end{equation}
where $n_e$ is the electron number density.\\
The imaginary part of the polarization tensor can be written as
\begin{equation}
    {\rm Im}\,\Pi_{\rm L/T}=-E (1-e^{-E/T})\Gamma_{\rm abs, L/T}\,,
\end{equation}
where $\Gamma_{\rm abs, L/T}$ is the absorptive width of SM photons
\begin{equation}
    \Gamma'_{\rm ibr, L/T}=(\epsilon_m)^2_{\rm L/T} \Gamma_{\rm ibr, L/T} \,.
\end{equation}

\begin{figure}[t!]
\includegraphics[width=1\columnwidth]{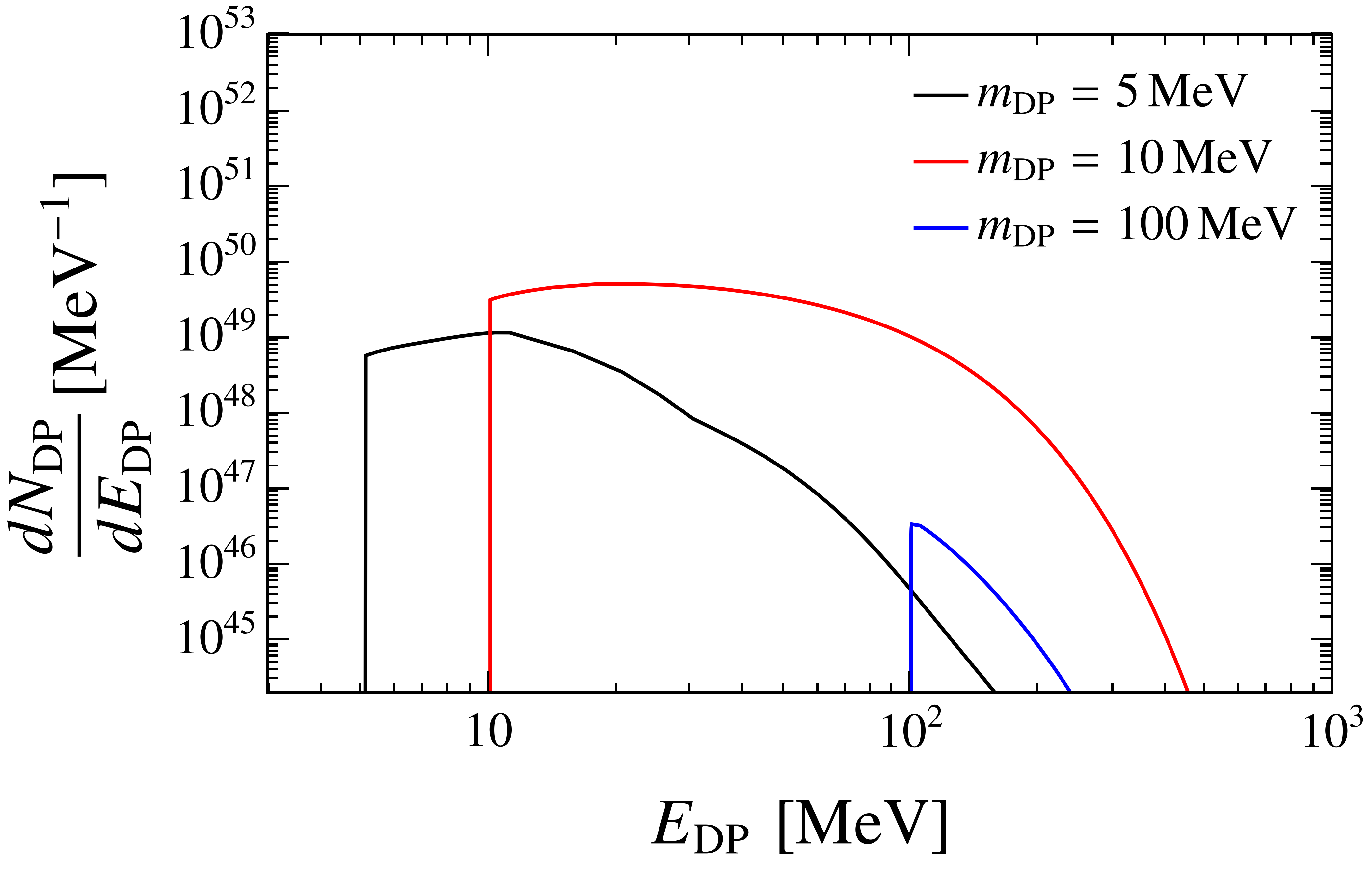}
\caption{DP production spectra via proton bremsstrahlung integrated over the SN volume and over the duration of the SN cooling phase. In this figure we set $\epsilon = 10^{-13}$.}
\label{fig:DPproduction}
\end{figure}

Remarkably, in the interior of the SN core ${{\rm Im}\,\Pi_{\rm L/T}\ll {\rm Re}\,\Pi_{\rm L/T}}$. Therefore, Eq.~\eqref{eq:pi} suggests that DP production is resonantly enhanced in the regions where ${\rm Re}\,\Pi_{\rm L/T}=m_{A'}$. This feature is directly encoded in the DP emission spectra reported in Fig.~\ref{fig:DPproduction}. We observe that DP production is efficient for masses $m_{\rm DP}\sim15\,$MeV, which is the common value assumed by the plasma frequency in the inner regions of the SN model considered in this work. Conversely, DP photon emission is sensibly suppressed for DP photon masses $m_{\rm DP}\gtrsim50$ MeV, where resonant effects are washed out and Boltzmann suppression affects the production mechanism considered.


\subsection{Positron spectra}

Integrating Eq.~\eqref{eq:flux} over a spherically symmetric SN model, we obtain the positron spectrum as
\begin{equation}
\begin{split}
    \frac{dN_{\rm pos}}{dE_{\rm pos}}&= \sum_{i}n_i B_i\Bigg[\int dt\,\int_0^\infty4\pi r^2 dr \,\frac{d^2 n_a}{dE_a\,dt}\\
    &\times\left(\epsilon_{II}\,e^{-r_{II}/\lambda_{i}(E_a, r)}+\epsilon_{I}\,e^{-r_{I}/\lambda_{i}(E_a, r)}\right)\Bigg]\Bigg|_{E_{a}=x_{i}E_{\rm pos}}\,,    
\end{split}
\label{eq:posspec}
\end{equation}
where $r$ is the radial coordinate from the center of the SN core, $t$ is integrated over the FIP emission time, i.e. $1-10$~s, $\lambda_{i} (E_a, r)$ is the mean-free-path for the $i$-th decay channel at a given FIP energy and location, which produces $n_{i}$ positrons with a branching ratio $B_{i}$ and average energy $E_{a}/x_{i}$. In the electrophilic axion case, only $a\to e^{+}e^{-}$ decays are allowed, therefore $n_{1}=2$, $B_{1}=1$ and $x_{i}=2$.
Moreover, following Ref.~\cite{DeRocco:2019njg} we fix
\begin{equation}
    r_{II}=10^{14}~{\rm cm}, \qquad  r_{I}=2\times 10^{12}~{\rm cm}\,,
\end{equation}
for the envelope radii of Type II and Ib/c SNe, while according to Ref.~\cite{Li:2010kd}, we take as average fractions of SNe of Type II and Ib/c 
\begin{equation}
    \epsilon_{II}=1-\epsilon_{I},\qquad \epsilon_{I}=0.33\,\ .
\end{equation}

\section{Secondary $\gamma$-ray production and comparison with MeV data}
\label{sec:results}

\subsection{Injection and propagation of $e^{\pm}$ from FIPs}

The decay into electrons and positrons from any FIP produced in SNe leads to a continuous diffuse sea of $e^+$$e^-$ pairs with energies of tens to hundreds of MeV, which are confined in the Galaxy for Myrs~\cite{DelaTorreLuque:2023huu, DelaTorreLuque:2023nhh}. Given that the rate of SN explosion is negligible compared to the propagation time of these particles, their injection can be considered smooth and continuous and following the spatial distribution of SN remnants.
Once the $e^+$$e^-$ are injected into the ISM, these particles interact with the Galactic environment, being scattered by magnetohydrodynamic fluctuations in the ISM plasma~\cite{Ginz&Syr, Ginzburg_H, 1998ApJ...509..212S, Strong:1998pw}, which makes them propagate diffusively in the Galaxy. During their propagation, their interaction with the Galactic magnetic field, gas and radiation fields produces secondary radiations and makes these particles lose energy. 

Following a similar procedure to Refs.~\cite{DelaTorreLuque:2023huu, DelaTorreLuque:2023olp}, we use a customized version~\cite{de_la_torre_luque_2023_10076728} of the {\tt DRAGON2} code~\cite{Evoli:2016xgn, Evoli:2017vim}, a dedicated CR propagation code prepared to simulate CR diffusion, accounting for all diffusion-reacceleration-advection-loss effects in the propagation of Galactic CRs~\cite{Ginz&Syr}.  We simulate electron-positron signals for different kinds of FIPs produced in SNe in the range of kinetic energies from $100$~eV to $5$~GeV, with an energy resolution of $5\%$.

\begin{figure*}[t!]
\includegraphics[width=0.49\textwidth]{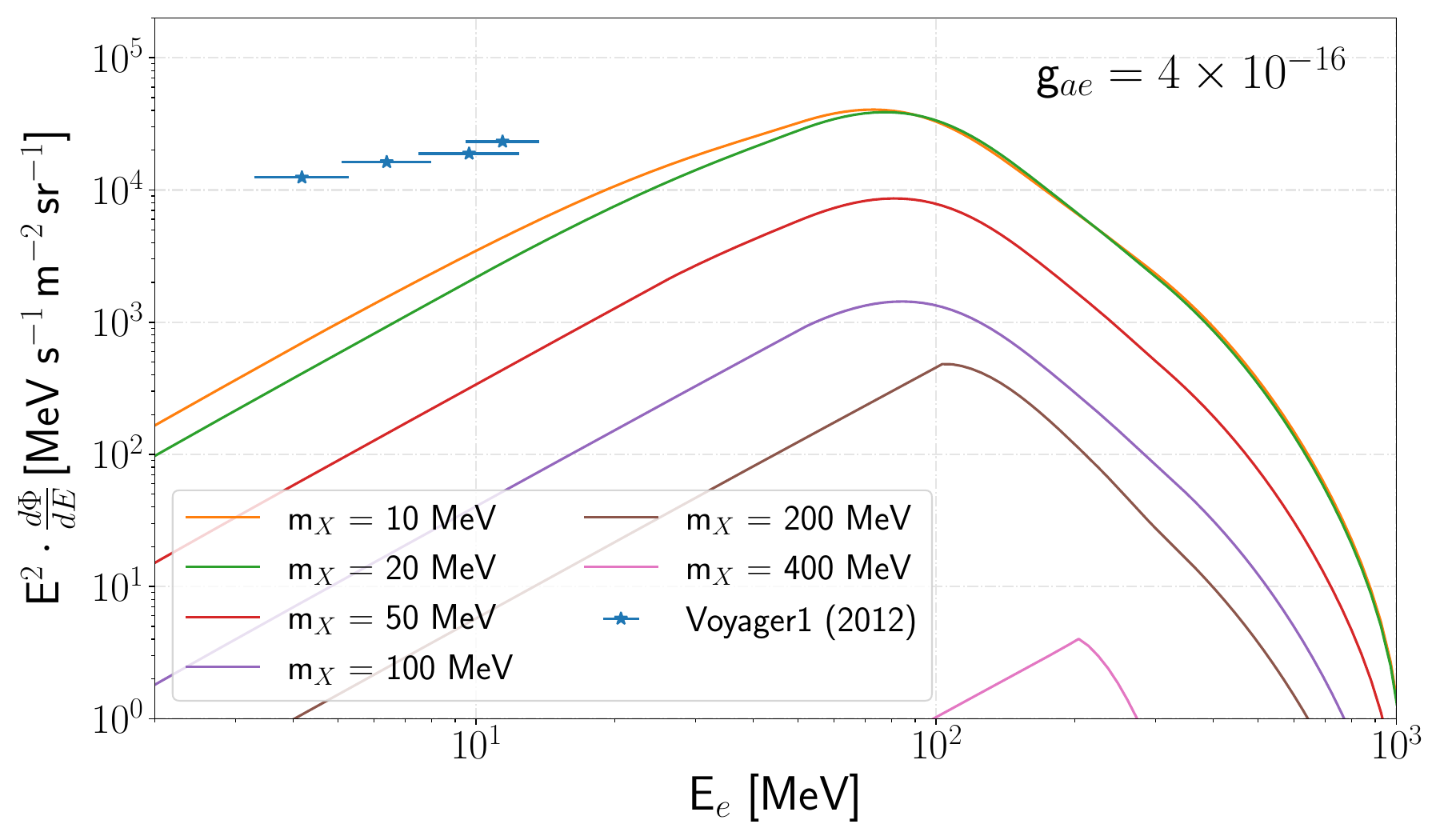}
\includegraphics[width=0.49\textwidth]{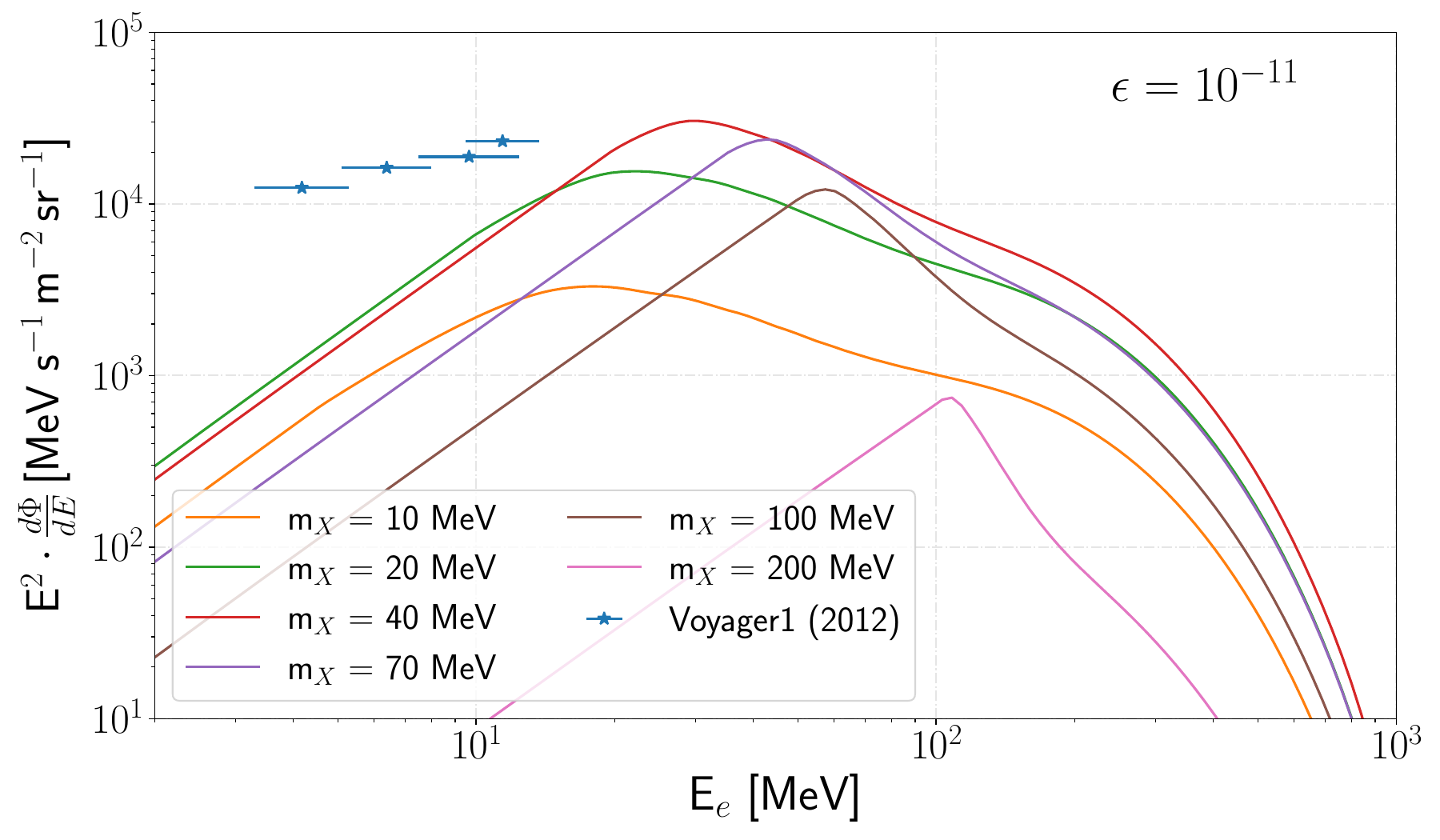}

\includegraphics[width=0.50\textwidth]{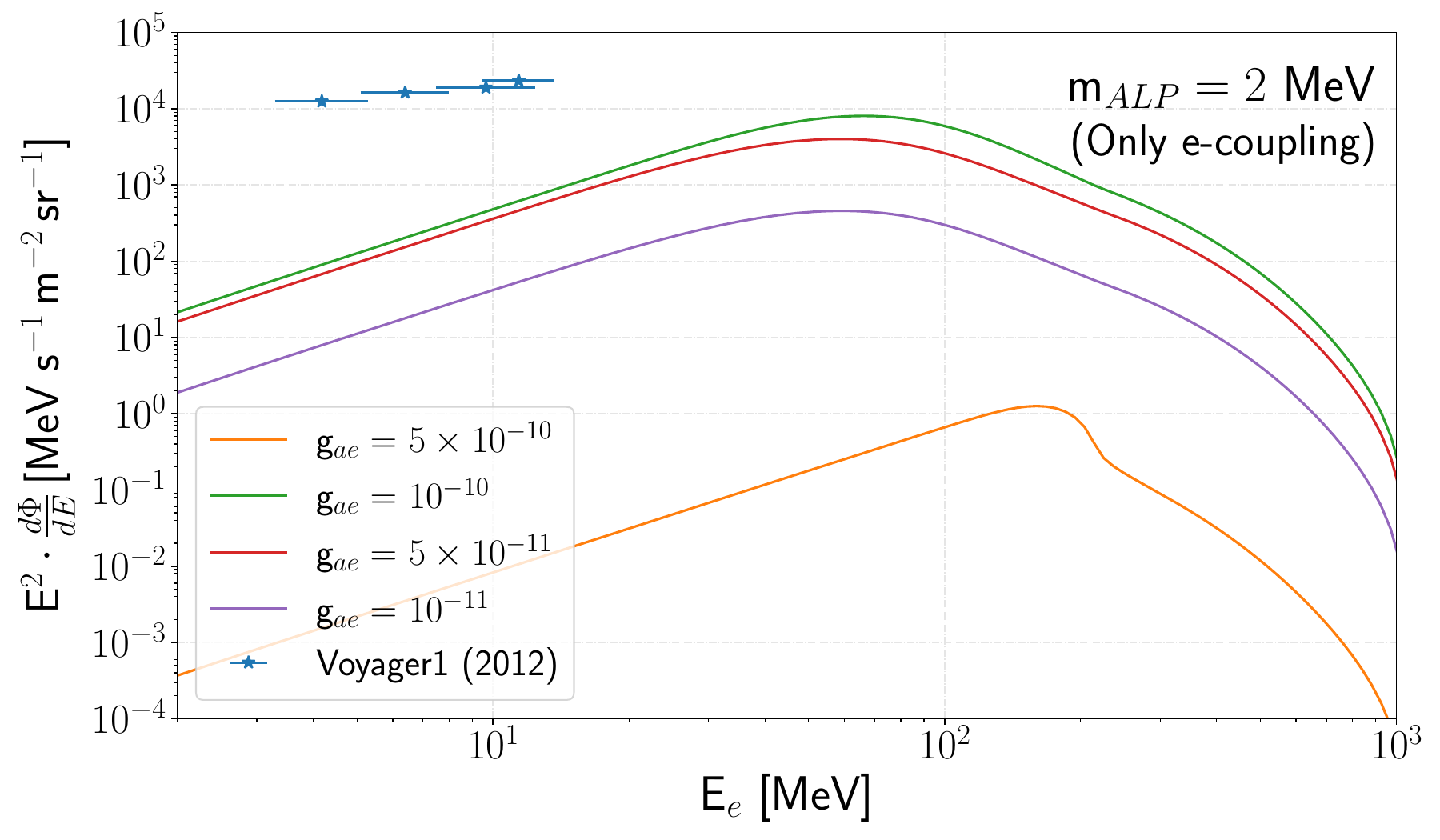}
\caption{Local diffuse $e^{\pm}$ flux predicted for different FIP models, compared to Voyager-1 measurements~\cite{stone2013voyager, cummings2016galactic}. The upper-left panel correspond to the case of ALPs coupling with both baryons and electrons (here we assume $g_{ap}=2\times10^{-11}$), while upper-right panel refer to DPs. Different colors depict the case of different masses. The bottom panel illustrates the diffuse fluxes predicted for ALPs with a mass of $2$~MeV coupling only to electrons. } 
\label{fig:Voy}
\end{figure*}

We compute the steady-state diffuse distribution (in space and energy) of the electrons and positrons produced from the different FIP models studied here by solving their propagation equation numerically. In particular, we follow the same procedure illustrated in Refs.~\cite{DelaTorreLuque:2023huu, DelaTorreLuque:2023olp, DelaTorreLuque:2024qms}, where we refer the reader for more details. The steady-state solution for the distribution of these particles in the Galaxy can be obtained by solving the differential diffusion-reacceleration equation imposing time-independent $e^{\pm}$ density in each point of the Galaxy ($\frac{dn_e}{dt} = 0$) 
    \begin{equation}
    \label{eq:CRtransport}
        - \nabla\cdot\left(D\vec{\nabla} f_e \right) - \frac{\partial}{\partial p_e} \left[\dot{p}_e f_e - p_e^2 D_{pp} \frac{\partial}{\partial p_e}\left(\frac{f_e}{p_e^2}\right) \right] = Q_e\;.
    \end{equation}
In this equation, $p_e$ is the momentum and therefore $f_e \equiv \frac{dn_e}{dp_e}$ is the density of $e^\pm$ per unit momentum {at a given position. Spatial diffusion is characterized by the diffusion coefficient $D$, which basically follows a power-law in rigidity. This diffusion coefficient can be precisely determined at GeV energies for analyses of CR nuclei~\cite{Evoli:2019wwu, Derome_2019, Luque:2021nxb}, however, uncertainties can be important at lower energies. We adopt a set of propagation parameters that allow us to reproduce all current CR observables at the GeV scale, adopting the spiral arm distribution of the gas. These are the propagation parameters used in Ref.~\cite{DelaTorreLuque:2023olp}, which reports detailed discussions on this choice.  We also examine the effect of uncertainties in these parameters below (see Fig.~\ref{fig:Uncerts} and the discussion around it).
In Eq.~\eqref{eq:CRtransport} we take into account all the sources of momentum losses $\dot{p}_e$ due to interactions with the Galactic environment, which are mainly synchrotron interactions with the Galactic magnetic field, inverse Compton scattering off CMB light and Galactic radiation fields, ionization of molecular gas, and Coulomb interactions and bremsstrahlung with the ISM plasma. Momentum diffusion (or reacceleration) is also included through the term $D_{pp} \propto \frac{v_{A}^2 \cdot p^2}{D}$. This term becomes the main source of uncertainty in the predicted $e^{\pm}$ spectrum for particles injected with energies below a few tens of MeV, as shown in Refs.~\cite{DelaTorreLuque:2023huu, DelaTorreLuque:2023nhh, DelaTorreLuque:2023olp}

Finally, the source term, $Q_e$, regulates the injection of $e^{\pm}$. It is described with a spatially dependent term featuring the spatial distribution of SNe, and the energy-dependent injection term, which is calculated imposing that the injected number of particles per unit energy is equal to the integral of the total flux density of particles over the volume of the Galaxy, as described in Ref.~\cite{DelaTorreLuque:2023huu}. In this calculation, we assume that SNe follows the Ferriere distribution~\cite{Ferriere:2007yq}, convolved with the Steiman-Cameron distribution~\cite{Steiman-Cameron:2010iuq} of the spiral arms (four-arm model). We have checked that the effect of using a different spatial distribution (concretely, the Lorimer distribution~\cite{Lorimer:2006qs}) leads to uncertainties always below $20\%$ in the predicted $e^{\pm}$ flux.
We have computed tables with the $e^{\pm}$ injection spectra from each class of FIP, which are used as input in the \verb|DRAGON2| code. These tables are available upon request to the authors.

We show our estimations of the propagated local spectrum of $e^{\pm}$, in the MeV range, in Fig.~\ref{fig:Voy}, compared to Voyager-1 measurements~\cite{Stone150, stone2013voyager, cummings2016galactic}. The upper panels correspond to the case of ALPs (coupling with both baryons and electrons) in the left, and DPs in the right. For the case of ALPs, here we set the production by fixing $g_{ap}=2\times10^{-11}$. These panels show that Voyager-1 observations cannot set very strong constraints for FIPs from SNe, given that their spectrum generally peaks around $\sim100$~MeV, similarly to what we showed for the case of sterile neutrinos in Ref.~\cite{DelaTorreLuque:2024zsr}. In all these FIP models we observe that the expected diffuse flux of $e^{\pm}$ in the Galaxy starts to decrease significantly for masses above $100$~MeV. 
As for the case of ALPs that only couple with electrons, the coupling needed to be constrained by Voyager-1 data (or secondary radiations) is so high ($g_{ae} > 10^{-9}$) that the ALPs cannot escape the SNe and they decay in their interior, therefore not producing any diffuse flux of $e^{\pm}$. This can be seen from the bottom panel of Fig.~\ref{fig:Voy}, where we show the local $e^{\pm}$ predicted for different coupling values, for a mass of $2$ MeV. In this picture, the decrease in flux at couplings larger than $g_{ae}\gtrsim10^{-10}$ is clearly evident between the green and orange curves with a strong flux suppression for $g_{ae}=5\times 10^{-10}$. 
We note that for larger masses, the fluxes at high g$_{ae}$ are even more suppressed.

\subsection{$\gamma$-ray signals}

\begin{figure*}[t!]
\includegraphics[width=0.49\textwidth]{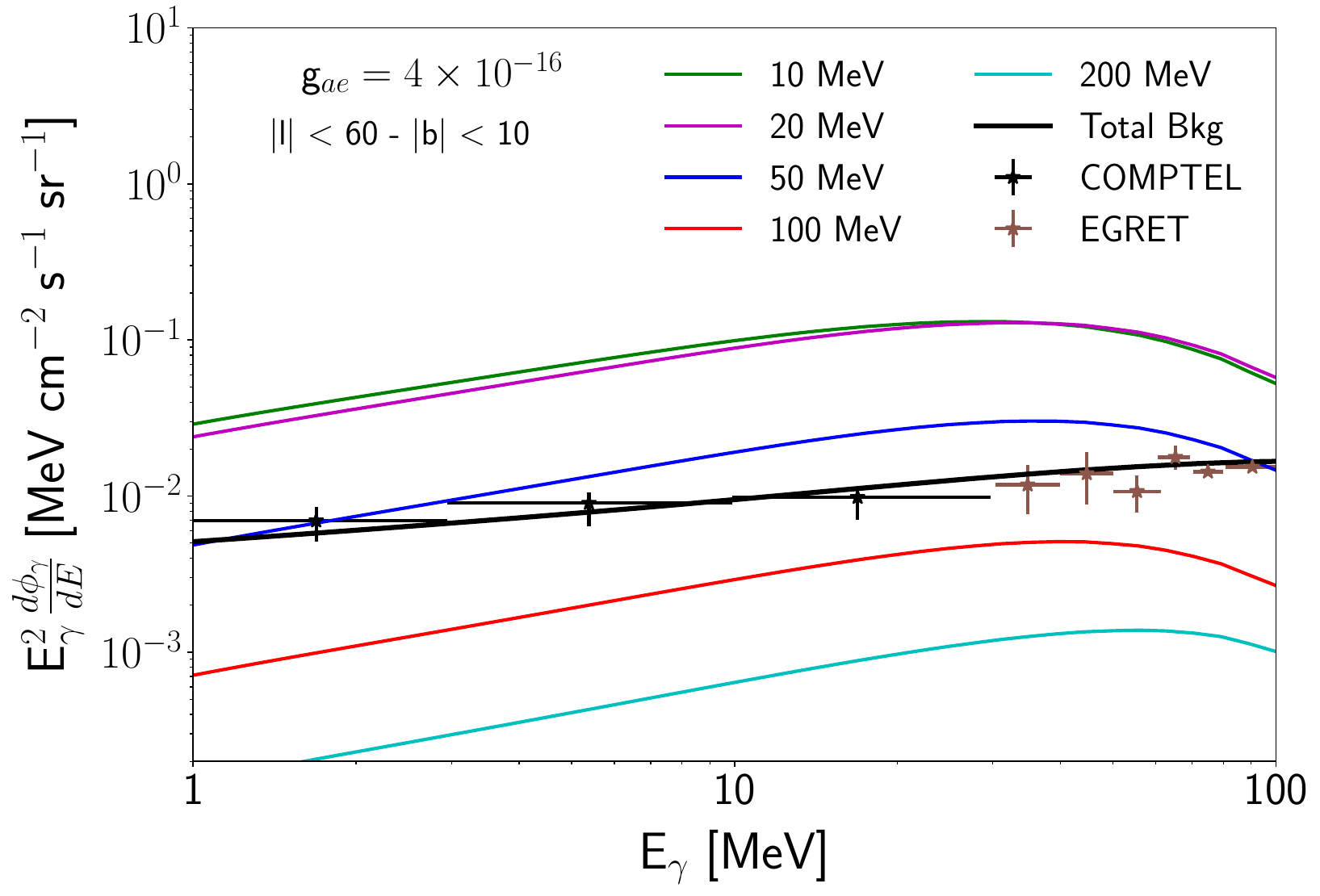}
\includegraphics[width=0.49\textwidth]{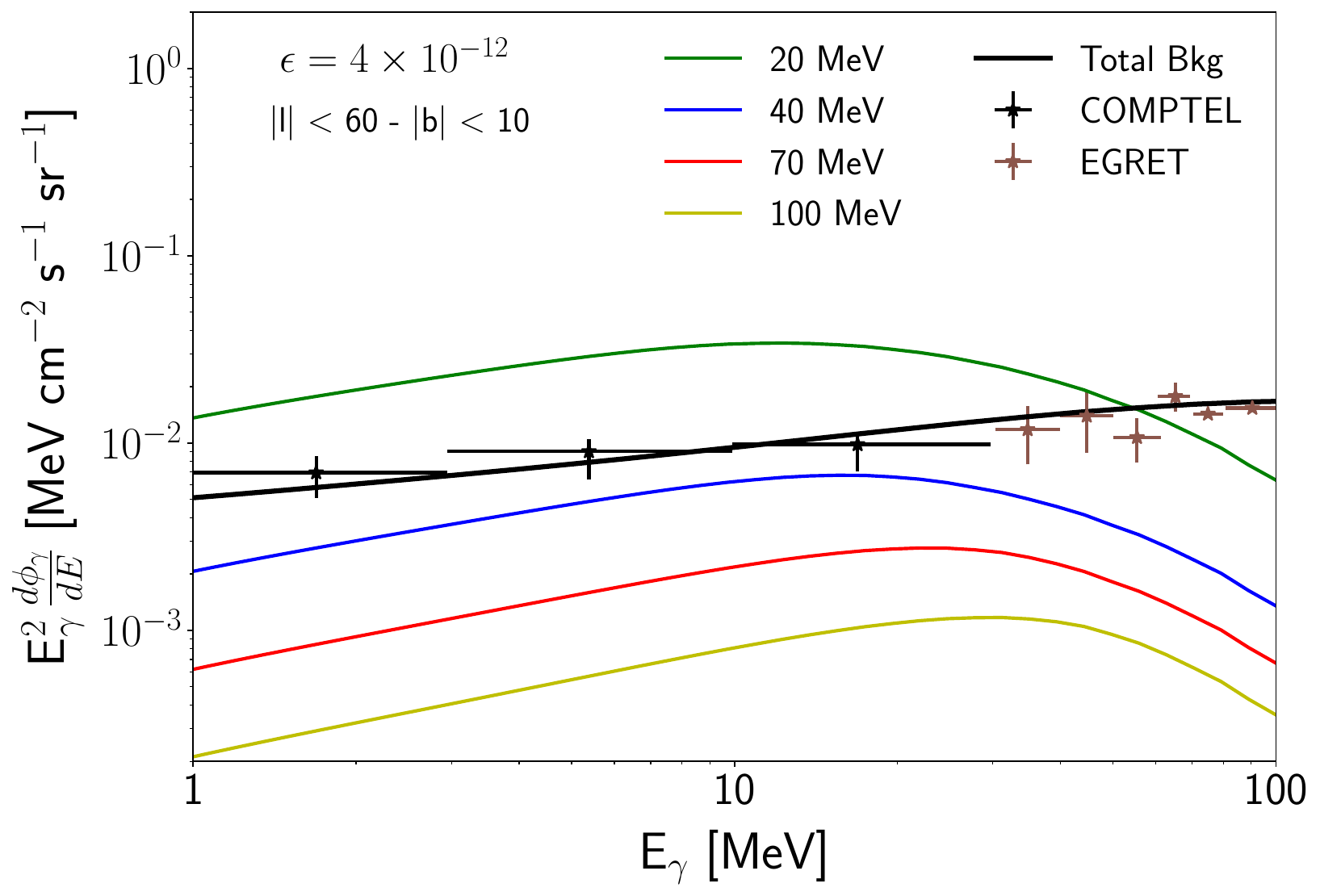}
\caption{IA emission from ALPs coupling to baryons and electrons (left panel) and DPs (right panel) produced in SNe, for several masses of these FIPs. For the ALP scenario, production is set by fixing the ALP-proton coupling at $g_{ap}=2\times10^{-11}$. The predicted signals are compared to COMPTEL~\cite{COMPTEL_6010, COMPTEL1994} and EGRET~\cite{EGRET} measurements of the diffuse $\gamma$-ray flux in the MeV range for a region of interest covering the Galactic plane ($|l| <60^{\circ}$ and $|b|<10^{\circ}$). The black line in both panels refers to the expected Galactic background emission (from Inverse-Compton and bremsstrahlung of CR electrons).} 
\label{fig:Comptel}
\end{figure*}

Given the limited data on low-energy CR electrons and positrons, previous works showed~\cite{Calore:2021klc, Calore:2021lih, DelaTorreLuque:2023huu} that the coupling of FIPs with electrons and positrons can be better probed through the secondary emissions of the positrons injected. The main reason is that CR positrons are not very abundant, since they are mainly produced from interactions of high-energy protons with the ISM gas. One of the most promising emissions to constrain the FIP coupling is the $511$~keV line emission produced when electrons in the ISM encounter the positrons injected by FIPs, once they have become thermal, and form the para-positronium bound state. However, recently, in Ref.~\cite{DelaTorreLuque:2024zsr} we showed that IA emission can be more effective to probe FIP production in the Galaxy. This emission stems from the direct annihilation of ISM electrons with relativistic positrons produced by FIPs and results in a continuum photon emission that peaks around a few tens of MeV, where backgrounds are expected to be very low. 

We calculate the $511$~keV diffuse line emission as we have recently done in Refs.~\cite{DelaTorreLuque:2023cef, DelaTorreLuque:2024zsr, DelaTorreLuque:2024wfz}, where we refer the reader for technical details (in particular, see Eq.~($2$)+ of Ref.~\cite{DelaTorreLuque:2024zsr} and the related discussion). In this calculation, we compute the diffuse $511$~keV line emission from the positrons injected by FIPs as proportional to the thermalized distribution of positrons which we assume to follow the steady-state solution for the injected positrons.
In addition, to account for the quick decrease in the free-electron density above and below the disk of the Galaxy we apply a scaling relation to the $511$~keV profiles, following the vertical distribution of free-electron density in the Galaxy, adopting the NE2001 model~\cite{Cordes:2003ik, Cordes:2002wz}. 

As discussed in Refs.~\cite{Calore:2021lih, DelaTorreLuque:2023nhh}, the morphology of the expected $511$~keV line emission from the FIP positrons cannot explain the high emission around the center of the Galaxy, leading instead to a flatter profile following the Galactic disk emission, as can be seen from Fig.~5 of Ref.~\cite{DelaTorreLuque:2023huu}.

In the case of the IA emission, we calculate this contribution following our previous work (see Eq.~(2) of Ref.~\cite{DelaTorreLuque:2024zsr}), that adopts a similar strategy as in Ref.~\cite{Beacom:2005qv}. This strategy consists of first evaluating the integrated $511$~keV line emission in a region of interest and then computing the continuum IA flux via the measured ratio of the para-positronium emission to IA emission at $511$~keV. In this way, the energy dependence of the IA emission depends on the probability of relativistic positrons interacting with a free-electron (described by the usual Dirac cross sections~\cite{Dirac}) and the time that a positron of a certain energy spends in the Galaxy before thermalizing (i.e. the energy loss rate), while the normalization is fixed by the $511$~keV emission in that region of interest. Full details are given in our companion work~\cite{DelaTorreLuque:2024zsr} as well as in a related recent work~\cite{DelaTorreLuque:2024wfz}.

Fig.~\ref{fig:Comptel} shows the IA emission from ALPs coupling to baryons and electrons (left panel) and DPs (right panel) produced in SNe, for several masses of these FIPs, and compared to COMPTEL~\cite{COMPTEL_6010, Comptel, COMPTEL1994} and EGRET~\cite{EGRET} measurements of the diffuse $\gamma$-ray flux in the MeV range. This region of interest provided the best constraints on the sterile neutrino mixing angle in Ref.~\cite{DelaTorreLuque:2024zsr}, among the few regions where COMPTEL data of the diffuse Galactic emission is available. This is because this is an extended region that covers the Galactic plane, where SNe can inject these particles. The morphology of the IA signals from ALPs and DPs is similar, although we note that for DPs this emission peaks around $20-30$~MeV, while the emission from ALPs peaks at slightly higher energies, around $\sim60$~MeV. These features suggest that COMPTEL can be considered a valuable experiment to look for these kinds of signals. Above $\sim60$~MeV we start to have a contribution from $\pi^0$ decays, making it more difficult to spot any signature from IA emission. Conversely, we note that EGRET is provided with very low spatial resolution, and it is likely to have an important contribution from unresolved sources in their observations of the diffuse Galactic flux. Therefore, we use EGRET data just for illustrative purposes and make use of COMPTEL data in the region $|l| <60^{\circ}$ and $|b|<10^{\circ}$ to derive constraints.
In this figure, we also show the expected background Galactic emission, which mainly consists of IC emission and bremsstrahlung  from CR electrons, which are the dominant $\gamma$-ray production processes below energies of a few hundred MeV. For this background, we use the electron model from Refs.~\cite{delaTorreLuque:2022vhm, DelaTorreLuque:2023zyd}, which is optimized to reproduce the electron and positron emission at Earth location, as well as the local $\gamma$-ray emissivity measured by the Fermi-LAT telescope, down to a few tens of MeV. We have checked that this model shows strong agreement with the data down to $100$~keV.

To better understand the morphology of the IA signals from different FIP models, we compare them in Fig.~\ref{fig:FIP_IA}. Here, we show in blue, the case of a sterile neutrino with coupling to $\tau$ particles $|U_{\tau-\nu_{4}}|^2 = 10^{-11}$, in green a DP with coupling $\epsilon = 2.5\times 10^{-11}$, in yellow an ALP with $g_{ae}=3\times 10^{-18}$ and $g_{ap}=2\times10^{-11}$  and, in maroon, a general FIP particle (described by Eq.~\eqref{eq:spectrum}) injecting $5.5\times10^{54}$ positrons per SN. These coupling values have been chosen to normalize them close to data in the region of COMPTEL measurements ($|l| <60^{\circ}$ and $|b|<10^{\circ}$) and refer to a $10$~MeV mass. In addition, for comparison, we include the signal from a source of monoenergetic positrons (with energy of $10$ MeV), as derived in Ref.~\cite{Beacom:2005qv}. This allows us to illustrate why the IA emission can be so constraining for FIPs: given that the energy of the positrons from these FIPs is always above a few tens of MeV, their IA emission is still very high up to these energies, where the $\gamma$-ray backgrounds at around $10$-$50$~MeV are very low, mainly produced from IC emission of CR electrons. In fact, the high IA emission from high-energy positrons (above a few MeV) served to exclude heavy dark matter particles as the origin of the anomalous $511$~keV line emission in Ref.~\cite{Beacom:2005qv} (but see also Refs.~\cite{DelaTorreLuque:2024wfz, DelaTorreLuque:2023cef}).
Furthermore, positrons from FIPs produced in SNe have energies close to $\sim100$~MeV, where the cross sections for this process is more efficient. In addition, at these energies, the positrons still propagate for very long times, increasing the probability of interacting with electrons in the ISM.

\begin{figure}[t!]
\includegraphics[width=0.48\textwidth]{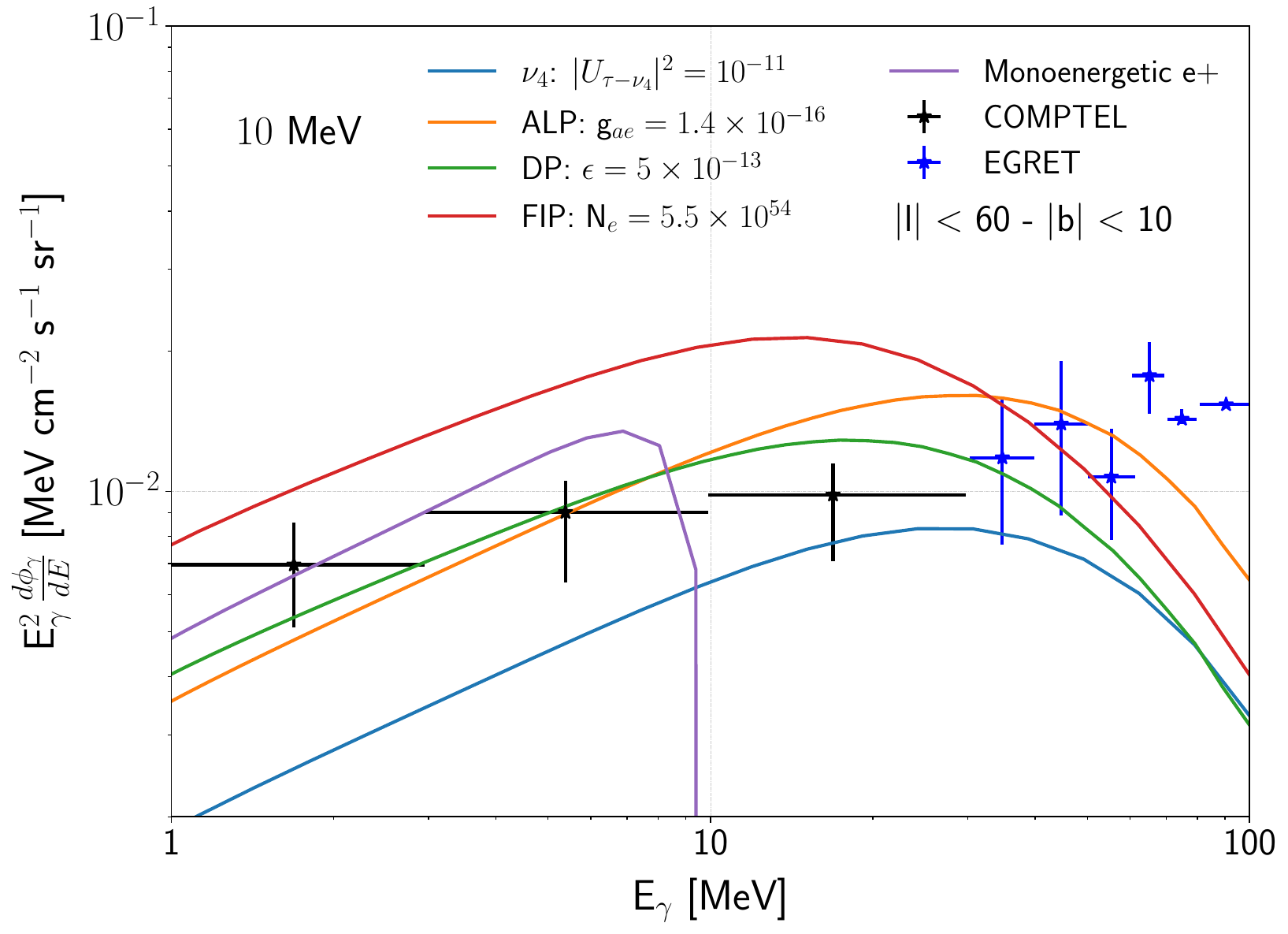}
\caption{Comparison of the morphology of the IA signals produced from different FIP models, in the region at $|l| <60^{\circ}$ and $|b|<10^{\circ}$, and for a mass of $10$~MeV. We show in blue, the case of a sterile neutrino with coupling to $\tau$ particles (for $|U_{\tau-\nu_{4}}|^2 = 10^{-11}$), in green a DP with coupling $\epsilon = 2.5\times 10^{-11}$, in yellow an ALP with $g_{ae}=3\times 10^{-18}$ and $g_{ap}=2\times10^{-11}$ and in maroon, a general FIP particle (described by Eq.~\eqref{eq:spectrum}) injecting $5.5\times10^{54}$ positrons per SN. In addition, for comparison, we include the signal from a source of monoenergetic positrons (with energy of $10$ MeV).} 
\label{fig:FIP_IA}
\end{figure}

To estimate the effect of uncertainties in propagation, we adopt the same strategy as used in Refs.~\cite{DelaTorreLuque:2023olp, DelaTorreLuque:2024qms}, where  we adopted two extreme diffusion scenarios, a more aggressive one that leads to stronger constraints and a very conservative one that leads to weaker constraints.  
As we showed in Refs.~\cite{DelaTorreLuque:2023olp, DelaTorreLuque:2023nhh}, the propagation parameters with a greater effect on the diffuse spectra produced from FIPs are the Alfvén velocity ($V_A$ parameter) that controls the level of diffuse reacceleration~\cite{1995ApJ...441..209H, Drury_1, seo1994stochastic, osborne1987cosmic} and the height ($H$) of the halo, which dictates the volume where CRs are confined and where FIPs produce electrons and positrons that can reach us. In the conservative setup, we set $H = 3$~kpc and $V_A=0$~km/s (i.e. no reacceleration), which produces a lower (and therefore more conservative) flux of $e^\pm$. In turn, the more aggressive setup is meant to increase the flux of $e^\pm$ from the decay of FIPs, and uses values of $H = 16$~kpc and $V_A=40$~km/s. These are at the extreme values within which we can still reproduce the current CR measurements, although they are not statistically favored~\cite{Weinrich_halo, delaTorreLuque:2022vhm, Luque:2021nxb, Evoli:2019wwu} \footnote{We consider that $V_A=40$~km/s is the maximum realistic value for $V_A$, since it already implies that most of the injected energy of CRs are coming from the perturbations of the interstellar plasma and not from supernova remnants, which will break the standard paradigm of CR propagation, see Ref.~\cite{Drudy_VA_Energetics, Drury_1}}
For comparison, we recall that our benchmark values are $H = 8$~kpc and $V_A=13.4$~km/s.
In the left panel of Fig.~\ref{fig:Uncerts}, we show the uncertainty in the prediction of the IA emission from ALPs couplings to baryons and electrons for different masses. Each band represents the uncertainty in our estimated flux directly related to uncertainties in the propagation of CRs. As we will discuss later, these uncertainties translate into a factor of a few uncertainty in the constraint of the FIP coupling. We observe similar uncertainties for all the FIP models that we are studying in this paper.

\begin{figure*}[t!]
\includegraphics[width=0.47\textwidth]{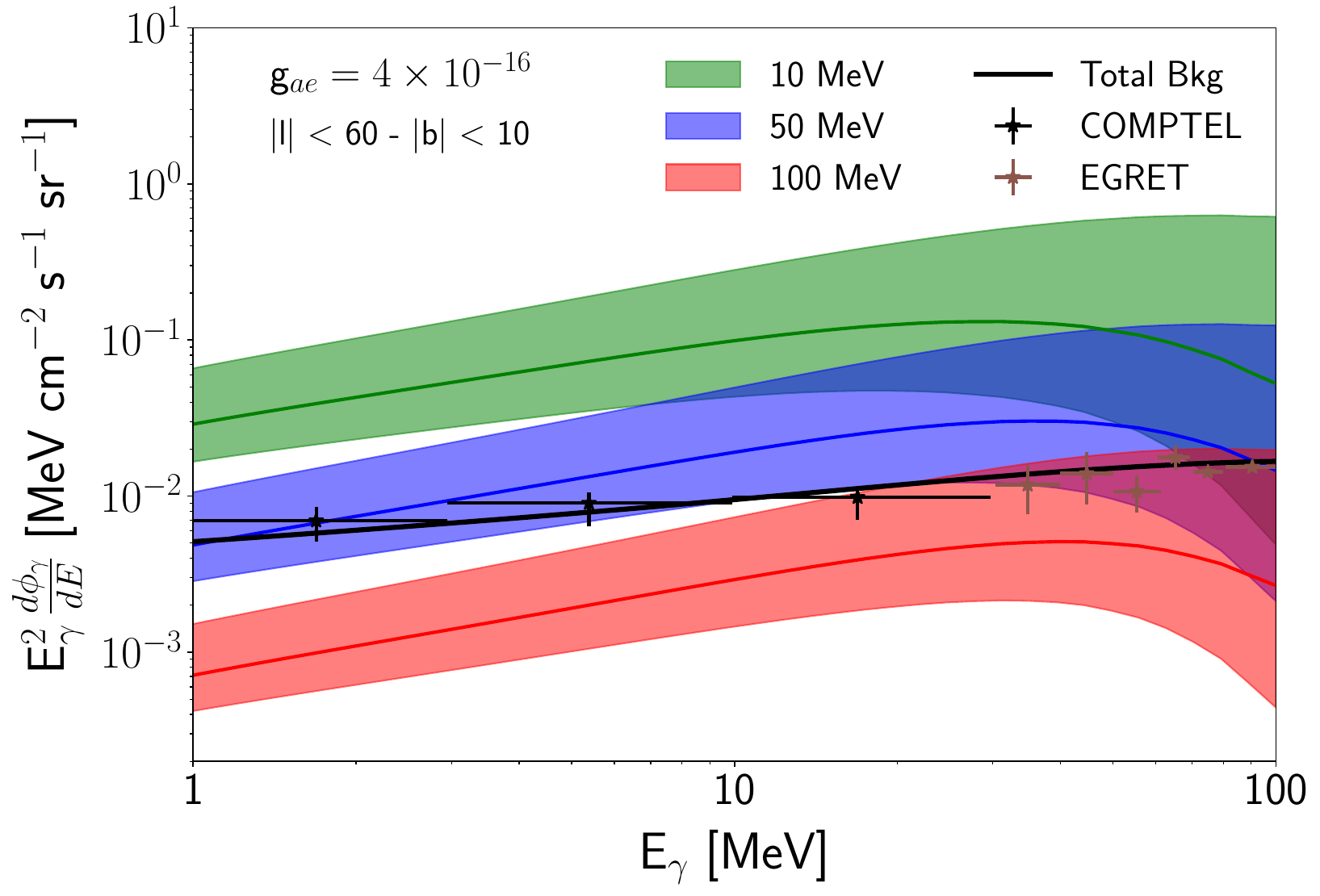}
\includegraphics[width=0.5\textwidth, height=0.242\textheight]{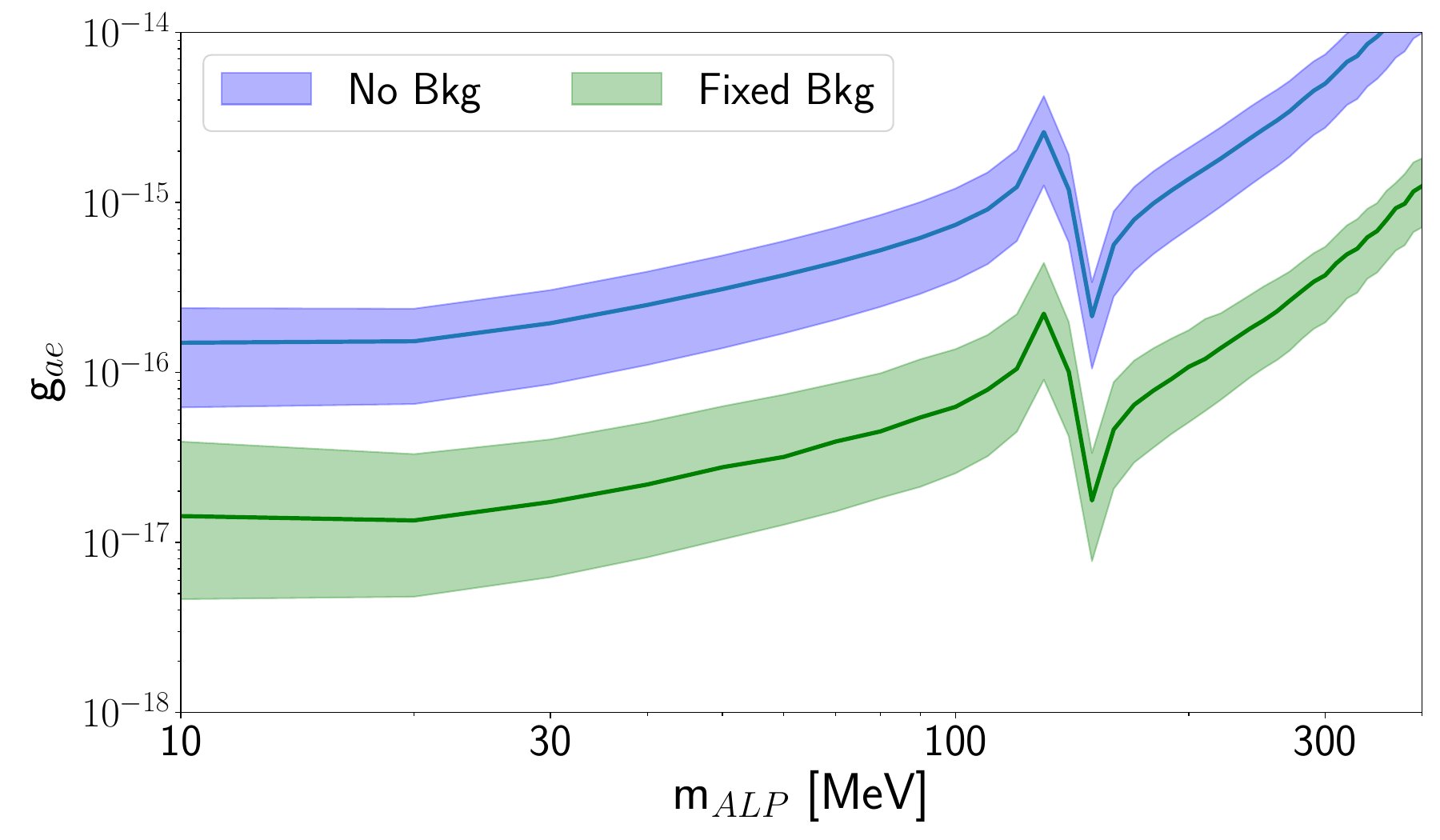}
\caption{Effect of propagation uncertainties on the predicted IA signals from ALPs produced in SNe (left panel) for a few different masses. For completeness, we also show here our reference background model (model prediction from Ref.~\cite{delaTorreLuque:2022vhm}) and COMPTEL and EGRET measurements. The right panel shows the effect of propagation uncertainties on the constraints (at $2\sigma$ confidence level) on the ALP coupling, in the conservative case, in which we do not include any background emission (``No Bkg''), and in the case in which we  account for background when deriving these limits labeled as ``Fixed Bkg'' in the legend of the figure.} 
\label{fig:Uncerts}
\end{figure*}

\subsection{Analysis of FIP $\gamma$-ray signals and constraints}
As mentioned above, we derive $2\sigma$ constraints on the coupling for the different FIP models by comparing their IA emission with the measured Galactic $\gamma$-ray diffuse emission measured by COMPTEL in the $|l| <60^{\circ}$ and $|b|<10^{\circ}$ region of the sky. This direct comparison gives us a first very conservative constraint. However, on top of the positron-induced $\gamma$-ray continuum signals, the Galactic IC and bremsstrahlung emission from CR electrons are expected to be the dominant source of $\gamma$-ray production below energies of a few hundreds of MeV. Therefore, we also derive more realistic constraints for different estimations of the background model. In a first case, we model the background as a simple power-law and get the constraint comparing the sum of the background plus the IA FIP signals. In a second case, we make use of a more refined background model. In particular, we use the electron model from Refs.~\cite{delaTorreLuque:2022vhm, DelaTorreLuque:2023zyd}, which is optimized to reproduce the electron and positron emission at Earth location, as well as the local $\gamma$ ray emissivity, down to a few tens of MeV. We have checked that this model of the $\gamma$-ray background emission shows a notable agreement with data down to $100$~keV, for different regions of interest. We will refer to this model as our reference model. Using this reference model, which is shown as a black line in Fig.~\ref{fig:Comptel}, we compute constraints not only for the case of a fixed background model, but also leaving the normalization of the background emission free in our analysis. Therefore, in addition to the conservative limits without any background, we obtain three other constraints: one assuming a power-law background, another one from the predicted background model explained above and a last one letting the normalization of this background model be free. 
\begin{figure}[h!]
\includegraphics[width=0.5\textwidth]{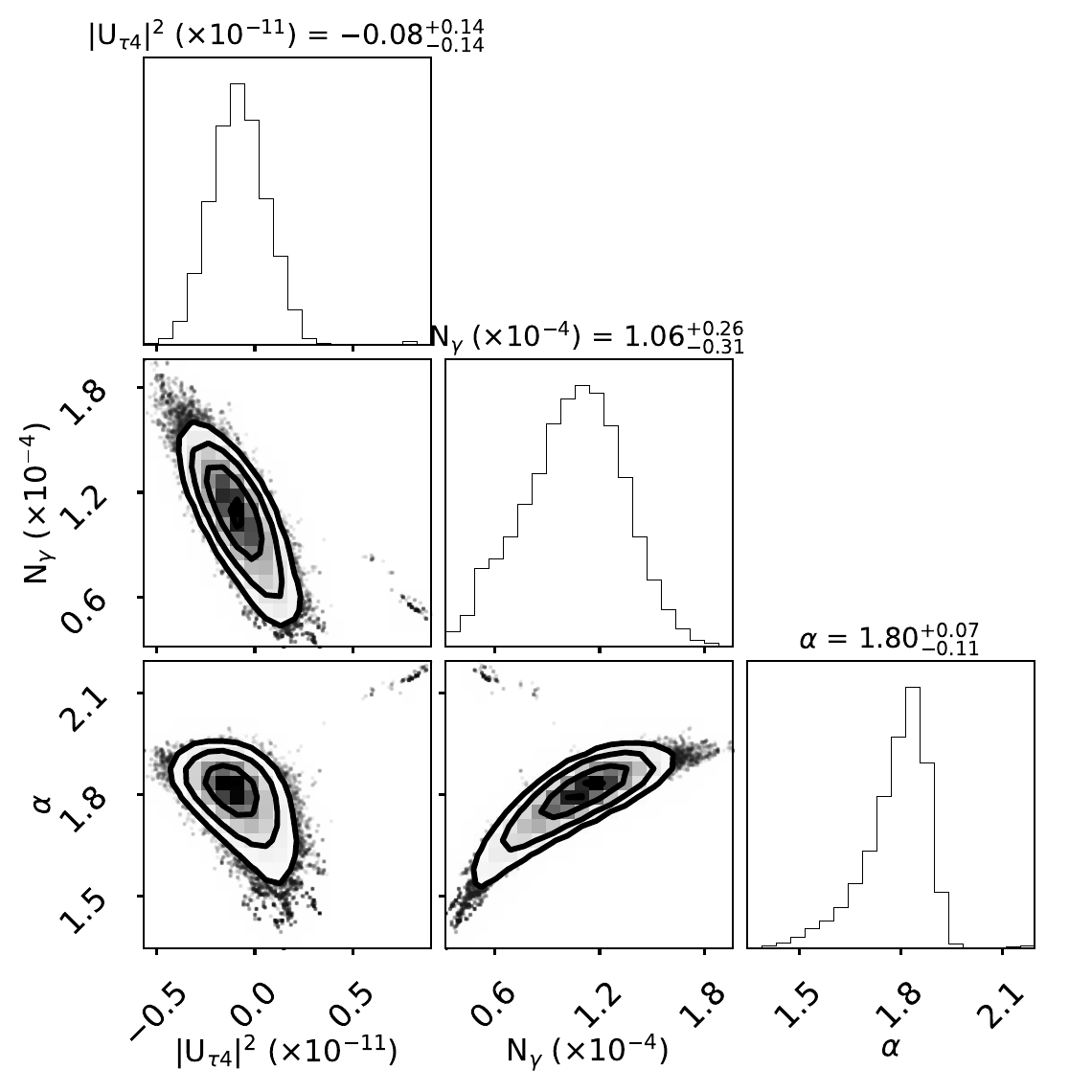}
\caption{Example of the probability distribution and credible intervals for the parameters included in the analysis of an IA signal from a $30$~MeV sterile neutrino when adding the background contribution described as a power-law (with free parameters the normalization at $10$ MeV, ``N$_\gamma$'', in units of MeV cm$^{-2}$ s$^{-1}$ sr$^{-1}$, and spectral index, ``$\alpha$''). 
On top of each panel, the best-fit values and their $1\sigma$ uncertainty are shown.} 
\label{fig:IA_Anal}
\end{figure}

To perform these fits and derive the constraints we rely on the Markov Chain Monte Carlo (MCMC) package \textit{Emcee}~\cite{emcee}, that is based on Bayesian inference, since this technique is more robust than conventional optimizers and less prone to finding false local minima. This analysis provides the probability distribution functions for every parameter to reproduce the data, and, thus, we obtain, as an output, the credible (confidence) intervals of each parameter included in the fit (normalization of the background, spectral index, coupling of the FIP, etc.). 
Having the credible intervals for each parameter, we obtain the limits at the $95\%$ confidence level on the FIP coupling.
A similar setup has been successfully used in different studies involving dark matter limits from antiprotons~\cite{DelaTorreLuque:2024ozf, Luque:2021ddh}, line searches in $\gamma$-ray data~\cite{DeLaTorreLuque:2023fyg} or even to study propagation of CRs~\cite{Luque:2021nxb}.
\begin{figure*}[t]
\mbox{\includegraphics[width=0.5\textwidth]{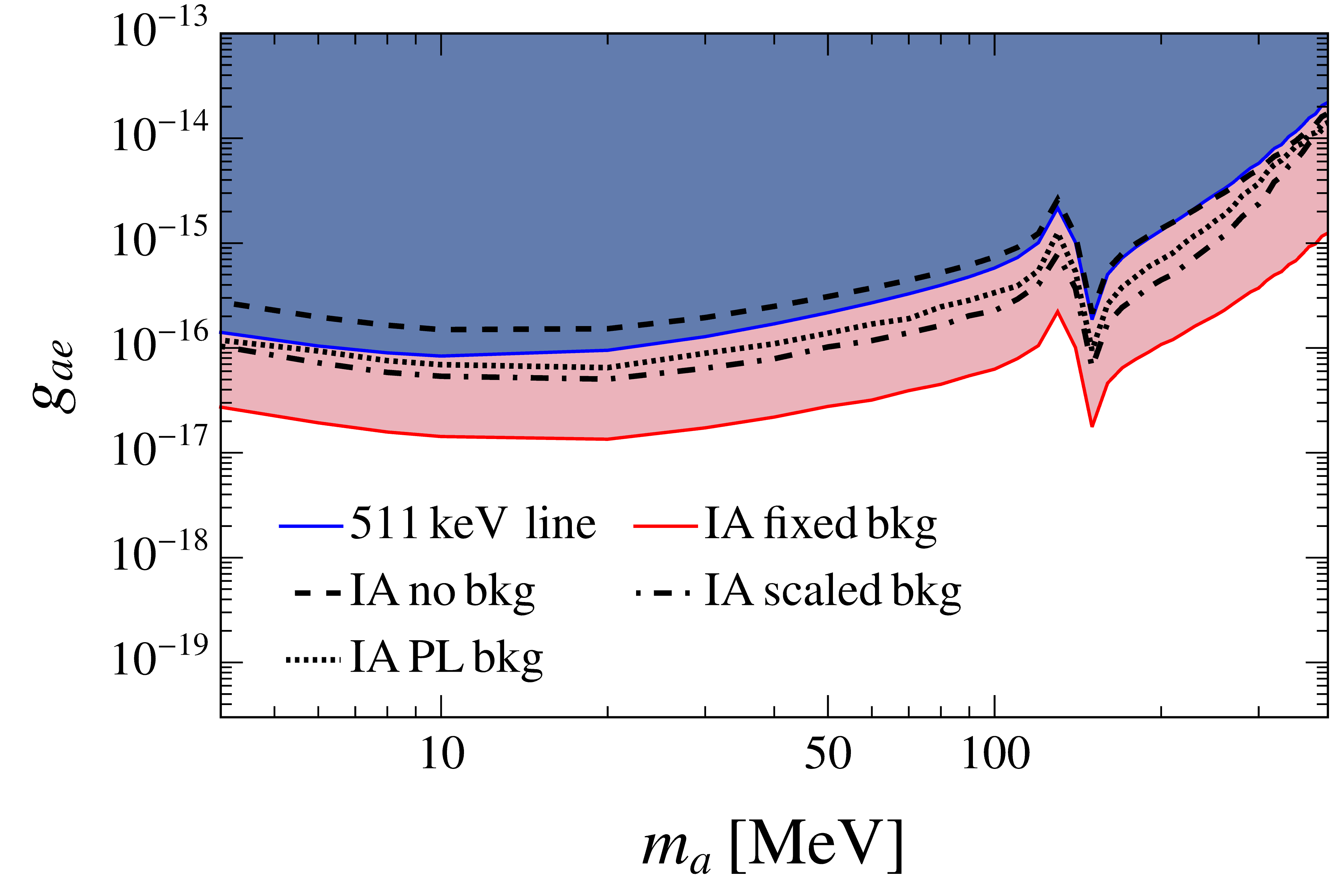}
\includegraphics[width=0.5\textwidth]{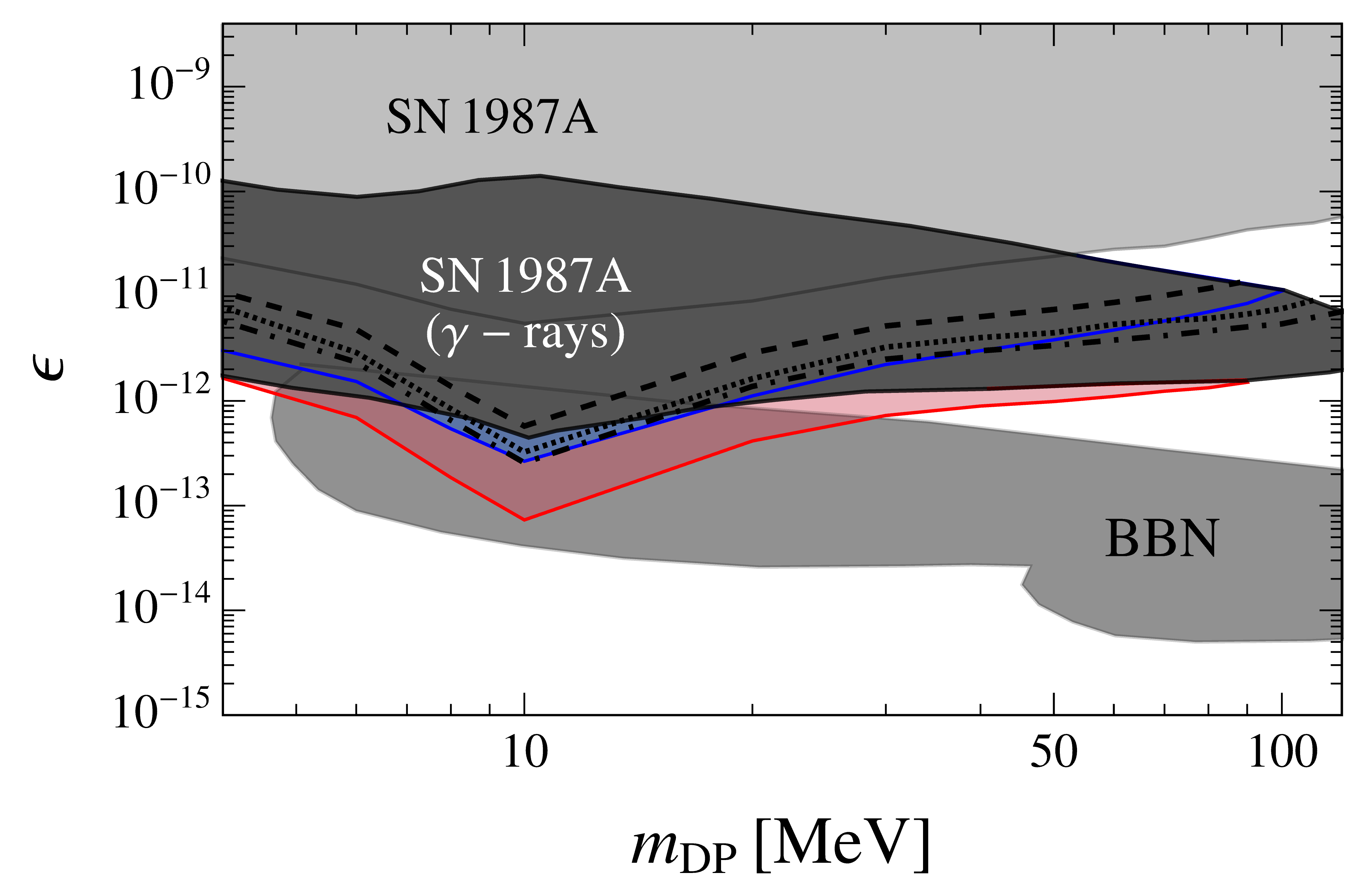}}
\mbox{\includegraphics[width=0.5\textwidth]{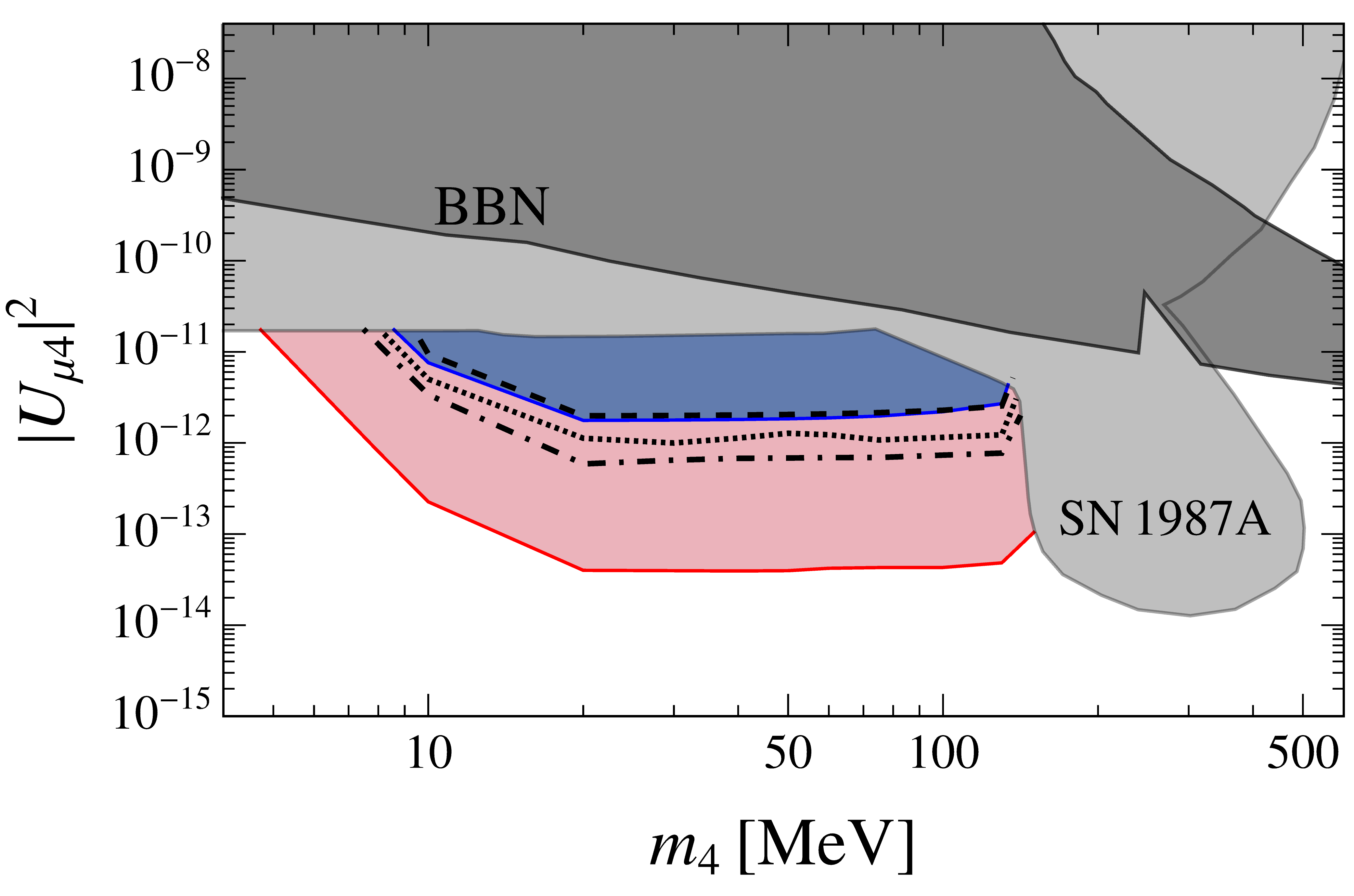}
\includegraphics[width=0.5\textwidth]{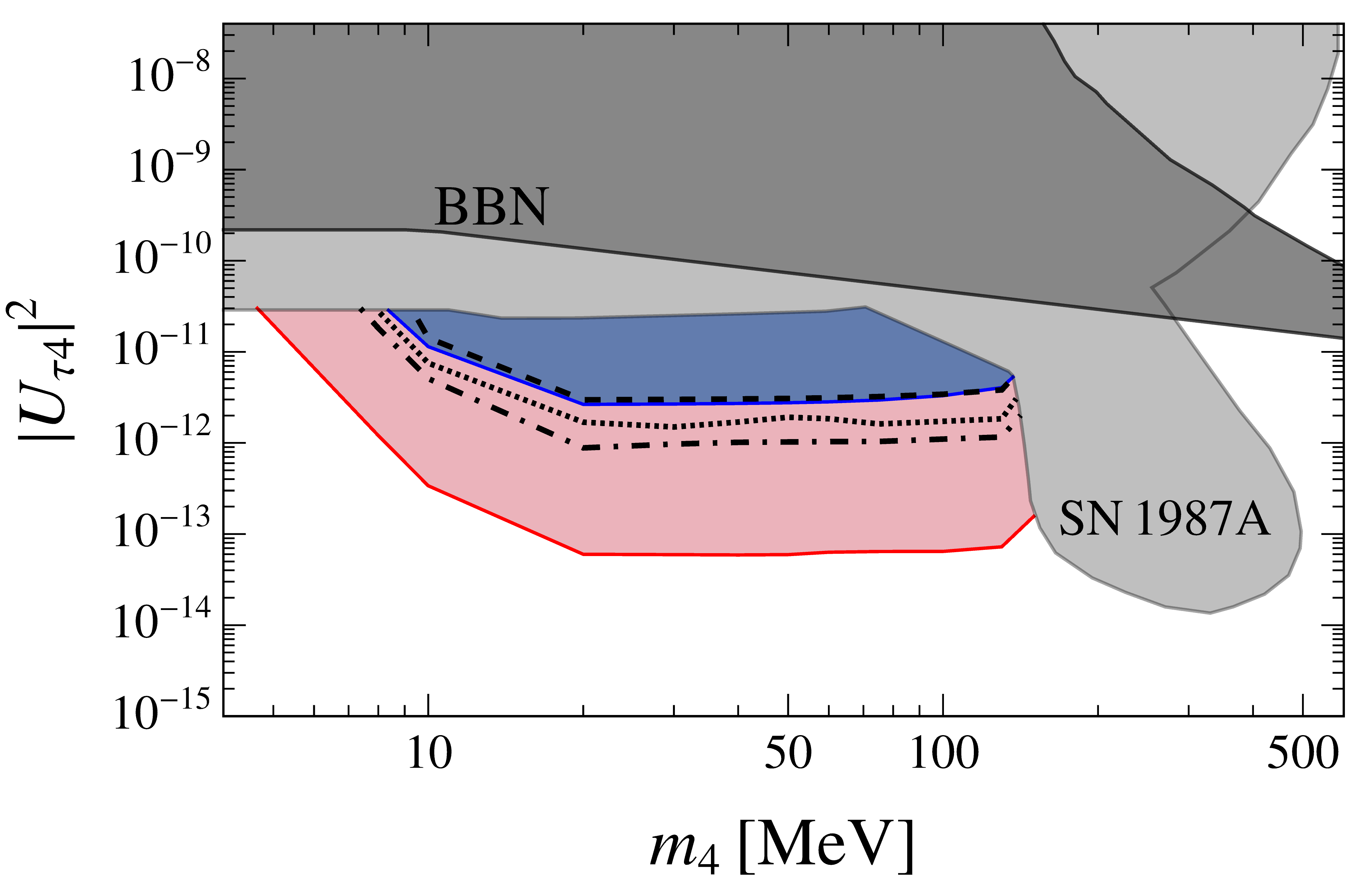}}
\caption{ Upper limits, at $2\sigma$ confidence level, for the case of ALPs coupled to nucleons and electrons with $g_{ap}=2\times10^{-11}$ (upper left panel), DPs (upper right panel), and sterile neutrinos (lower panels). We show the limits obtained in the different analyses performed: Including our background model (``fixed bkg''), with the same background model but letting it's normalization be free in the fit (``scaled bkg''), including a background described as a simple power-law (``PL bkg'') and, the most conservative case, without accounting for any background (``no bkg''). For completeness, we also include the limit obtained from the analysis of the $511$~keV emission. SN 1987A and BBN constraints on the DP parameter space have been introduced in Refs.~\cite{Chang:2016ntp,DeRocco:2019njg,Sung:2019xie} and Refs.~\cite{Redondo:2008ec,Fradette:2014sza,Li:2020roy}, respectively. Analogously, other bounds on the sterile neutrino parameter space are placed by observations of SN 1987A~\cite{Carenza:2023old} and cosmological constraints~\cite{Sabti:2020yrt,Boyarsky:2020dzc,Mastrototaro:2021wzl}.}
\label{fig:All_Lims}
\end{figure*}

Firstly, we investigated if there is any evidence for IA signals needed on top of the background, to reproduce COMPTEL data. In all the background configurations and FIP models, we found that the IA signals do not lead to a significant improvement of the fit to COMPTEL data, meaning that the background suffices to explain these observations. In the case of the adopted reference background model, we find that the best-fit normalization is always within $10\%$ of the original prediction. As an example, we illustrate, in the form of a triangle plot, the posterior distributions obtained in the case of the signal from a $30$~MeV sterile neutrino assuming a power-law background, in Fig.~\ref{fig:IA_Anal}. See more details in the label of the figure. In this example, one can see that the best-fit coupling of the signal is consistent with no coupling, but the upper limit (at $2\sigma$ confidence) still allows non-negligible couplings.

The right panel of Fig.~\ref{fig:Uncerts} shows the limits obtained for ALPs coupling to baryons and electrons in the conservative case (i.e. including no background -- ``No Bkg'') and in the most optimistic case (obtained with our reference background -- ``Fixed Bkg''). The bands correspond to the uncertainties associated with the propagation parameters, as discussed previously. Referring to the discussion in Sec.~\ref{subsec:NucProduction}, we warn the reader about the behavior of the constraints around the range $m_a\simeq135\,\mathrm{MeV}$ and $m_a \simeq147\,\mathrm{MeV}$, where resonant effects on the induced ALP-photon coupling could affect the results of the analysis. In particular, around $m_a\simeq135\,\mathrm{MeV}$ positron injection is suppressed since all the ALPs tend to decay in photons. Thus, our constraints relax. Conversely, around $m_a\simeq147\,\mathrm{MeV}$ electronic decays are enhanced by the suppression of the induced ALP-photon coupling and our limits get more stringent.
As we see, the difference in the limits obtained when accounting for the background contribution is much higher than the uncertainties in the limits associated with the propagation parameters.

In Fig.~\ref{fig:All_Lims} we show the limits derived for all the FIP models discussed in this work. Here, we also include, for comparison, the limit obtained from the $511$~keV emission (shown in all panels as a blue shaded region) produced by FIPs, similar to what was done in our companion work~Ref.\cite{DelaTorreLuque:2024zsr}. The $511$~keV emission at the disk from astrophysical sources is expected to dominate the emission away from the bulge and including this emission would lead to much stronger constraints. However, understanding the $511$~keV background emission is much beyond our scope and currently there is no such accurate model that we can include in our analysis to obtain stronger limits~\cite{DelaTorreLuque:2024wfz}. Therefore, we limit ourselves to extract conservative constraints as a reference. 

In the upper left panel of Fig.~\ref{fig:All_Lims} we consider an ALP coupling to baryons and electrons, for a fixed nucleon coupling of $g_{ap}=2\times 10^{-11}$, while the upper-right panel refer to the DP case. Finally, lower panels display our results for the case of sterile neutrinos coupling with muons and tauons.
Moreover, we consider scenarios with several background models. This includes a standard background model (“fixed bkg” shown as a red solid curve), with the same background model but letting its normalization be free in the fit (“scaled bkg” shown as a black dot-dashed curve), including a background described as a simple power-law (“PL bkg” shown as a black dotted curve) and, the most conservative case, without accounting for any background (“no bkg” shown as a black dashed curve). For completeness, we also include the limit obtained from the analysis of the 511 keV emission. We consider FIP masses from 4 MeV up to a few hundreds of MeV in all the cases considered. We observe, for every model the constraints from the IA emission including background are always stronger than those from the $511$~keV line, which has recently been recognized to be one of the strongest astrophysical observables to probe the electron coupling of FIPs produced in SNe. In particular, we observe that, including the background emission predicted from  Refs.~\cite{delaTorreLuque:2022vhm, DelaTorreLuque:2023zyd}, IA can rule out $U_{\mu4}\gtrsim4\times10^{-14}$ and $U_{\tau4}\gtrsim6\times10^{-14}$ for all sterile neutrinos with masses $10\,\mathrm{MeV}\lesssim m_a\lesssim150\,\mathrm{MeV}$. This region of the parameter space nicely complements the SN cooling constraint set in Ref.~\cite{Carenza:2023old}. In the case of DP with masses $5\,\mathrm{MeV}\lesssim m_{\rm DP}\lesssim 50\,\mathrm{MeV}$, constraints induced by the non-observation of any signature related to IA can reinforce by a factor a few astrophysical limits set by the non-observation of $\gamma$-rays in coincidence to SN1987A~\cite{Chang:2016ntp,Sung:2019xie,DeRocco:2019njg}~(see also constraints from fireball formation in coincidence with the GW170817 neutron star merger event introduced in Ref.~\cite{Diamond:2021ekg}), excluding mixing parameters down to $\epsilon\gtrsim 7\times 10^{-14}$ when including background and down to $\epsilon\gtrsim 6 \times 10^{-13}$ without background. In this context, IA annihilation has to be considered only a complementary probe of this region of the DP parameter space which is extensively excluded by cosmological surveys~\cite{Redondo:2008ec,Fradette:2014sza,Li:2020roy}. Finally, the top-left panel displays the case of ALPs coupled to both nucleons and electrons. By fixing the ALP-proton coupling at $g_{ap}=2\times10^{-11}$ the analysis of IA annihilation signatures allows us to excluded ALP electron couplings down to $g_{ae}\lesssim10^{-17}$ for $10\,\mathrm{MeV}\lesssim m_a \lesssim 100\,\mathrm{MeV}$. Again here, we note the presence of resonant effects around $135$-$147$~MeV, explained above already. 
In order to make a direct comparison to bounds present in the previous literature, in Fig.~\ref{fig:ALPs_gapvsgae} we show the IA bounds for ALPs in the $g_{ae}$ vs $g_{ap}$ plane for a fixed and representative ALP mass of 100 MeV. The red hatched regions indicate areas excluded due to the positron IA arguments we make in this work, while the blue regions are excluded based on the non-observation of the 511 keV line from the Galactic center. The hatched areas represent the uncertainty bands for these limits, ranging from the most conservative to the most optimistic scenarios discussed in Ref.~\cite{DelaTorreLuque:2024zsr}. Additionally, the gray regions show areas excluded by other arguments, such as the SN cooling bound and the loop-induced $\gamma$-decay bound on $g_{ae}$, as discussed in Ref.~\cite{Ferreira:2022xlw}. Furthermore, we include constraints from energy deposition in the SN envelope and the 511-keV limit from extra-Galactic SNe, as introduced in Ref.~\cite{Lella:2022uwi}. We note that even the most conservative scenario we consider is stronger than the 511 keV line for this 100 MeV ALP and excludes a region of parameter space that is complimentary to other aforementioned constraints, particularly at nucleon couplings $g_{ap}\lesssim 10^{-11}$ and over electron couplings $10^{-18}\lesssim g_{ae} \lesssim 10^{-11}$. This underscores the important of IA at ``higher" masses to probe ALP parameter space.

\begin{figure}[t]
\includegraphics[width=0.5\textwidth]{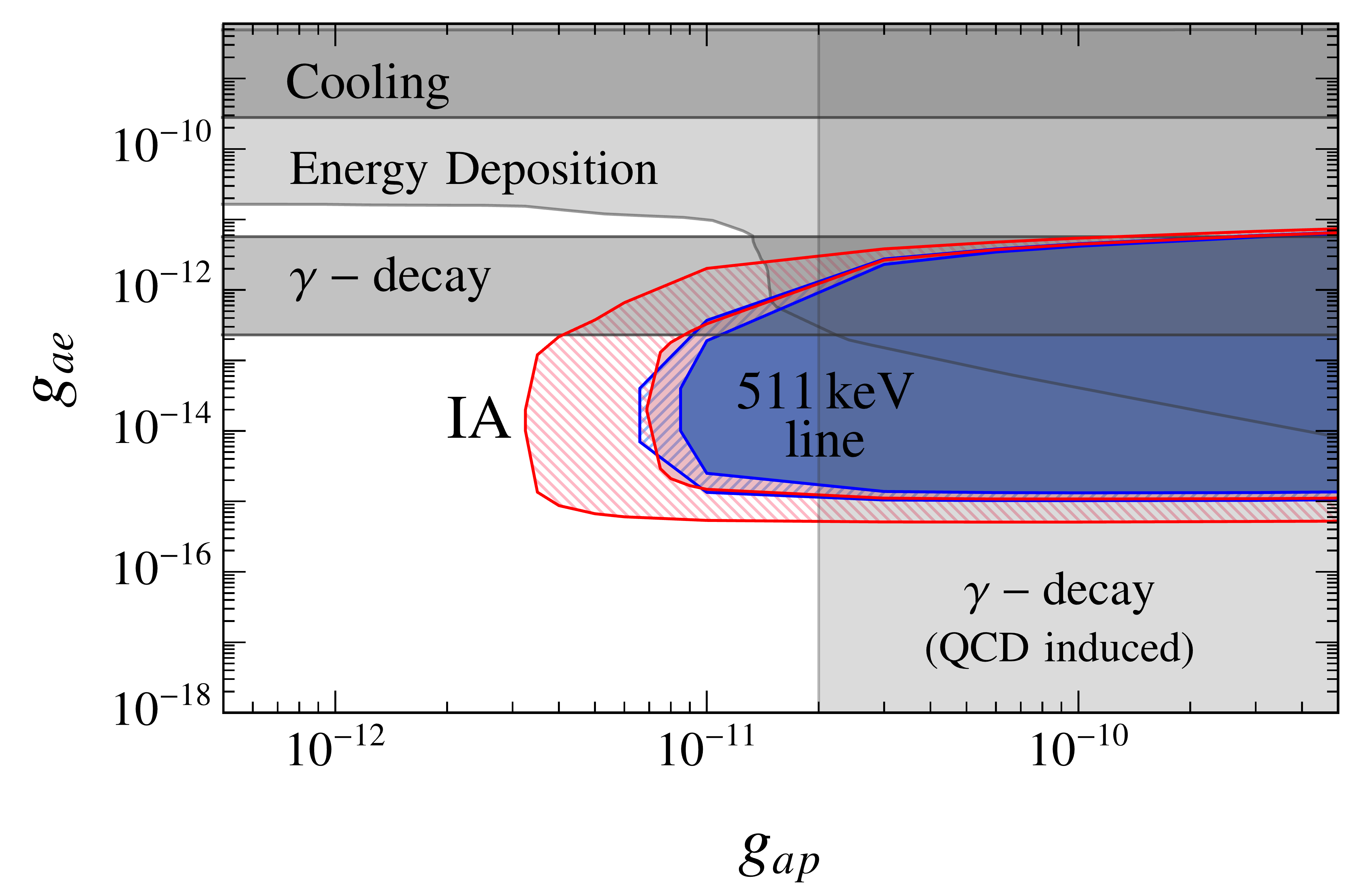}

\caption{IA annihilation bounds in the $g_{ae}$ vs $g_{ap}$ plane for an ALP mass $m_a=100\,$MeV. We report in red, regions of the parameter space excluded by the arguments related to positron IA, while blue regions are excluded by the non-observation of the 511 keV line. Hatched regions display the uncertainty bands for these limits, ranging from the most conservative to the most optimistic cases discussed in Ref.~\cite{DelaTorreLuque:2024zsr}. We also report in gray, regions of the parameter space excluded by other arguments, namely the SN cooling bound and the loop-induced $\gamma$-decay bound on $g_{ae}$ discussed in Ref.~\cite{Ferreira:2022xlw}. Moreover, we also display constraints from energy deposition in the SN envelope~\cite{Caputo:2022mah}, adapted from Ref.~\cite{Lella:2022uwi}, and the $\gamma$-decay constraint related to the ALP-photon coupling induced by QCD effects, taken from Ref.~\cite{Lella:2024dmx}.} 
\label{fig:ALPs_gapvsgae}
\end{figure}

\section{Discussion and conclusion}
\label{sec:Conclusion}

In this work, we discuss constraints on electrophilic Feebly Interacting Particles (FIPs) produced in Galactic Supernova (SN) explosions during the Milky Way's lifetime. FIPs with masses up to hundreds of MeV are efficiently produced in SN explosions and escape the star, subsequently decaying into electron-positron pairs. Injecting energetic leptons in the Galaxy produces secondary photon fluxes that might be detectable and shed light on the nature and properties of FIPs. Building upon the study conducted in Ref.~\cite{DelaTorreLuque:2024zsr}, we discuss the IA signal produced when energetic FIP-produced positrons encounter the free-electrons in the interstellar medium, which produces a peculiar $\gamma$-ray signal at tens of MeV. In particular, we analyzed possible signatures related to positron IA induced by the SN emission of axions (coupled to electrons and/or nucleons), sterile neutrinos and dark photons from COMPTEL/EGRET measurements of the diffuse $\gamma$-ray background. 

We observe that the IA bound is the most constraining astrophysical probe for MeV-scale axions coupled to nucleons and electrons, reaching extremely small electron couplings that cannot be probed otherwise. As already pointed out in Ref.~\cite{DelaTorreLuque:2024zsr}, sterile neutrinos mixed with mu and tau flavors can be probed down to $|U_{x4}|^{2}\sim10^{-13}$ for $x=\{\mu,\tau\}$, making IA bounds the strongest ones in the $10$-$200$~MeV mass range. Also in the case of DPs, the IA phenomenology allows one to cover a large portion of the parameter space, providing complementary constraints to cosmological probes.

In conclusion, we show that IA can be considered a novel and powerful $\gamma$-ray emission mechanism, making it an important tool for probing new physics and placing stringent constraints over a wide range of positron emitters.

\acknowledgements
\section{Acknowledgements}
This article/publication is based upon work from COST Action COSMIC WISPers CA21106, supported by COST (European Cooperation in Science and Technology). SB is supported by the Science and Technology Facilities Council (STFC) under grants ST/X000753/1 and ST/T00679X/1. The work of PC is supported by the European Research Council under Grant No.~742104 and by the Swedish Research Council (VR) under grants  2018-03641, 2019-02337 and 2022-04283.
PDL is supported by the Juan de la Cierva JDC2022-048916-I grant, funded by MCIU/AEI/10.13039/501100011033 European Union ``NextGenerationEU"/PRTR. The work of PDL is also supported by the grants PID2021-125331NB-I00 and CEX2020-001007-S, both funded by MCIN/AEI/10.13039/501100011033 and by ``ERDF A way of making Europe''. PDL also acknowledges the MultiDark Network, ref. RED2022-134411-T. This project used computing resources from the Swedish National Infrastructure for Computing (SNIC) under project Nos. 2021/3-42, 2021/6-326, 2021-1-24 and 2022/3-27 partially funded by the Swedish Research Council through grant no. 2018-05973.
The work of LM is supported by the Italian Istituto Nazionale di Fisica Nucleare (INFN) through the ``QGSKY'' project and by Ministero dell'Universit\`a e Ricerca (MUR).
The work of AL was partially supported by the research grant number 2022E2J4RK ``PANTHEON: Perspectives in Astroparticle and
Neutrino THEory with Old and New messengers" under the program PRIN 2022 funded by the Italian Ministero dell’Universit\`a e della Ricerca (MUR). 
This work is (partially) supported
by ICSC – Centro Nazionale di Ricerca in High Performance Computing. For the purpose of open access, the authors have applied a Creative Commons Attribution (CC BY) license to any Author Accepted Manuscript version arising from this submission. The data supporting the findings of this study are available within the paper. No experimental datasets were generated by this research.

\bibliographystyle{apsrev4-1}
\bibliography{references.bib}

\begin{thebibliography}{138}%
\makeatletter
\providecommand \@ifxundefined [1]{%
 \@ifx{#1\undefined}
}%
\providecommand \@ifnum [1]{%
 \ifnum #1\expandafter \@firstoftwo
 \else \expandafter \@secondoftwo
 \fi
}%
\providecommand \@ifx [1]{%
 \ifx #1\expandafter \@firstoftwo
 \else \expandafter \@secondoftwo
 \fi
}%
\providecommand \natexlab [1]{#1}%
\providecommand \enquote  [1]{``#1''}%
\providecommand \bibnamefont  [1]{#1}%
\providecommand \bibfnamefont [1]{#1}%
\providecommand \citenamefont [1]{#1}%
\providecommand \href@noop [0]{\@secondoftwo}%
\providecommand \href [0]{\begingroup \@sanitize@url \@href}%
\providecommand \@href[1]{\@@startlink{#1}\@@href}%
\providecommand \@@href[1]{\endgroup#1\@@endlink}%
\providecommand \@sanitize@url [0]{\catcode `\\12\catcode `\$12\catcode
  `\&12\catcode `\#12\catcode `\^12\catcode `\_12\catcode `\%12\relax}%
\providecommand \@@startlink[1]{}%
\providecommand \@@endlink[0]{}%
\providecommand \url  [0]{\begingroup\@sanitize@url \@url }%
\providecommand \@url [1]{\endgroup\@href {#1}{\urlprefix }}%
\providecommand \urlprefix  [0]{URL }%
\providecommand \Eprint [0]{\href }%
\providecommand \doibase [0]{http://dx.doi.org/}%
\providecommand \selectlanguage [0]{\@gobble}%
\providecommand \bibinfo  [0]{\@secondoftwo}%
\providecommand \bibfield  [0]{\@secondoftwo}%
\providecommand \translation [1]{[#1]}%
\providecommand \BibitemOpen [0]{}%
\providecommand \bibitemStop [0]{}%
\providecommand \bibitemNoStop [0]{.\EOS\space}%
\providecommand \EOS [0]{\spacefactor3000\relax}%
\providecommand \BibitemShut  [1]{\csname bibitem#1\endcsname}%
\let\auto@bib@innerbib\@empty
\bibitem [{\citenamefont {Raffelt}(1996)}]{Raffelt:1996wa}%
  \BibitemOpen
  \bibfield  {author} {\bibinfo {author} {\bibfnamefont {G.~G.}\ \bibnamefont
  {Raffelt}},\ }\href@noop {} {}\ (\bibinfo {year} {1996})\BibitemShut
  {NoStop}%
\bibitem [{\citenamefont {Mirizzi}\ \emph {et~al.}(2016)\citenamefont
  {Mirizzi}, \citenamefont {Tamborra}, \citenamefont {Janka}, \citenamefont
  {Saviano}, \citenamefont {Scholberg}, \citenamefont {Bollig}, \citenamefont
  {Hudepohl},\ and\ \citenamefont {Chakraborty}}]{Mirizzi:2015eza}%
  \BibitemOpen
  \bibfield  {author} {\bibinfo {author} {\bibfnamefont {A.}~\bibnamefont
  {Mirizzi}}, \bibinfo {author} {\bibfnamefont {I.}~\bibnamefont {Tamborra}},
  \bibinfo {author} {\bibfnamefont {H.-T.}\ \bibnamefont {Janka}}, \bibinfo
  {author} {\bibfnamefont {N.}~\bibnamefont {Saviano}}, \bibinfo {author}
  {\bibfnamefont {K.}~\bibnamefont {Scholberg}}, \bibinfo {author}
  {\bibfnamefont {R.}~\bibnamefont {Bollig}}, \bibinfo {author} {\bibfnamefont
  {L.}~\bibnamefont {Hudepohl}}, \ and\ \bibinfo {author} {\bibfnamefont
  {S.}~\bibnamefont {Chakraborty}},\ }\href {\doibase
  10.1393/ncr/i2016-10120-8} {\bibfield  {journal} {\bibinfo  {journal} {Riv.
  Nuovo Cim.}\ }\textbf {\bibinfo {volume} {39}},\ \bibinfo {pages} {1}
  (\bibinfo {year} {2016})},\ \Eprint {http://arxiv.org/abs/1508.00785}
  {arXiv:1508.00785 [astro-ph.HE]} \BibitemShut {NoStop}%
\bibitem [{\citenamefont {Horiuchi}\ and\ \citenamefont
  {Kneller}(2018)}]{Horiuchi:2018ofe}%
  \BibitemOpen
  \bibfield  {author} {\bibinfo {author} {\bibfnamefont {S.}~\bibnamefont
  {Horiuchi}}\ and\ \bibinfo {author} {\bibfnamefont {J.~P.}\ \bibnamefont
  {Kneller}},\ }\href {\doibase 10.1088/1361-6471/aaa90a} {\bibfield  {journal}
  {\bibinfo  {journal} {J. Phys. G}\ }\textbf {\bibinfo {volume} {45}},\
  \bibinfo {pages} {043002} (\bibinfo {year} {2018})},\ \Eprint
  {http://arxiv.org/abs/1709.01515} {arXiv:1709.01515 [astro-ph.HE]}
  \BibitemShut {NoStop}%
\bibitem [{\citenamefont {Caputo}\ and\ \citenamefont
  {Raffelt}(2024)}]{Caputo:2024oqc}%
  \BibitemOpen
  \bibfield  {author} {\bibinfo {author} {\bibfnamefont {A.}~\bibnamefont
  {Caputo}}\ and\ \bibinfo {author} {\bibfnamefont {G.}~\bibnamefont
  {Raffelt}},\ }\href {\doibase 10.22323/1.454.0041} {\bibfield  {journal}
  {\bibinfo  {journal} {PoS}\ }\textbf {\bibinfo {volume} {COSMICWISPers}},\
  \bibinfo {pages} {041} (\bibinfo {year} {2024})},\ \Eprint
  {http://arxiv.org/abs/2401.13728} {arXiv:2401.13728 [hep-ph]} \BibitemShut
  {NoStop}%
\bibitem [{\citenamefont {Raffelt}\ and\ \citenamefont
  {Seckel}(1988)}]{Raffelt:1987yt}%
  \BibitemOpen
  \bibfield  {author} {\bibinfo {author} {\bibfnamefont {G.}~\bibnamefont
  {Raffelt}}\ and\ \bibinfo {author} {\bibfnamefont {D.}~\bibnamefont
  {Seckel}},\ }\href {\doibase 10.1103/PhysRevLett.60.1793} {\bibfield
  {journal} {\bibinfo  {journal} {Phys. Rev. Lett.}\ }\textbf {\bibinfo
  {volume} {60}},\ \bibinfo {pages} {1793} (\bibinfo {year}
  {1988})}\BibitemShut {NoStop}%
\bibitem [{\citenamefont {Keil}\ \emph {et~al.}(1997)\citenamefont {Keil},
  \citenamefont {Janka}, \citenamefont {Schramm}, \citenamefont {Sigl},
  \citenamefont {Turner},\ and\ \citenamefont {Ellis}}]{Keil:1996ju}%
  \BibitemOpen
  \bibfield  {author} {\bibinfo {author} {\bibfnamefont {W.}~\bibnamefont
  {Keil}}, \bibinfo {author} {\bibfnamefont {H.-T.}\ \bibnamefont {Janka}},
  \bibinfo {author} {\bibfnamefont {D.~N.}\ \bibnamefont {Schramm}}, \bibinfo
  {author} {\bibfnamefont {G.}~\bibnamefont {Sigl}}, \bibinfo {author}
  {\bibfnamefont {M.~S.}\ \bibnamefont {Turner}}, \ and\ \bibinfo {author}
  {\bibfnamefont {J.~R.}\ \bibnamefont {Ellis}},\ }\href {\doibase
  10.1103/PhysRevD.56.2419} {\bibfield  {journal} {\bibinfo  {journal} {Phys.
  Rev. D}\ }\textbf {\bibinfo {volume} {56}},\ \bibinfo {pages} {2419}
  (\bibinfo {year} {1997})},\ \Eprint {http://arxiv.org/abs/astro-ph/9612222}
  {arXiv:astro-ph/9612222} \BibitemShut {NoStop}%
\bibitem [{\citenamefont {Chang}\ \emph {et~al.}(2018)\citenamefont {Chang},
  \citenamefont {Essig},\ and\ \citenamefont {McDermott}}]{Chang:2018rso}%
  \BibitemOpen
  \bibfield  {author} {\bibinfo {author} {\bibfnamefont {J.~H.}\ \bibnamefont
  {Chang}}, \bibinfo {author} {\bibfnamefont {R.}~\bibnamefont {Essig}}, \ and\
  \bibinfo {author} {\bibfnamefont {S.~D.}\ \bibnamefont {McDermott}},\ }\href
  {\doibase 10.1007/JHEP09(2018)051} {\bibfield  {journal} {\bibinfo  {journal}
  {JHEP}\ }\textbf {\bibinfo {volume} {09}},\ \bibinfo {pages} {051} (\bibinfo
  {year} {2018})},\ \Eprint {http://arxiv.org/abs/1803.00993} {arXiv:1803.00993
  [hep-ph]} \BibitemShut {NoStop}%
\bibitem [{\citenamefont {Carenza}\ \emph {et~al.}(2019)\citenamefont
  {Carenza}, \citenamefont {Fischer}, \citenamefont {Giannotti}, \citenamefont
  {Guo}, \citenamefont {Mart\'\i{}nez-Pinedo},\ and\ \citenamefont
  {Mirizzi}}]{Carenza:2019pxu}%
  \BibitemOpen
  \bibfield  {author} {\bibinfo {author} {\bibfnamefont {P.}~\bibnamefont
  {Carenza}}, \bibinfo {author} {\bibfnamefont {T.}~\bibnamefont {Fischer}},
  \bibinfo {author} {\bibfnamefont {M.}~\bibnamefont {Giannotti}}, \bibinfo
  {author} {\bibfnamefont {G.}~\bibnamefont {Guo}}, \bibinfo {author}
  {\bibfnamefont {G.}~\bibnamefont {Mart\'\i{}nez-Pinedo}}, \ and\ \bibinfo
  {author} {\bibfnamefont {A.}~\bibnamefont {Mirizzi}},\ }\href {\doibase
  10.1088/1475-7516/2019/10/016} {\bibfield  {journal} {\bibinfo  {journal}
  {JCAP}\ }\textbf {\bibinfo {volume} {10}},\ \bibinfo {pages} {016} (\bibinfo
  {year} {2019})},\ \bibinfo {note} {[Erratum: JCAP 05, E01 (2020)]},\ \Eprint
  {http://arxiv.org/abs/1906.11844} {arXiv:1906.11844 [hep-ph]} \BibitemShut
  {NoStop}%
\bibitem [{\citenamefont {Carenza}\ \emph {et~al.}(2021)\citenamefont
  {Carenza}, \citenamefont {Fore}, \citenamefont {Giannotti}, \citenamefont
  {Mirizzi},\ and\ \citenamefont {Reddy}}]{Carenza:2020cis}%
  \BibitemOpen
  \bibfield  {author} {\bibinfo {author} {\bibfnamefont {P.}~\bibnamefont
  {Carenza}}, \bibinfo {author} {\bibfnamefont {B.}~\bibnamefont {Fore}},
  \bibinfo {author} {\bibfnamefont {M.}~\bibnamefont {Giannotti}}, \bibinfo
  {author} {\bibfnamefont {A.}~\bibnamefont {Mirizzi}}, \ and\ \bibinfo
  {author} {\bibfnamefont {S.}~\bibnamefont {Reddy}},\ }\href {\doibase
  10.1103/PhysRevLett.126.071102} {\bibfield  {journal} {\bibinfo  {journal}
  {Phys. Rev. Lett.}\ }\textbf {\bibinfo {volume} {126}},\ \bibinfo {pages}
  {071102} (\bibinfo {year} {2021})},\ \Eprint
  {http://arxiv.org/abs/2010.02943} {arXiv:2010.02943 [hep-ph]} \BibitemShut
  {NoStop}%
\bibitem [{\citenamefont {Caputo}\ \emph
  {et~al.}(2022{\natexlab{a}})\citenamefont {Caputo}, \citenamefont {Raffelt},\
  and\ \citenamefont {Vitagliano}}]{Caputo:2022rca}%
  \BibitemOpen
  \bibfield  {author} {\bibinfo {author} {\bibfnamefont {A.}~\bibnamefont
  {Caputo}}, \bibinfo {author} {\bibfnamefont {G.}~\bibnamefont {Raffelt}}, \
  and\ \bibinfo {author} {\bibfnamefont {E.}~\bibnamefont {Vitagliano}},\
  }\href {\doibase 10.1088/1475-7516/2022/08/045} {\bibfield  {journal}
  {\bibinfo  {journal} {JCAP}\ }\textbf {\bibinfo {volume} {08}},\ \bibinfo
  {pages} {045} (\bibinfo {year} {2022}{\natexlab{a}})},\ \Eprint
  {http://arxiv.org/abs/2204.11862} {arXiv:2204.11862 [astro-ph.SR]}
  \BibitemShut {NoStop}%
\bibitem [{\citenamefont {Caputo}\ \emph
  {et~al.}(2022{\natexlab{b}})\citenamefont {Caputo}, \citenamefont {Raffelt},\
  and\ \citenamefont {Vitagliano}}]{Caputo:2021rux}%
  \BibitemOpen
  \bibfield  {author} {\bibinfo {author} {\bibfnamefont {A.}~\bibnamefont
  {Caputo}}, \bibinfo {author} {\bibfnamefont {G.}~\bibnamefont {Raffelt}}, \
  and\ \bibinfo {author} {\bibfnamefont {E.}~\bibnamefont {Vitagliano}},\
  }\href {\doibase 10.1103/PhysRevD.105.035022} {\bibfield  {journal} {\bibinfo
   {journal} {Phys. Rev. D}\ }\textbf {\bibinfo {volume} {105}},\ \bibinfo
  {pages} {035022} (\bibinfo {year} {2022}{\natexlab{b}})},\ \Eprint
  {http://arxiv.org/abs/2109.03244} {arXiv:2109.03244 [hep-ph]} \BibitemShut
  {NoStop}%
\bibitem [{\citenamefont {Kolb}\ \emph {et~al.}(1996)\citenamefont {Kolb},
  \citenamefont {Mohapatra},\ and\ \citenamefont {Teplitz}}]{Kolb:1996pa}%
  \BibitemOpen
  \bibfield  {author} {\bibinfo {author} {\bibfnamefont {E.~W.}\ \bibnamefont
  {Kolb}}, \bibinfo {author} {\bibfnamefont {R.~N.}\ \bibnamefont {Mohapatra}},
  \ and\ \bibinfo {author} {\bibfnamefont {V.~L.}\ \bibnamefont {Teplitz}},\
  }\href {\doibase 10.1103/PhysRevLett.77.3066} {\bibfield  {journal} {\bibinfo
   {journal} {Phys. Rev. Lett.}\ }\textbf {\bibinfo {volume} {77}},\ \bibinfo
  {pages} {3066} (\bibinfo {year} {1996})},\ \Eprint
  {http://arxiv.org/abs/hep-ph/9605350} {arXiv:hep-ph/9605350} \BibitemShut
  {NoStop}%
\bibitem [{\citenamefont {Raffelt}\ and\ \citenamefont
  {Zhou}(2011)}]{Raffelt:2011nc}%
  \BibitemOpen
  \bibfield  {author} {\bibinfo {author} {\bibfnamefont {G.~G.}\ \bibnamefont
  {Raffelt}}\ and\ \bibinfo {author} {\bibfnamefont {S.}~\bibnamefont {Zhou}},\
  }\href {\doibase 10.1103/PhysRevD.83.093014} {\bibfield  {journal} {\bibinfo
  {journal} {Phys. Rev. D}\ }\textbf {\bibinfo {volume} {83}},\ \bibinfo
  {pages} {093014} (\bibinfo {year} {2011})},\ \Eprint
  {http://arxiv.org/abs/1102.5124} {arXiv:1102.5124 [hep-ph]} \BibitemShut
  {NoStop}%
\bibitem [{\citenamefont {Mastrototaro}\ \emph {et~al.}(2020)\citenamefont
  {Mastrototaro}, \citenamefont {Mirizzi}, \citenamefont {Serpico},\ and\
  \citenamefont {Esmaili}}]{Mastrototaro:2019vug}%
  \BibitemOpen
  \bibfield  {author} {\bibinfo {author} {\bibfnamefont {L.}~\bibnamefont
  {Mastrototaro}}, \bibinfo {author} {\bibfnamefont {A.}~\bibnamefont
  {Mirizzi}}, \bibinfo {author} {\bibfnamefont {P.~D.}\ \bibnamefont
  {Serpico}}, \ and\ \bibinfo {author} {\bibfnamefont {A.}~\bibnamefont
  {Esmaili}},\ }\href {\doibase 10.1088/1475-7516/2020/01/010} {\bibfield
  {journal} {\bibinfo  {journal} {JCAP}\ }\textbf {\bibinfo {volume} {01}},\
  \bibinfo {pages} {010} (\bibinfo {year} {2020})},\ \Eprint
  {http://arxiv.org/abs/1910.10249} {arXiv:1910.10249 [hep-ph]} \BibitemShut
  {NoStop}%
\bibitem [{\citenamefont {Carenza}\ \emph {et~al.}(2024)\citenamefont
  {Carenza}, \citenamefont {Lucente}, \citenamefont {Mastrototaro},
  \citenamefont {Mirizzi},\ and\ \citenamefont {Serpico}}]{Carenza:2023old}%
  \BibitemOpen
  \bibfield  {author} {\bibinfo {author} {\bibfnamefont {P.}~\bibnamefont
  {Carenza}}, \bibinfo {author} {\bibfnamefont {G.}~\bibnamefont {Lucente}},
  \bibinfo {author} {\bibfnamefont {L.}~\bibnamefont {Mastrototaro}}, \bibinfo
  {author} {\bibfnamefont {A.}~\bibnamefont {Mirizzi}}, \ and\ \bibinfo
  {author} {\bibfnamefont {P.~D.}\ \bibnamefont {Serpico}},\ }\href {\doibase
  10.1103/PhysRevD.109.063010} {\bibfield  {journal} {\bibinfo  {journal}
  {Phys. Rev. D}\ }\textbf {\bibinfo {volume} {109}},\ \bibinfo {pages}
  {063010} (\bibinfo {year} {2024})},\ \Eprint
  {http://arxiv.org/abs/2311.00033} {arXiv:2311.00033 [hep-ph]} \BibitemShut
  {NoStop}%
\bibitem [{\citenamefont {Akita}\ \emph {et~al.}(2024)\citenamefont {Akita},
  \citenamefont {Im}, \citenamefont {Masud},\ and\ \citenamefont
  {Yun}}]{Akita:2023iwq}%
  \BibitemOpen
  \bibfield  {author} {\bibinfo {author} {\bibfnamefont {K.}~\bibnamefont
  {Akita}}, \bibinfo {author} {\bibfnamefont {S.~H.}\ \bibnamefont {Im}},
  \bibinfo {author} {\bibfnamefont {M.}~\bibnamefont {Masud}}, \ and\ \bibinfo
  {author} {\bibfnamefont {S.}~\bibnamefont {Yun}},\ }\href {\doibase
  10.1007/JHEP07(2024)057} {\bibfield  {journal} {\bibinfo  {journal} {JHEP}\
  }\textbf {\bibinfo {volume} {07}},\ \bibinfo {pages} {057} (\bibinfo {year}
  {2024})},\ \Eprint {http://arxiv.org/abs/2312.13627} {arXiv:2312.13627
  [hep-ph]} \BibitemShut {NoStop}%
\bibitem [{\citenamefont {Chang}\ \emph {et~al.}(2017)\citenamefont {Chang},
  \citenamefont {Essig},\ and\ \citenamefont {McDermott}}]{Chang:2016ntp}%
  \BibitemOpen
  \bibfield  {author} {\bibinfo {author} {\bibfnamefont {J.~H.}\ \bibnamefont
  {Chang}}, \bibinfo {author} {\bibfnamefont {R.}~\bibnamefont {Essig}}, \ and\
  \bibinfo {author} {\bibfnamefont {S.~D.}\ \bibnamefont {McDermott}},\ }\href
  {\doibase 10.1007/JHEP01(2017)107} {\bibfield  {journal} {\bibinfo  {journal}
  {JHEP}\ }\textbf {\bibinfo {volume} {01}},\ \bibinfo {pages} {107} (\bibinfo
  {year} {2017})},\ \Eprint {http://arxiv.org/abs/1611.03864} {arXiv:1611.03864
  [hep-ph]} \BibitemShut {NoStop}%
\bibitem [{\citenamefont {Linden}\ \emph {et~al.}(2024)\citenamefont {Linden},
  \citenamefont {Nguyen},\ and\ \citenamefont {Tait}}]{Linden:2024fby}%
  \BibitemOpen
  \bibfield  {author} {\bibinfo {author} {\bibfnamefont {T.}~\bibnamefont
  {Linden}}, \bibinfo {author} {\bibfnamefont {T.~T.~Q.}\ \bibnamefont
  {Nguyen}}, \ and\ \bibinfo {author} {\bibfnamefont {T.~M.~P.}\ \bibnamefont
  {Tait}},\ }\href@noop {} {\  (\bibinfo {year} {2024})},\ \Eprint
  {http://arxiv.org/abs/2406.19445} {arXiv:2406.19445 [hep-ph]} \BibitemShut
  {NoStop}%
\bibitem [{\citenamefont {Dev}\ \emph {et~al.}(2020)\citenamefont {Dev},
  \citenamefont {Mohapatra},\ and\ \citenamefont {Zhang}}]{Dev:2020eam}%
  \BibitemOpen
  \bibfield  {author} {\bibinfo {author} {\bibfnamefont {P.~S.~B.}\
  \bibnamefont {Dev}}, \bibinfo {author} {\bibfnamefont {R.~N.}\ \bibnamefont
  {Mohapatra}}, \ and\ \bibinfo {author} {\bibfnamefont {Y.}~\bibnamefont
  {Zhang}},\ }\href {\doibase 10.1088/1475-7516/2020/08/003} {\bibfield
  {journal} {\bibinfo  {journal} {JCAP}\ }\textbf {\bibinfo {volume} {08}},\
  \bibinfo {pages} {003} (\bibinfo {year} {2020})},\ \bibinfo {note} {[Erratum:
  JCAP 11, E01 (2020)]},\ \Eprint {http://arxiv.org/abs/2005.00490}
  {arXiv:2005.00490 [hep-ph]} \BibitemShut {NoStop}%
\bibitem [{\citenamefont {Balaji}\ \emph {et~al.}(2022)\citenamefont {Balaji},
  \citenamefont {Dev}, \citenamefont {Silk},\ and\ \citenamefont
  {Zhang}}]{Balaji:2022noj}%
  \BibitemOpen
  \bibfield  {author} {\bibinfo {author} {\bibfnamefont {S.}~\bibnamefont
  {Balaji}}, \bibinfo {author} {\bibfnamefont {P.~S.~B.}\ \bibnamefont {Dev}},
  \bibinfo {author} {\bibfnamefont {J.}~\bibnamefont {Silk}}, \ and\ \bibinfo
  {author} {\bibfnamefont {Y.}~\bibnamefont {Zhang}},\ }\href {\doibase
  10.1088/1475-7516/2022/12/024} {\bibfield  {journal} {\bibinfo  {journal}
  {JCAP}\ }\textbf {\bibinfo {volume} {12}},\ \bibinfo {pages} {024} (\bibinfo
  {year} {2022})},\ \Eprint {http://arxiv.org/abs/2205.01669} {arXiv:2205.01669
  [hep-ph]} \BibitemShut {NoStop}%
\bibitem [{\citenamefont {Camalich}\ \emph {et~al.}(2021)\citenamefont
  {Camalich}, \citenamefont {Terol-Calvo}, \citenamefont {Tolos},\ and\
  \citenamefont {Ziegler}}]{Camalich:2020wac}%
  \BibitemOpen
  \bibfield  {author} {\bibinfo {author} {\bibfnamefont {J.~M.}\ \bibnamefont
  {Camalich}}, \bibinfo {author} {\bibfnamefont {J.}~\bibnamefont
  {Terol-Calvo}}, \bibinfo {author} {\bibfnamefont {L.}~\bibnamefont {Tolos}},
  \ and\ \bibinfo {author} {\bibfnamefont {R.}~\bibnamefont {Ziegler}},\ }\href
  {\doibase 10.1103/PhysRevD.103.L121301} {\bibfield  {journal} {\bibinfo
  {journal} {Phys. Rev. D}\ }\textbf {\bibinfo {volume} {103}},\ \bibinfo
  {pages} {L121301} (\bibinfo {year} {2021})},\ \Eprint
  {http://arxiv.org/abs/2012.11632} {arXiv:2012.11632 [hep-ph]} \BibitemShut
  {NoStop}%
\bibitem [{\citenamefont {Hannestad}\ \emph {et~al.}(2007)\citenamefont
  {Hannestad}, \citenamefont {Raffelt},\ and\ \citenamefont
  {Wong}}]{Hannestad:2007ys}%
  \BibitemOpen
  \bibfield  {author} {\bibinfo {author} {\bibfnamefont {S.}~\bibnamefont
  {Hannestad}}, \bibinfo {author} {\bibfnamefont {G.}~\bibnamefont {Raffelt}},
  \ and\ \bibinfo {author} {\bibfnamefont {Y.~Y.~Y.}\ \bibnamefont {Wong}},\
  }\href {\doibase 10.1103/PhysRevD.76.121701} {\bibfield  {journal} {\bibinfo
  {journal} {Phys. Rev. D}\ }\textbf {\bibinfo {volume} {76}},\ \bibinfo
  {pages} {121701} (\bibinfo {year} {2007})},\ \Eprint
  {http://arxiv.org/abs/0708.1404} {arXiv:0708.1404 [hep-ph]} \BibitemShut
  {NoStop}%
\bibitem [{\citenamefont {Antel}\ \emph {et~al.}(2023)\citenamefont {Antel}
  \emph {et~al.}}]{Antel:2023hkf}%
  \BibitemOpen
  \bibfield  {author} {\bibinfo {author} {\bibfnamefont {C.}~\bibnamefont
  {Antel}} \emph {et~al.},\ }\href {\doibase 10.1140/epjc/s10052-023-12168-5}
  {\bibfield  {journal} {\bibinfo  {journal} {Eur. Phys. J. C}\ }\textbf
  {\bibinfo {volume} {83}},\ \bibinfo {pages} {1122} (\bibinfo {year}
  {2023})},\ \Eprint {http://arxiv.org/abs/2305.01715} {arXiv:2305.01715
  [hep-ph]} \BibitemShut {NoStop}%
\bibitem [{\citenamefont {De~la Torre~Luque}\ \emph
  {et~al.}(2024{\natexlab{a}})\citenamefont {De~la Torre~Luque}, \citenamefont
  {Balaji}, \citenamefont {Carenza},\ and\ \citenamefont
  {Mastrototaro}}]{DelaTorreLuque:2024zsr}%
  \BibitemOpen
  \bibfield  {author} {\bibinfo {author} {\bibfnamefont {P.}~\bibnamefont
  {De~la Torre~Luque}}, \bibinfo {author} {\bibfnamefont {S.}~\bibnamefont
  {Balaji}}, \bibinfo {author} {\bibfnamefont {P.}~\bibnamefont {Carenza}}, \
  and\ \bibinfo {author} {\bibfnamefont {L.}~\bibnamefont {Mastrototaro}},\
  }\href@noop {} {\  (\bibinfo {year} {2024}{\natexlab{a}})},\ \Eprint
  {http://arxiv.org/abs/2405.08482} {arXiv:2405.08482 [hep-ph]} \BibitemShut
  {NoStop}%
\bibitem [{\citenamefont {De~la Torre~Luque}\ \emph
  {et~al.}(2024{\natexlab{b}})\citenamefont {De~la Torre~Luque}, \citenamefont
  {Balaji},\ and\ \citenamefont {Carenza}}]{DelaTorreLuque:2023nhh}%
  \BibitemOpen
  \bibfield  {author} {\bibinfo {author} {\bibfnamefont {P.}~\bibnamefont
  {De~la Torre~Luque}}, \bibinfo {author} {\bibfnamefont {S.}~\bibnamefont
  {Balaji}}, \ and\ \bibinfo {author} {\bibfnamefont {P.}~\bibnamefont
  {Carenza}},\ }\href {\doibase 10.1103/PhysRevD.109.L101305} {\bibfield
  {journal} {\bibinfo  {journal} {Phys. Rev. D}\ }\textbf {\bibinfo {volume}
  {109}},\ \bibinfo {pages} {L101305} (\bibinfo {year} {2024}{\natexlab{b}})},\
  \Eprint {http://arxiv.org/abs/2307.13728} {arXiv:2307.13728 [hep-ph]}
  \BibitemShut {NoStop}%
\bibitem [{\citenamefont {De~la Torre~Luque}\ \emph
  {et~al.}(2024{\natexlab{c}})\citenamefont {De~la Torre~Luque}, \citenamefont
  {Balaji},\ and\ \citenamefont {Carenza}}]{DelaTorreLuque:2023huu}%
  \BibitemOpen
  \bibfield  {author} {\bibinfo {author} {\bibfnamefont {P.}~\bibnamefont
  {De~la Torre~Luque}}, \bibinfo {author} {\bibfnamefont {S.}~\bibnamefont
  {Balaji}}, \ and\ \bibinfo {author} {\bibfnamefont {P.}~\bibnamefont
  {Carenza}},\ }\href {\doibase 10.1103/PhysRevD.109.103028} {\bibfield
  {journal} {\bibinfo  {journal} {Phys. Rev. D}\ }\textbf {\bibinfo {volume}
  {109}},\ \bibinfo {pages} {103028} (\bibinfo {year} {2024}{\natexlab{c}})},\
  \Eprint {http://arxiv.org/abs/2307.13731} {arXiv:2307.13731 [hep-ph]}
  \BibitemShut {NoStop}%
\bibitem [{\citenamefont {Calore}\ \emph {et~al.}(2022)\citenamefont {Calore},
  \citenamefont {Carenza}, \citenamefont {Giannotti}, \citenamefont {Jaeckel},
  \citenamefont {Lucente}, \citenamefont {Mastrototaro},\ and\ \citenamefont
  {Mirizzi}}]{Calore:2021lih}%
  \BibitemOpen
  \bibfield  {author} {\bibinfo {author} {\bibfnamefont {F.}~\bibnamefont
  {Calore}}, \bibinfo {author} {\bibfnamefont {P.}~\bibnamefont {Carenza}},
  \bibinfo {author} {\bibfnamefont {M.}~\bibnamefont {Giannotti}}, \bibinfo
  {author} {\bibfnamefont {J.}~\bibnamefont {Jaeckel}}, \bibinfo {author}
  {\bibfnamefont {G.}~\bibnamefont {Lucente}}, \bibinfo {author} {\bibfnamefont
  {L.}~\bibnamefont {Mastrototaro}}, \ and\ \bibinfo {author} {\bibfnamefont
  {A.}~\bibnamefont {Mirizzi}},\ }\href {\doibase 10.1103/PhysRevD.105.063026}
  {\bibfield  {journal} {\bibinfo  {journal} {Phys. Rev. D}\ }\textbf {\bibinfo
  {volume} {105}},\ \bibinfo {pages} {063026} (\bibinfo {year} {2022})},\
  \Eprint {http://arxiv.org/abs/2112.08382} {arXiv:2112.08382 [hep-ph]}
  \BibitemShut {NoStop}%
\bibitem [{\citenamefont {Calore}\ \emph {et~al.}(2021)\citenamefont {Calore},
  \citenamefont {Carenza}, \citenamefont {Giannotti}, \citenamefont {Jaeckel},
  \citenamefont {Lucente},\ and\ \citenamefont {Mirizzi}}]{Calore:2021klc}%
  \BibitemOpen
  \bibfield  {author} {\bibinfo {author} {\bibfnamefont {F.}~\bibnamefont
  {Calore}}, \bibinfo {author} {\bibfnamefont {P.}~\bibnamefont {Carenza}},
  \bibinfo {author} {\bibfnamefont {M.}~\bibnamefont {Giannotti}}, \bibinfo
  {author} {\bibfnamefont {J.}~\bibnamefont {Jaeckel}}, \bibinfo {author}
  {\bibfnamefont {G.}~\bibnamefont {Lucente}}, \ and\ \bibinfo {author}
  {\bibfnamefont {A.}~\bibnamefont {Mirizzi}},\ }\href {\doibase
  10.1103/PhysRevD.104.043016} {\bibfield  {journal} {\bibinfo  {journal}
  {Phys. Rev. D}\ }\textbf {\bibinfo {volume} {104}},\ \bibinfo {pages}
  {043016} (\bibinfo {year} {2021})},\ \Eprint
  {http://arxiv.org/abs/2107.02186} {arXiv:2107.02186 [hep-ph]} \BibitemShut
  {NoStop}%
\bibitem [{\citenamefont {Foster}\ \emph {et~al.}(2021)\citenamefont {Foster},
  \citenamefont {Kongsore}, \citenamefont {Dessert}, \citenamefont {Park},
  \citenamefont {Rodd}, \citenamefont {Cranmer},\ and\ \citenamefont
  {Safdi}}]{Foster:2021ngm}%
  \BibitemOpen
  \bibfield  {author} {\bibinfo {author} {\bibfnamefont {J.~W.}\ \bibnamefont
  {Foster}}, \bibinfo {author} {\bibfnamefont {M.}~\bibnamefont {Kongsore}},
  \bibinfo {author} {\bibfnamefont {C.}~\bibnamefont {Dessert}}, \bibinfo
  {author} {\bibfnamefont {Y.}~\bibnamefont {Park}}, \bibinfo {author}
  {\bibfnamefont {N.~L.}\ \bibnamefont {Rodd}}, \bibinfo {author}
  {\bibfnamefont {K.}~\bibnamefont {Cranmer}}, \ and\ \bibinfo {author}
  {\bibfnamefont {B.~R.}\ \bibnamefont {Safdi}},\ }\href {\doibase
  10.1103/PhysRevLett.127.051101} {\bibfield  {journal} {\bibinfo  {journal}
  {Phys. Rev. Lett.}\ }\textbf {\bibinfo {volume} {127}},\ \bibinfo {pages}
  {051101} (\bibinfo {year} {2021})},\ \Eprint
  {http://arxiv.org/abs/2102.02207} {arXiv:2102.02207 [astro-ph.CO]}
  \BibitemShut {NoStop}%
\bibitem [{\citenamefont {Bouchet}\ \emph {et~al.}(2008)\citenamefont
  {Bouchet}, \citenamefont {Jourdain}, \citenamefont {Roques}, \citenamefont
  {Strong}, \citenamefont {Diehl}, \citenamefont {Lebrun},\ and\ \citenamefont
  {Terrier}}]{Bouchet:2008rp}%
  \BibitemOpen
  \bibfield  {author} {\bibinfo {author} {\bibfnamefont {L.}~\bibnamefont
  {Bouchet}}, \bibinfo {author} {\bibfnamefont {E.}~\bibnamefont {Jourdain}},
  \bibinfo {author} {\bibfnamefont {J.~P.}\ \bibnamefont {Roques}}, \bibinfo
  {author} {\bibfnamefont {A.}~\bibnamefont {Strong}}, \bibinfo {author}
  {\bibfnamefont {R.}~\bibnamefont {Diehl}}, \bibinfo {author} {\bibfnamefont
  {F.}~\bibnamefont {Lebrun}}, \ and\ \bibinfo {author} {\bibfnamefont
  {R.}~\bibnamefont {Terrier}},\ }\href {\doibase 10.1086/529489} {\bibfield
  {journal} {\bibinfo  {journal} {Astrophys. J.}\ }\textbf {\bibinfo {volume}
  {679}},\ \bibinfo {pages} {1315} (\bibinfo {year} {2008})},\ \Eprint
  {http://arxiv.org/abs/0801.2086} {arXiv:0801.2086 [astro-ph]} \BibitemShut
  {NoStop}%
\bibitem [{\citenamefont {Bouchet}\ \emph {et~al.}(2010)\citenamefont
  {Bouchet}, \citenamefont {Roques},\ and\ \citenamefont
  {Jourdain}}]{Bouchet:2010dj}%
  \BibitemOpen
  \bibfield  {author} {\bibinfo {author} {\bibfnamefont {L.}~\bibnamefont
  {Bouchet}}, \bibinfo {author} {\bibfnamefont {J.-P.}\ \bibnamefont {Roques}},
  \ and\ \bibinfo {author} {\bibfnamefont {E.}~\bibnamefont {Jourdain}},\
  }\href {\doibase 10.1088/0004-637X/720/2/1772} {\bibfield  {journal}
  {\bibinfo  {journal} {Astrophys. J.}\ }\textbf {\bibinfo {volume} {720}},\
  \bibinfo {pages} {1772} (\bibinfo {year} {2010})},\ \Eprint
  {http://arxiv.org/abs/1007.4753} {arXiv:1007.4753 [astro-ph.HE]} \BibitemShut
  {NoStop}%
\bibitem [{\citenamefont {Siegert}\ \emph {et~al.}(2016)\citenamefont
  {Siegert}, \citenamefont {Diehl}, \citenamefont {Khachatryan}, \citenamefont
  {Krause}, \citenamefont {Guglielmetti}, \citenamefont {Greiner},
  \citenamefont {Strong},\ and\ \citenamefont {Zhang}}]{Siegert:2015knp}%
  \BibitemOpen
  \bibfield  {author} {\bibinfo {author} {\bibfnamefont {T.}~\bibnamefont
  {Siegert}}, \bibinfo {author} {\bibfnamefont {R.}~\bibnamefont {Diehl}},
  \bibinfo {author} {\bibfnamefont {G.}~\bibnamefont {Khachatryan}}, \bibinfo
  {author} {\bibfnamefont {M.~G.~H.}\ \bibnamefont {Krause}}, \bibinfo {author}
  {\bibfnamefont {F.}~\bibnamefont {Guglielmetti}}, \bibinfo {author}
  {\bibfnamefont {J.}~\bibnamefont {Greiner}}, \bibinfo {author} {\bibfnamefont
  {A.~W.}\ \bibnamefont {Strong}}, \ and\ \bibinfo {author} {\bibfnamefont
  {X.}~\bibnamefont {Zhang}},\ }\href {\doibase 10.1051/0004-6361/201527510}
  {\bibfield  {journal} {\bibinfo  {journal} {Astron. Astrophys.}\ }\textbf
  {\bibinfo {volume} {586}},\ \bibinfo {pages} {A84} (\bibinfo {year}
  {2016})},\ \Eprint {http://arxiv.org/abs/1512.00325} {arXiv:1512.00325
  [astro-ph.HE]} \BibitemShut {NoStop}%
\bibitem [{\citenamefont {Berteaud}\ \emph {et~al.}(2022)\citenamefont
  {Berteaud}, \citenamefont {Calore}, \citenamefont {Iguaz}, \citenamefont
  {Serpico},\ and\ \citenamefont {Siegert}}]{Berteaud:2022tws}%
  \BibitemOpen
  \bibfield  {author} {\bibinfo {author} {\bibfnamefont {J.}~\bibnamefont
  {Berteaud}}, \bibinfo {author} {\bibfnamefont {F.}~\bibnamefont {Calore}},
  \bibinfo {author} {\bibfnamefont {J.}~\bibnamefont {Iguaz}}, \bibinfo
  {author} {\bibfnamefont {P.}~\bibnamefont {Serpico}}, \ and\ \bibinfo
  {author} {\bibfnamefont {T.}~\bibnamefont {Siegert}},\ }\href {\doibase
  10.1103/physrevd.106.023030} {\bibfield  {journal} {\bibinfo  {journal}
  {Physical Review D}\ }\textbf {\bibinfo {volume} {106}} (\bibinfo {year}
  {2022}),\ 10.1103/physrevd.106.023030}\BibitemShut {NoStop}%
\bibitem [{\citenamefont {Sreekumar}\ \emph {et~al.}(1997)\citenamefont
  {Sreekumar}, \citenamefont {Stecker},\ and\ \citenamefont
  {Kappadath}}]{Sreekumar:1997yg}%
  \BibitemOpen
  \bibfield  {author} {\bibinfo {author} {\bibfnamefont {P.}~\bibnamefont
  {Sreekumar}}, \bibinfo {author} {\bibfnamefont {F.~W.}\ \bibnamefont
  {Stecker}}, \ and\ \bibinfo {author} {\bibfnamefont {S.~C.}\ \bibnamefont
  {Kappadath}},\ }\href {\doibase 10.1063/1.54124} {\bibfield  {journal}
  {\bibinfo  {journal} {AIP Conf. Proc.}\ }\textbf {\bibinfo {volume} {410}},\
  \bibinfo {pages} {344} (\bibinfo {year} {1997})},\ \Eprint
  {http://arxiv.org/abs/astro-ph/9709258} {arXiv:astro-ph/9709258} \BibitemShut
  {NoStop}%
\bibitem [{\citenamefont {Strong}\ \emph
  {et~al.}(2004{\natexlab{a}})\citenamefont {Strong}, \citenamefont
  {Moskalenko},\ and\ \citenamefont {Reimer}}]{Strong:2004de}%
  \BibitemOpen
  \bibfield  {author} {\bibinfo {author} {\bibfnamefont {A.~W.}\ \bibnamefont
  {Strong}}, \bibinfo {author} {\bibfnamefont {I.~V.}\ \bibnamefont
  {Moskalenko}}, \ and\ \bibinfo {author} {\bibfnamefont {O.}~\bibnamefont
  {Reimer}},\ }\href {\doibase 10.1086/423193} {\bibfield  {journal} {\bibinfo
  {journal} {Astrophys. J.}\ }\textbf {\bibinfo {volume} {613}},\ \bibinfo
  {pages} {962} (\bibinfo {year} {2004}{\natexlab{a}})},\ \Eprint
  {http://arxiv.org/abs/astro-ph/0406254} {arXiv:astro-ph/0406254} \BibitemShut
  {NoStop}%
\bibitem [{\citenamefont {{Stecker}}(1969)}]{Stecker}%
  \BibitemOpen
  \bibfield  {author} {\bibinfo {author} {\bibfnamefont {F.~W.}\ \bibnamefont
  {{Stecker}}},\ }\href {\doibase 10.1007/BF00704862} {\bibfield  {journal}
  {\bibinfo  {journal} {apss}\ }\textbf {\bibinfo {volume} {3}},\ \bibinfo
  {pages} {579} (\bibinfo {year} {1969})}\BibitemShut {NoStop}%
\bibitem [{\citenamefont {Stone}\ \emph
  {et~al.}(2013{\natexlab{a}})\citenamefont {Stone}, \citenamefont {Cummings},
  \citenamefont {McDonald}, \citenamefont {Heikkila}, \citenamefont {Lal},\
  and\ \citenamefont {Webber}}]{Stone:2013zlg}%
  \BibitemOpen
  \bibfield  {author} {\bibinfo {author} {\bibfnamefont {E.~C.}\ \bibnamefont
  {Stone}}, \bibinfo {author} {\bibfnamefont {A.~C.}\ \bibnamefont {Cummings}},
  \bibinfo {author} {\bibfnamefont {F.~B.}\ \bibnamefont {McDonald}}, \bibinfo
  {author} {\bibfnamefont {B.~C.}\ \bibnamefont {Heikkila}}, \bibinfo {author}
  {\bibfnamefont {N.}~\bibnamefont {Lal}}, \ and\ \bibinfo {author}
  {\bibfnamefont {W.~R.}\ \bibnamefont {Webber}},\ }\href {\doibase
  10.1126/science.1236408} {\bibfield  {journal} {\bibinfo  {journal}
  {Science}\ }\textbf {\bibinfo {volume} {341}},\ \bibinfo {pages} {1236408}
  (\bibinfo {year} {2013}{\natexlab{a}})}\BibitemShut {NoStop}%
\bibitem [{SNa()}]{SNarchive}%
  \BibitemOpen
  \href@noop {} {\enquote {\bibinfo {title} {{Garching core-collapse supernova
  research archive}},}\ }\bibinfo {howpublished}
  {\url{https://wwwmpa.mpa-garching.mpg.de/ccsnarchive//}}\BibitemShut
  {NoStop}%
\bibitem [{\citenamefont {Bollig}\ \emph {et~al.}(2020)\citenamefont {Bollig},
  \citenamefont {DeRocco}, \citenamefont {Graham},\ and\ \citenamefont
  {Janka}}]{Bollig:2020xdr}%
  \BibitemOpen
  \bibfield  {author} {\bibinfo {author} {\bibfnamefont {R.}~\bibnamefont
  {Bollig}}, \bibinfo {author} {\bibfnamefont {W.}~\bibnamefont {DeRocco}},
  \bibinfo {author} {\bibfnamefont {P.~W.}\ \bibnamefont {Graham}}, \ and\
  \bibinfo {author} {\bibfnamefont {H.-T.}\ \bibnamefont {Janka}},\ }\href
  {\doibase 10.1103/PhysRevLett.125.051104} {\bibfield  {journal} {\bibinfo
  {journal} {Phys. Rev. Lett.}\ }\textbf {\bibinfo {volume} {125}},\ \bibinfo
  {pages} {051104} (\bibinfo {year} {2020})},\ \bibinfo {note} {[Erratum:
  Phys.Rev.Lett. 126, 189901 (2021)]},\ \Eprint
  {http://arxiv.org/abs/2005.07141} {arXiv:2005.07141 [hep-ph]} \BibitemShut
  {NoStop}%
\bibitem [{\citenamefont {Caputo}\ \emph
  {et~al.}(2022{\natexlab{c}})\citenamefont {Caputo}, \citenamefont {Janka},
  \citenamefont {Raffelt},\ and\ \citenamefont {Vitagliano}}]{Caputo:2022mah}%
  \BibitemOpen
  \bibfield  {author} {\bibinfo {author} {\bibfnamefont {A.}~\bibnamefont
  {Caputo}}, \bibinfo {author} {\bibfnamefont {H.-T.}\ \bibnamefont {Janka}},
  \bibinfo {author} {\bibfnamefont {G.}~\bibnamefont {Raffelt}}, \ and\
  \bibinfo {author} {\bibfnamefont {E.}~\bibnamefont {Vitagliano}},\ }\href
  {\doibase 10.1103/PhysRevLett.128.221103} {\bibfield  {journal} {\bibinfo
  {journal} {Phys. Rev. Lett.}\ }\textbf {\bibinfo {volume} {128}},\ \bibinfo
  {pages} {221103} (\bibinfo {year} {2022}{\natexlab{c}})},\ \Eprint
  {http://arxiv.org/abs/2201.09890} {arXiv:2201.09890 [astro-ph.HE]}
  \BibitemShut {NoStop}%
\bibitem [{\citenamefont {Lella}\ \emph {et~al.}(2023)\citenamefont {Lella},
  \citenamefont {Carenza}, \citenamefont {Lucente}, \citenamefont {Giannotti},\
  and\ \citenamefont {Mirizzi}}]{Lella:2022uwi}%
  \BibitemOpen
  \bibfield  {author} {\bibinfo {author} {\bibfnamefont {A.}~\bibnamefont
  {Lella}}, \bibinfo {author} {\bibfnamefont {P.}~\bibnamefont {Carenza}},
  \bibinfo {author} {\bibfnamefont {G.}~\bibnamefont {Lucente}}, \bibinfo
  {author} {\bibfnamefont {M.}~\bibnamefont {Giannotti}}, \ and\ \bibinfo
  {author} {\bibfnamefont {A.}~\bibnamefont {Mirizzi}},\ }\href {\doibase
  10.1103/PhysRevD.107.103017} {\bibfield  {journal} {\bibinfo  {journal}
  {Phys. Rev. D}\ }\textbf {\bibinfo {volume} {107}},\ \bibinfo {pages}
  {103017} (\bibinfo {year} {2023})},\ \Eprint
  {http://arxiv.org/abs/2211.13760} {arXiv:2211.13760 [hep-ph]} \BibitemShut
  {NoStop}%
\bibitem [{\citenamefont {Lella}\ \emph
  {et~al.}(2024{\natexlab{a}})\citenamefont {Lella}, \citenamefont {Carenza},
  \citenamefont {Co'}, \citenamefont {Lucente}, \citenamefont {Giannotti},
  \citenamefont {Mirizzi},\ and\ \citenamefont {Rauscher}}]{Lella:2023bfb}%
  \BibitemOpen
  \bibfield  {author} {\bibinfo {author} {\bibfnamefont {A.}~\bibnamefont
  {Lella}}, \bibinfo {author} {\bibfnamefont {P.}~\bibnamefont {Carenza}},
  \bibinfo {author} {\bibfnamefont {G.}~\bibnamefont {Co'}}, \bibinfo {author}
  {\bibfnamefont {G.}~\bibnamefont {Lucente}}, \bibinfo {author} {\bibfnamefont
  {M.}~\bibnamefont {Giannotti}}, \bibinfo {author} {\bibfnamefont
  {A.}~\bibnamefont {Mirizzi}}, \ and\ \bibinfo {author} {\bibfnamefont
  {T.}~\bibnamefont {Rauscher}},\ }\href {\doibase 10.1103/PhysRevD.109.023001}
  {\bibfield  {journal} {\bibinfo  {journal} {Phys. Rev. D}\ }\textbf {\bibinfo
  {volume} {109}},\ \bibinfo {pages} {023001} (\bibinfo {year}
  {2024}{\natexlab{a}})},\ \Eprint {http://arxiv.org/abs/2306.01048}
  {arXiv:2306.01048 [hep-ph]} \BibitemShut {NoStop}%
\bibitem [{\citenamefont {Lella}\ \emph
  {et~al.}(2024{\natexlab{b}})\citenamefont {Lella}, \citenamefont {Calore},
  \citenamefont {Carenza}, \citenamefont {Eckner}, \citenamefont {Giannotti},
  \citenamefont {Lucente},\ and\ \citenamefont {Mirizzi}}]{Lella:2024hfk}%
  \BibitemOpen
  \bibfield  {author} {\bibinfo {author} {\bibfnamefont {A.}~\bibnamefont
  {Lella}}, \bibinfo {author} {\bibfnamefont {F.}~\bibnamefont {Calore}},
  \bibinfo {author} {\bibfnamefont {P.}~\bibnamefont {Carenza}}, \bibinfo
  {author} {\bibfnamefont {C.}~\bibnamefont {Eckner}}, \bibinfo {author}
  {\bibfnamefont {M.}~\bibnamefont {Giannotti}}, \bibinfo {author}
  {\bibfnamefont {G.}~\bibnamefont {Lucente}}, \ and\ \bibinfo {author}
  {\bibfnamefont {A.}~\bibnamefont {Mirizzi}},\ }\href@noop {} {\  (\bibinfo
  {year} {2024}{\natexlab{b}})},\ \Eprint {http://arxiv.org/abs/2405.02395}
  {arXiv:2405.02395 [hep-ph]} \BibitemShut {NoStop}%
\bibitem [{\citenamefont {Lella}\ \emph
  {et~al.}(2024{\natexlab{c}})\citenamefont {Lella}, \citenamefont
  {Ravensburg}, \citenamefont {Carenza},\ and\ \citenamefont
  {Marsh}}]{Lella:2024dmx}%
  \BibitemOpen
  \bibfield  {author} {\bibinfo {author} {\bibfnamefont {A.}~\bibnamefont
  {Lella}}, \bibinfo {author} {\bibfnamefont {E.}~\bibnamefont {Ravensburg}},
  \bibinfo {author} {\bibfnamefont {P.}~\bibnamefont {Carenza}}, \ and\
  \bibinfo {author} {\bibfnamefont {M.~C.~D.}\ \bibnamefont {Marsh}},\ }\href
  {\doibase 10.1103/PhysRevD.110.043019} {\bibfield  {journal} {\bibinfo
  {journal} {Phys. Rev. D}\ }\textbf {\bibinfo {volume} {110}},\ \bibinfo
  {pages} {043019} (\bibinfo {year} {2024}{\natexlab{c}})},\ \Eprint
  {http://arxiv.org/abs/2405.00153} {arXiv:2405.00153 [hep-ph]} \BibitemShut
  {NoStop}%
\bibitem [{\citenamefont {Manzari}\ \emph {et~al.}(2024)\citenamefont
  {Manzari}, \citenamefont {Park}, \citenamefont {Safdi},\ and\ \citenamefont
  {Savoray}}]{Manzari:2024jns}%
  \BibitemOpen
  \bibfield  {author} {\bibinfo {author} {\bibfnamefont {C.~A.}\ \bibnamefont
  {Manzari}}, \bibinfo {author} {\bibfnamefont {Y.}~\bibnamefont {Park}},
  \bibinfo {author} {\bibfnamefont {B.~R.}\ \bibnamefont {Safdi}}, \ and\
  \bibinfo {author} {\bibfnamefont {I.}~\bibnamefont {Savoray}},\ }\href
  {\doibase 10.1103/PhysRevLett.133.211002} {\bibfield  {journal} {\bibinfo
  {journal} {Phys. Rev. Lett.}\ }\textbf {\bibinfo {volume} {133}},\ \bibinfo
  {pages} {211002} (\bibinfo {year} {2024})},\ \Eprint
  {http://arxiv.org/abs/2405.19393} {arXiv:2405.19393 [hep-ph]} \BibitemShut
  {NoStop}%
\bibitem [{\citenamefont {Sukhbold}\ \emph {et~al.}(2018)\citenamefont
  {Sukhbold}, \citenamefont {Woosley},\ and\ \citenamefont
  {Heger}}]{Sukhbold:2017cnt}%
  \BibitemOpen
  \bibfield  {author} {\bibinfo {author} {\bibfnamefont {T.}~\bibnamefont
  {Sukhbold}}, \bibinfo {author} {\bibfnamefont {S.}~\bibnamefont {Woosley}}, \
  and\ \bibinfo {author} {\bibfnamefont {A.}~\bibnamefont {Heger}},\ }\href
  {\doibase 10.3847/1538-4357/aac2da} {\bibfield  {journal} {\bibinfo
  {journal} {Astrophys. J.}\ }\textbf {\bibinfo {volume} {860}},\ \bibinfo
  {pages} {93} (\bibinfo {year} {2018})},\ \Eprint
  {http://arxiv.org/abs/1710.03243} {arXiv:1710.03243 [astro-ph.HE]}
  \BibitemShut {NoStop}%
\bibitem [{\citenamefont {Rampp}\ and\ \citenamefont
  {Janka}(2002)}]{Rampp:2002bq}%
  \BibitemOpen
  \bibfield  {author} {\bibinfo {author} {\bibfnamefont {M.}~\bibnamefont
  {Rampp}}\ and\ \bibinfo {author} {\bibfnamefont {H.~T.}\ \bibnamefont
  {Janka}},\ }\href {\doibase 10.1051/0004-6361:20021398} {\bibfield  {journal}
  {\bibinfo  {journal} {Astron. Astrophys.}\ }\textbf {\bibinfo {volume}
  {396}},\ \bibinfo {pages} {361} (\bibinfo {year} {2002})},\ \Eprint
  {http://arxiv.org/abs/astro-ph/0203101} {arXiv:astro-ph/0203101} \BibitemShut
  {NoStop}%
\bibitem [{\citenamefont {Buras}\ \emph {et~al.}(2006)\citenamefont {Buras},
  \citenamefont {Rampp}, \citenamefont {Janka},\ and\ \citenamefont
  {Kifonidis}}]{Buras:2005rp}%
  \BibitemOpen
  \bibfield  {author} {\bibinfo {author} {\bibfnamefont {R.}~\bibnamefont
  {Buras}}, \bibinfo {author} {\bibfnamefont {M.}~\bibnamefont {Rampp}},
  \bibinfo {author} {\bibfnamefont {H.~T.}\ \bibnamefont {Janka}}, \ and\
  \bibinfo {author} {\bibfnamefont {K.}~\bibnamefont {Kifonidis}},\ }\href
  {\doibase 10.1051/0004-6361:20053783} {\bibfield  {journal} {\bibinfo
  {journal} {Astron. Astrophys.}\ }\textbf {\bibinfo {volume} {447}},\ \bibinfo
  {pages} {1049} (\bibinfo {year} {2006})},\ \Eprint
  {http://arxiv.org/abs/astro-ph/0507135} {arXiv:astro-ph/0507135} \BibitemShut
  {NoStop}%
\bibitem [{\citenamefont {Janka}(2012)}]{Janka:2012wk}%
  \BibitemOpen
  \bibfield  {author} {\bibinfo {author} {\bibfnamefont {H.-T.}\ \bibnamefont
  {Janka}},\ }\href {\doibase 10.1146/annurev-nucl-102711-094901} {\bibfield
  {journal} {\bibinfo  {journal} {Ann. Rev. Nucl. Part. Sci.}\ }\textbf
  {\bibinfo {volume} {62}},\ \bibinfo {pages} {407} (\bibinfo {year} {2012})},\
  \Eprint {http://arxiv.org/abs/1206.2503} {arXiv:1206.2503 [astro-ph.SR]}
  \BibitemShut {NoStop}%
\bibitem [{\citenamefont {Bollig}\ \emph {et~al.}(2017)\citenamefont {Bollig},
  \citenamefont {Janka}, \citenamefont {Lohs}, \citenamefont {Martinez-Pinedo},
  \citenamefont {Horowitz},\ and\ \citenamefont {Melson}}]{Bollig:2017lki}%
  \BibitemOpen
  \bibfield  {author} {\bibinfo {author} {\bibfnamefont {R.}~\bibnamefont
  {Bollig}}, \bibinfo {author} {\bibfnamefont {H.~T.}\ \bibnamefont {Janka}},
  \bibinfo {author} {\bibfnamefont {A.}~\bibnamefont {Lohs}}, \bibinfo {author}
  {\bibfnamefont {G.}~\bibnamefont {Martinez-Pinedo}}, \bibinfo {author}
  {\bibfnamefont {C.~J.}\ \bibnamefont {Horowitz}}, \ and\ \bibinfo {author}
  {\bibfnamefont {T.}~\bibnamefont {Melson}},\ }\href {\doibase
  10.1103/PhysRevLett.119.242702} {\bibfield  {journal} {\bibinfo  {journal}
  {Phys. Rev. Lett.}\ }\textbf {\bibinfo {volume} {119}},\ \bibinfo {pages}
  {242702} (\bibinfo {year} {2017})},\ \Eprint
  {http://arxiv.org/abs/1706.04630} {arXiv:1706.04630 [astro-ph.HE]}
  \BibitemShut {NoStop}%
\bibitem [{\citenamefont {Fiorillo}\ \emph {et~al.}(2023)\citenamefont
  {Fiorillo}, \citenamefont {Heinlein}, \citenamefont {Janka}, \citenamefont
  {Raffelt}, \citenamefont {Vitagliano},\ and\ \citenamefont
  {Bollig}}]{Fiorillo:2023frv}%
  \BibitemOpen
  \bibfield  {author} {\bibinfo {author} {\bibfnamefont {D.~F.~G.}\
  \bibnamefont {Fiorillo}}, \bibinfo {author} {\bibfnamefont {M.}~\bibnamefont
  {Heinlein}}, \bibinfo {author} {\bibfnamefont {H.-T.}\ \bibnamefont {Janka}},
  \bibinfo {author} {\bibfnamefont {G.}~\bibnamefont {Raffelt}}, \bibinfo
  {author} {\bibfnamefont {E.}~\bibnamefont {Vitagliano}}, \ and\ \bibinfo
  {author} {\bibfnamefont {R.}~\bibnamefont {Bollig}},\ }\href {\doibase
  10.1103/PhysRevD.108.083040} {\bibfield  {journal} {\bibinfo  {journal}
  {Phys. Rev. D}\ }\textbf {\bibinfo {volume} {108}},\ \bibinfo {pages}
  {083040} (\bibinfo {year} {2023})},\ \Eprint
  {http://arxiv.org/abs/2308.01403} {arXiv:2308.01403 [astro-ph.HE]}
  \BibitemShut {NoStop}%
\bibitem [{\citenamefont {Rembiasz}\ \emph {et~al.}(2018)\citenamefont
  {Rembiasz}, \citenamefont {Obergaulinger}, \citenamefont {Masip},
  \citenamefont {P\'erez-Garc\'\i{}a}, \citenamefont {Aloy},\ and\
  \citenamefont {Albertus}}]{Rembiasz:2018lok}%
  \BibitemOpen
  \bibfield  {author} {\bibinfo {author} {\bibfnamefont {T.}~\bibnamefont
  {Rembiasz}}, \bibinfo {author} {\bibfnamefont {M.}~\bibnamefont
  {Obergaulinger}}, \bibinfo {author} {\bibfnamefont {M.}~\bibnamefont
  {Masip}}, \bibinfo {author} {\bibfnamefont {M.~A.}\ \bibnamefont
  {P\'erez-Garc\'\i{}a}}, \bibinfo {author} {\bibfnamefont {M.-A.}\
  \bibnamefont {Aloy}}, \ and\ \bibinfo {author} {\bibfnamefont
  {C.}~\bibnamefont {Albertus}},\ }\href {\doibase 10.1103/PhysRevD.98.103010}
  {\bibfield  {journal} {\bibinfo  {journal} {Phys. Rev. D}\ }\textbf {\bibinfo
  {volume} {98}},\ \bibinfo {pages} {103010} (\bibinfo {year} {2018})},\
  \Eprint {http://arxiv.org/abs/1806.03300} {arXiv:1806.03300 [astro-ph.HE]}
  \BibitemShut {NoStop}%
\bibitem [{\citenamefont {Fischer}\ \emph {et~al.}(2021)\citenamefont
  {Fischer}, \citenamefont {Carenza}, \citenamefont {Fore}, \citenamefont
  {Giannotti}, \citenamefont {Mirizzi},\ and\ \citenamefont
  {Reddy}}]{Fischer:2021jfm}%
  \BibitemOpen
  \bibfield  {author} {\bibinfo {author} {\bibfnamefont {T.}~\bibnamefont
  {Fischer}}, \bibinfo {author} {\bibfnamefont {P.}~\bibnamefont {Carenza}},
  \bibinfo {author} {\bibfnamefont {B.}~\bibnamefont {Fore}}, \bibinfo {author}
  {\bibfnamefont {M.}~\bibnamefont {Giannotti}}, \bibinfo {author}
  {\bibfnamefont {A.}~\bibnamefont {Mirizzi}}, \ and\ \bibinfo {author}
  {\bibfnamefont {S.}~\bibnamefont {Reddy}},\ }\href {\doibase
  10.1103/PhysRevD.104.103012} {\bibfield  {journal} {\bibinfo  {journal}
  {Phys. Rev. D}\ }\textbf {\bibinfo {volume} {104}},\ \bibinfo {pages}
  {103012} (\bibinfo {year} {2021})},\ \Eprint
  {http://arxiv.org/abs/2108.13726} {arXiv:2108.13726 [hep-ph]} \BibitemShut
  {NoStop}%
\bibitem [{\citenamefont {Betranhandy}\ and\ \citenamefont
  {O'Connor}(2022)}]{Betranhandy:2022bvr}%
  \BibitemOpen
  \bibfield  {author} {\bibinfo {author} {\bibfnamefont {A.}~\bibnamefont
  {Betranhandy}}\ and\ \bibinfo {author} {\bibfnamefont {E.}~\bibnamefont
  {O'Connor}},\ }\href {\doibase 10.1103/PhysRevD.106.063019} {\bibfield
  {journal} {\bibinfo  {journal} {Phys. Rev. D}\ }\textbf {\bibinfo {volume}
  {106}},\ \bibinfo {pages} {063019} (\bibinfo {year} {2022})},\ \Eprint
  {http://arxiv.org/abs/2204.00503} {arXiv:2204.00503 [astro-ph.HE]}
  \BibitemShut {NoStop}%
\bibitem [{\citenamefont {Mori}\ \emph {et~al.}(2024)\citenamefont {Mori},
  \citenamefont {Takiwaki}, \citenamefont {Kotake},\ and\ \citenamefont
  {Horiuchi}}]{Mori:2024vrf}%
  \BibitemOpen
  \bibfield  {author} {\bibinfo {author} {\bibfnamefont {K.}~\bibnamefont
  {Mori}}, \bibinfo {author} {\bibfnamefont {T.}~\bibnamefont {Takiwaki}},
  \bibinfo {author} {\bibfnamefont {K.}~\bibnamefont {Kotake}}, \ and\ \bibinfo
  {author} {\bibfnamefont {S.}~\bibnamefont {Horiuchi}},\ }\href {\doibase
  10.1103/PhysRevD.110.023031} {\bibfield  {journal} {\bibinfo  {journal}
  {Phys. Rev. D}\ }\textbf {\bibinfo {volume} {110}},\ \bibinfo {pages}
  {023031} (\bibinfo {year} {2024})},\ \Eprint
  {http://arxiv.org/abs/2402.14333} {arXiv:2402.14333 [astro-ph.HE]}
  \BibitemShut {NoStop}%
\bibitem [{\citenamefont {Carenza}\ and\ \citenamefont
  {Lucente}(2021{\natexlab{a}})}]{Carenza:2021osu}%
  \BibitemOpen
  \bibfield  {author} {\bibinfo {author} {\bibfnamefont {P.}~\bibnamefont
  {Carenza}}\ and\ \bibinfo {author} {\bibfnamefont {G.}~\bibnamefont
  {Lucente}},\ }\href {\doibase 10.1103/PhysRevD.103.123024} {\bibfield
  {journal} {\bibinfo  {journal} {Phys. Rev. D}\ }\textbf {\bibinfo {volume}
  {103}},\ \bibinfo {pages} {123024} (\bibinfo {year} {2021}{\natexlab{a}})},\
  \Eprint {http://arxiv.org/abs/2104.09524} {arXiv:2104.09524 [hep-ph]}
  \BibitemShut {NoStop}%
\bibitem [{\citenamefont {Carenza}\ and\ \citenamefont
  {Lucente}(2021{\natexlab{b}})}]{Carenza:2021pcm}%
  \BibitemOpen
  \bibfield  {author} {\bibinfo {author} {\bibfnamefont {P.}~\bibnamefont
  {Carenza}}\ and\ \bibinfo {author} {\bibfnamefont {G.}~\bibnamefont
  {Lucente}},\ }\href {\doibase 10.1103/PhysRevD.104.103007} {\bibfield
  {journal} {\bibinfo  {journal} {Phys. Rev. D}\ }\textbf {\bibinfo {volume}
  {104}},\ \bibinfo {pages} {103007} (\bibinfo {year} {2021}{\natexlab{b}})},\
  \bibinfo {note} {[Erratum: Phys.Rev.D 110, 049901 (2024)]},\ \Eprint
  {http://arxiv.org/abs/2107.12393} {arXiv:2107.12393 [hep-ph]} \BibitemShut
  {NoStop}%
\bibitem [{\citenamefont {Di~Luzio}\ \emph {et~al.}(2020)\citenamefont
  {Di~Luzio}, \citenamefont {Giannotti}, \citenamefont {Nardi},\ and\
  \citenamefont {Visinelli}}]{DiLuzio:2020wdo}%
  \BibitemOpen
  \bibfield  {author} {\bibinfo {author} {\bibfnamefont {L.}~\bibnamefont
  {Di~Luzio}}, \bibinfo {author} {\bibfnamefont {M.}~\bibnamefont {Giannotti}},
  \bibinfo {author} {\bibfnamefont {E.}~\bibnamefont {Nardi}}, \ and\ \bibinfo
  {author} {\bibfnamefont {L.}~\bibnamefont {Visinelli}},\ }\href {\doibase
  10.1016/j.physrep.2020.06.002} {\bibfield  {journal} {\bibinfo  {journal}
  {Phys. Rept.}\ }\textbf {\bibinfo {volume} {870}},\ \bibinfo {pages} {1}
  (\bibinfo {year} {2020})},\ \Eprint {http://arxiv.org/abs/2003.01100}
  {arXiv:2003.01100 [hep-ph]} \BibitemShut {NoStop}%
\bibitem [{\citenamefont {Chang}\ and\ \citenamefont
  {Choi}(1993)}]{Chang:1993gm}%
  \BibitemOpen
  \bibfield  {author} {\bibinfo {author} {\bibfnamefont {S.}~\bibnamefont
  {Chang}}\ and\ \bibinfo {author} {\bibfnamefont {K.}~\bibnamefont {Choi}},\
  }\href {\doibase 10.1016/0370-2693(93)90656-3} {\bibfield  {journal}
  {\bibinfo  {journal} {Phys. Lett. B}\ }\textbf {\bibinfo {volume} {316}},\
  \bibinfo {pages} {51} (\bibinfo {year} {1993})},\ \Eprint
  {http://arxiv.org/abs/hep-ph/9306216} {arXiv:hep-ph/9306216} \BibitemShut
  {NoStop}%
\bibitem [{\citenamefont {Choi}\ \emph {et~al.}(2022)\citenamefont {Choi},
  \citenamefont {Kim}, \citenamefont {Seong},\ and\ \citenamefont
  {Shin}}]{Choi:2021ign}%
  \BibitemOpen
  \bibfield  {author} {\bibinfo {author} {\bibfnamefont {K.}~\bibnamefont
  {Choi}}, \bibinfo {author} {\bibfnamefont {H.~J.}\ \bibnamefont {Kim}},
  \bibinfo {author} {\bibfnamefont {H.}~\bibnamefont {Seong}}, \ and\ \bibinfo
  {author} {\bibfnamefont {C.~S.}\ \bibnamefont {Shin}},\ }\href {\doibase
  10.1007/JHEP02(2022)143} {\bibfield  {journal} {\bibinfo  {journal} {JHEP}\
  }\textbf {\bibinfo {volume} {02}},\ \bibinfo {pages} {143} (\bibinfo {year}
  {2022})},\ \Eprint {http://arxiv.org/abs/2110.01972} {arXiv:2110.01972
  [hep-ph]} \BibitemShut {NoStop}%
\bibitem [{\citenamefont {Workman}\ \emph {et~al.}(2022)\citenamefont {Workman}
  \emph {et~al.}}]{ParticleDataGroup:2022pth}%
  \BibitemOpen
  \bibfield  {author} {\bibinfo {author} {\bibfnamefont {R.~L.}\ \bibnamefont
  {Workman}} \emph {et~al.} (\bibinfo {collaboration} {Particle Data Group}),\
  }\href {\doibase 10.1093/ptep/ptac097} {\bibfield  {journal} {\bibinfo
  {journal} {PTEP}\ }\textbf {\bibinfo {volume} {2022}},\ \bibinfo {pages}
  {083C01} (\bibinfo {year} {2022})}\BibitemShut {NoStop}%
\bibitem [{\citenamefont {Grilli~di Cortona}\ \emph {et~al.}(2016)\citenamefont
  {Grilli~di Cortona}, \citenamefont {Hardy}, \citenamefont {Pardo~Vega},\ and\
  \citenamefont {Villadoro}}]{GrillidiCortona:2015jxo}%
  \BibitemOpen
  \bibfield  {author} {\bibinfo {author} {\bibfnamefont {G.}~\bibnamefont
  {Grilli~di Cortona}}, \bibinfo {author} {\bibfnamefont {E.}~\bibnamefont
  {Hardy}}, \bibinfo {author} {\bibfnamefont {J.}~\bibnamefont {Pardo~Vega}}, \
  and\ \bibinfo {author} {\bibfnamefont {G.}~\bibnamefont {Villadoro}},\ }\href
  {\doibase 10.1007/JHEP01(2016)034} {\bibfield  {journal} {\bibinfo  {journal}
  {JHEP}\ }\textbf {\bibinfo {volume} {01}},\ \bibinfo {pages} {034} (\bibinfo
  {year} {2016})},\ \Eprint {http://arxiv.org/abs/1511.02867} {arXiv:1511.02867
  [hep-ph]} \BibitemShut {NoStop}%
\bibitem [{\citenamefont {Carena}\ and\ \citenamefont
  {Peccei}(1989)}]{Carena:1988kr}%
  \BibitemOpen
  \bibfield  {author} {\bibinfo {author} {\bibfnamefont {M.}~\bibnamefont
  {Carena}}\ and\ \bibinfo {author} {\bibfnamefont {R.~D.}\ \bibnamefont
  {Peccei}},\ }\href {\doibase 10.1103/PhysRevD.40.652} {\bibfield  {journal}
  {\bibinfo  {journal} {Phys. Rev. D}\ }\textbf {\bibinfo {volume} {40}},\
  \bibinfo {pages} {652} (\bibinfo {year} {1989})}\BibitemShut {NoStop}%
\bibitem [{\citenamefont {Brinkmann}\ and\ \citenamefont
  {Turner}(1988)}]{Brinkmann:1988vi}%
  \BibitemOpen
  \bibfield  {author} {\bibinfo {author} {\bibfnamefont {R.~P.}\ \bibnamefont
  {Brinkmann}}\ and\ \bibinfo {author} {\bibfnamefont {M.~S.}\ \bibnamefont
  {Turner}},\ }\href {\doibase 10.1103/PhysRevD.38.2338} {\bibfield  {journal}
  {\bibinfo  {journal} {Phys. Rev. D}\ }\textbf {\bibinfo {volume} {38}},\
  \bibinfo {pages} {2338} (\bibinfo {year} {1988})}\BibitemShut {NoStop}%
\bibitem [{\citenamefont {Raffelt}\ and\ \citenamefont
  {Seckel}(1995)}]{Raffelt:1993ix}%
  \BibitemOpen
  \bibfield  {author} {\bibinfo {author} {\bibfnamefont {G.}~\bibnamefont
  {Raffelt}}\ and\ \bibinfo {author} {\bibfnamefont {D.}~\bibnamefont
  {Seckel}},\ }\href {\doibase 10.1103/PhysRevD.52.1780} {\bibfield  {journal}
  {\bibinfo  {journal} {Phys. Rev. D}\ }\textbf {\bibinfo {volume} {52}},\
  \bibinfo {pages} {1780} (\bibinfo {year} {1995})},\ \Eprint
  {http://arxiv.org/abs/astro-ph/9312019} {arXiv:astro-ph/9312019} \BibitemShut
  {NoStop}%
\bibitem [{\citenamefont {Stoica}\ \emph {et~al.}(2009)\citenamefont {Stoica},
  \citenamefont {Pastrav}, \citenamefont {Horvath},\ and\ \citenamefont
  {Allen}}]{Stoica:2009zh}%
  \BibitemOpen
  \bibfield  {author} {\bibinfo {author} {\bibfnamefont {S.}~\bibnamefont
  {Stoica}}, \bibinfo {author} {\bibfnamefont {B.}~\bibnamefont {Pastrav}},
  \bibinfo {author} {\bibfnamefont {J.~E.}\ \bibnamefont {Horvath}}, \ and\
  \bibinfo {author} {\bibfnamefont {M.~P.}\ \bibnamefont {Allen}},\ }\href
  {\doibase 10.1016/j.nuclphysa.2009.07.007} {\bibfield  {journal} {\bibinfo
  {journal} {Nucl. Phys. A}\ }\textbf {\bibinfo {volume} {828}},\ \bibinfo
  {pages} {439} (\bibinfo {year} {2009})},\ \bibinfo {note} {[Erratum:
  Nucl.Phys.A 832, 148 (2010)]},\ \Eprint {http://arxiv.org/abs/0906.3134}
  {arXiv:0906.3134 [nucl-th]} \BibitemShut {NoStop}%
\bibitem [{\citenamefont {Ericson}\ and\ \citenamefont
  {Mathiot}(1989)}]{Ericson:1988wr}%
  \BibitemOpen
  \bibfield  {author} {\bibinfo {author} {\bibfnamefont {T.~E.~O.}\
  \bibnamefont {Ericson}}\ and\ \bibinfo {author} {\bibfnamefont {J.~F.}\
  \bibnamefont {Mathiot}},\ }\href {\doibase 10.1016/0370-2693(89)91103-9}
  {\bibfield  {journal} {\bibinfo  {journal} {Phys. Lett. B}\ }\textbf
  {\bibinfo {volume} {219}},\ \bibinfo {pages} {507} (\bibinfo {year}
  {1989})}\BibitemShut {NoStop}%
\bibitem [{\citenamefont {Raffelt}\ and\ \citenamefont
  {Seckel}(1991)}]{Raffelt:1991pw}%
  \BibitemOpen
  \bibfield  {author} {\bibinfo {author} {\bibfnamefont {G.}~\bibnamefont
  {Raffelt}}\ and\ \bibinfo {author} {\bibfnamefont {D.}~\bibnamefont
  {Seckel}},\ }\href {\doibase 10.1103/PhysRevLett.67.2605} {\bibfield
  {journal} {\bibinfo  {journal} {Phys. Rev. Lett.}\ }\textbf {\bibinfo
  {volume} {67}},\ \bibinfo {pages} {2605} (\bibinfo {year}
  {1991})}\BibitemShut {NoStop}%
\bibitem [{\citenamefont {Janka}\ \emph {et~al.}(1996)\citenamefont {Janka},
  \citenamefont {Keil}, \citenamefont {Raffelt},\ and\ \citenamefont
  {Seckel}}]{Janka:1995ir}%
  \BibitemOpen
  \bibfield  {author} {\bibinfo {author} {\bibfnamefont {H.-T.}\ \bibnamefont
  {Janka}}, \bibinfo {author} {\bibfnamefont {W.}~\bibnamefont {Keil}},
  \bibinfo {author} {\bibfnamefont {G.}~\bibnamefont {Raffelt}}, \ and\
  \bibinfo {author} {\bibfnamefont {D.}~\bibnamefont {Seckel}},\ }\href
  {\doibase 10.1103/PhysRevLett.76.2621} {\bibfield  {journal} {\bibinfo
  {journal} {Phys. Rev. Lett.}\ }\textbf {\bibinfo {volume} {76}},\ \bibinfo
  {pages} {2621} (\bibinfo {year} {1996})},\ \Eprint
  {http://arxiv.org/abs/astro-ph/9507023} {arXiv:astro-ph/9507023} \BibitemShut
  {NoStop}%
\bibitem [{\citenamefont {Springmann}\ \emph {et~al.}(2024)\citenamefont
  {Springmann}, \citenamefont {Stadlbauer}, \citenamefont {Stelzl},\ and\
  \citenamefont {Weiler}}]{Springmann:2024mjp}%
  \BibitemOpen
  \bibfield  {author} {\bibinfo {author} {\bibfnamefont {K.}~\bibnamefont
  {Springmann}}, \bibinfo {author} {\bibfnamefont {M.}~\bibnamefont
  {Stadlbauer}}, \bibinfo {author} {\bibfnamefont {S.}~\bibnamefont {Stelzl}},
  \ and\ \bibinfo {author} {\bibfnamefont {A.}~\bibnamefont {Weiler}},\
  }\href@noop {} {\  (\bibinfo {year} {2024})},\ \Eprint
  {http://arxiv.org/abs/2410.10945} {arXiv:2410.10945 [hep-ph]} \BibitemShut
  {NoStop}%
\bibitem [{\citenamefont {Turner}(1992)}]{Turner:1991ax}%
  \BibitemOpen
  \bibfield  {author} {\bibinfo {author} {\bibfnamefont {M.~S.}\ \bibnamefont
  {Turner}},\ }\href {\doibase 10.1103/PhysRevD.45.1066} {\bibfield  {journal}
  {\bibinfo  {journal} {Phys. Rev. D}\ }\textbf {\bibinfo {volume} {45}},\
  \bibinfo {pages} {1066} (\bibinfo {year} {1992})}\BibitemShut {NoStop}%
\bibitem [{\citenamefont {Fore}\ and\ \citenamefont
  {Reddy}(2020)}]{Fore:2019wib}%
  \BibitemOpen
  \bibfield  {author} {\bibinfo {author} {\bibfnamefont {B.}~\bibnamefont
  {Fore}}\ and\ \bibinfo {author} {\bibfnamefont {S.}~\bibnamefont {Reddy}},\
  }\href {\doibase 10.1103/PhysRevC.101.035809} {\bibfield  {journal} {\bibinfo
   {journal} {Phys. Rev. C}\ }\textbf {\bibinfo {volume} {101}},\ \bibinfo
  {pages} {035809} (\bibinfo {year} {2020})},\ \Eprint
  {http://arxiv.org/abs/1911.02632} {arXiv:1911.02632 [astro-ph.HE]}
  \BibitemShut {NoStop}%
\bibitem [{\citenamefont {Ho}\ \emph {et~al.}(2023)\citenamefont {Ho},
  \citenamefont {Kim}, \citenamefont {Ko},\ and\ \citenamefont
  {Park}}]{Ho:2022oaw}%
  \BibitemOpen
  \bibfield  {author} {\bibinfo {author} {\bibfnamefont {S.-Y.}\ \bibnamefont
  {Ho}}, \bibinfo {author} {\bibfnamefont {J.}~\bibnamefont {Kim}}, \bibinfo
  {author} {\bibfnamefont {P.}~\bibnamefont {Ko}}, \ and\ \bibinfo {author}
  {\bibfnamefont {J.-h.}\ \bibnamefont {Park}},\ }\href {\doibase
  10.1103/PhysRevD.107.075002} {\bibfield  {journal} {\bibinfo  {journal}
  {Phys. Rev. D}\ }\textbf {\bibinfo {volume} {107}},\ \bibinfo {pages}
  {075002} (\bibinfo {year} {2023})},\ \Eprint
  {http://arxiv.org/abs/2212.01155} {arXiv:2212.01155 [hep-ph]} \BibitemShut
  {NoStop}%
\bibitem [{\citenamefont {Carenza}(2023)}]{Carenza:2023lci}%
  \BibitemOpen
  \bibfield  {author} {\bibinfo {author} {\bibfnamefont {P.}~\bibnamefont
  {Carenza}},\ }\href {\doibase 10.1140/epjp/s13360-023-04484-2} {\bibfield
  {journal} {\bibinfo  {journal} {Eur. Phys. J. Plus}\ }\textbf {\bibinfo
  {volume} {138}},\ \bibinfo {pages} {836} (\bibinfo {year} {2023})},\ \Eprint
  {http://arxiv.org/abs/2309.14798} {arXiv:2309.14798 [hep-ph]} \BibitemShut
  {NoStop}%
\bibitem [{\citenamefont {Altmann}\ \emph {et~al.}(1995)\citenamefont
  {Altmann}, \citenamefont {von Feilitzsch}, \citenamefont {Hagner},
  \citenamefont {Oberauer}, \citenamefont {Declais},\ and\ \citenamefont
  {Kajfasz}}]{Altmann:1995bw}%
  \BibitemOpen
  \bibfield  {author} {\bibinfo {author} {\bibfnamefont {M.}~\bibnamefont
  {Altmann}}, \bibinfo {author} {\bibfnamefont {F.}~\bibnamefont {von
  Feilitzsch}}, \bibinfo {author} {\bibfnamefont {C.}~\bibnamefont {Hagner}},
  \bibinfo {author} {\bibfnamefont {L.}~\bibnamefont {Oberauer}}, \bibinfo
  {author} {\bibfnamefont {Y.}~\bibnamefont {Declais}}, \ and\ \bibinfo
  {author} {\bibfnamefont {E.}~\bibnamefont {Kajfasz}},\ }\href {\doibase
  10.1007/BF01566670} {\bibfield  {journal} {\bibinfo  {journal} {Z. Phys. C}\
  }\textbf {\bibinfo {volume} {68}},\ \bibinfo {pages} {221} (\bibinfo {year}
  {1995})}\BibitemShut {NoStop}%
\bibitem [{\citenamefont {Suliga}\ \emph {et~al.}(2020)\citenamefont {Suliga},
  \citenamefont {Tamborra},\ and\ \citenamefont {Wu}}]{Suliga:2020vpz}%
  \BibitemOpen
  \bibfield  {author} {\bibinfo {author} {\bibfnamefont {A.~M.}\ \bibnamefont
  {Suliga}}, \bibinfo {author} {\bibfnamefont {I.}~\bibnamefont {Tamborra}}, \
  and\ \bibinfo {author} {\bibfnamefont {M.-R.}\ \bibnamefont {Wu}},\ }\href
  {\doibase 10.1088/1475-7516/2020/08/018} {\bibfield  {journal} {\bibinfo
  {journal} {JCAP}\ }\textbf {\bibinfo {volume} {08}},\ \bibinfo {pages} {018}
  (\bibinfo {year} {2020})},\ \Eprint {http://arxiv.org/abs/2004.11389}
  {arXiv:2004.11389 [astro-ph.HE]} \BibitemShut {NoStop}%
\bibitem [{\citenamefont {Arg\"uelles}\ \emph {et~al.}(2019)\citenamefont
  {Arg\"uelles}, \citenamefont {Brdar},\ and\ \citenamefont
  {Kopp}}]{Arguelles:2016uwb}%
  \BibitemOpen
  \bibfield  {author} {\bibinfo {author} {\bibfnamefont {C.~A.}\ \bibnamefont
  {Arg\"uelles}}, \bibinfo {author} {\bibfnamefont {V.}~\bibnamefont {Brdar}},
  \ and\ \bibinfo {author} {\bibfnamefont {J.}~\bibnamefont {Kopp}},\ }\href
  {\doibase 10.1103/PhysRevD.99.043012} {\bibfield  {journal} {\bibinfo
  {journal} {Phys. Rev. D}\ }\textbf {\bibinfo {volume} {99}},\ \bibinfo
  {pages} {043012} (\bibinfo {year} {2019})},\ \Eprint
  {http://arxiv.org/abs/1605.00654} {arXiv:1605.00654 [hep-ph]} \BibitemShut
  {NoStop}%
\bibitem [{\citenamefont {Alekhin}\ \emph {et~al.}(2016)\citenamefont {Alekhin}
  \emph {et~al.}}]{Alekhin:2015byh}%
  \BibitemOpen
  \bibfield  {author} {\bibinfo {author} {\bibfnamefont {S.}~\bibnamefont
  {Alekhin}} \emph {et~al.},\ }\href {\doibase 10.1088/0034-4885/79/12/124201}
  {\bibfield  {journal} {\bibinfo  {journal} {Rept. Prog. Phys.}\ }\textbf
  {\bibinfo {volume} {79}},\ \bibinfo {pages} {124201} (\bibinfo {year}
  {2016})},\ \Eprint {http://arxiv.org/abs/1504.04855} {arXiv:1504.04855
  [hep-ph]} \BibitemShut {NoStop}%
\bibitem [{\citenamefont {Okun}(1982)}]{Okun:1982xi}%
  \BibitemOpen
  \bibfield  {author} {\bibinfo {author} {\bibfnamefont {L.~B.}\ \bibnamefont
  {Okun}},\ }\href@noop {} {\bibfield  {journal} {\bibinfo  {journal} {Sov.
  Phys. JETP}\ }\textbf {\bibinfo {volume} {56}},\ \bibinfo {pages} {502}
  (\bibinfo {year} {1982})}\BibitemShut {NoStop}%
\bibitem [{\citenamefont {Holdom}(1986)}]{Holdom:1985ag}%
  \BibitemOpen
  \bibfield  {author} {\bibinfo {author} {\bibfnamefont {B.}~\bibnamefont
  {Holdom}},\ }\href {\doibase 10.1016/0370-2693(86)91377-8} {\bibfield
  {journal} {\bibinfo  {journal} {Phys. Lett. B}\ }\textbf {\bibinfo {volume}
  {166}},\ \bibinfo {pages} {196} (\bibinfo {year} {1986})}\BibitemShut
  {NoStop}%
\bibitem [{\citenamefont {Nguyen}\ \emph {et~al.}(2024)\citenamefont {Nguyen},
  \citenamefont {John}, \citenamefont {Linden},\ and\ \citenamefont
  {Tait}}]{Nguyen:2024kwy}%
  \BibitemOpen
  \bibfield  {author} {\bibinfo {author} {\bibfnamefont {T.~T.~Q.}\
  \bibnamefont {Nguyen}}, \bibinfo {author} {\bibfnamefont {I.}~\bibnamefont
  {John}}, \bibinfo {author} {\bibfnamefont {T.}~\bibnamefont {Linden}}, \ and\
  \bibinfo {author} {\bibfnamefont {T.~M.~P.}\ \bibnamefont {Tait}},\
  }\href@noop {} {\  (\bibinfo {year} {2024})},\ \Eprint
  {http://arxiv.org/abs/2412.00180} {arXiv:2412.00180 [hep-ph]} \BibitemShut
  {NoStop}%
\bibitem [{\citenamefont {Foot}\ and\ \citenamefont {He}(1991)}]{Foot:1991kb}%
  \BibitemOpen
  \bibfield  {author} {\bibinfo {author} {\bibfnamefont {R.}~\bibnamefont
  {Foot}}\ and\ \bibinfo {author} {\bibfnamefont {X.-G.}\ \bibnamefont {He}},\
  }\href {\doibase 10.1016/0370-2693(91)90901-2} {\bibfield  {journal}
  {\bibinfo  {journal} {Phys. Lett. B}\ }\textbf {\bibinfo {volume} {267}},\
  \bibinfo {pages} {509} (\bibinfo {year} {1991})}\BibitemShut {NoStop}%
\bibitem [{\citenamefont {Kazanas}\ \emph {et~al.}(2014)\citenamefont
  {Kazanas}, \citenamefont {Mohapatra}, \citenamefont {Nussinov}, \citenamefont
  {Teplitz},\ and\ \citenamefont {Zhang}}]{Kazanas:2014mca}%
  \BibitemOpen
  \bibfield  {author} {\bibinfo {author} {\bibfnamefont {D.}~\bibnamefont
  {Kazanas}}, \bibinfo {author} {\bibfnamefont {R.~N.}\ \bibnamefont
  {Mohapatra}}, \bibinfo {author} {\bibfnamefont {S.}~\bibnamefont {Nussinov}},
  \bibinfo {author} {\bibfnamefont {V.~L.}\ \bibnamefont {Teplitz}}, \ and\
  \bibinfo {author} {\bibfnamefont {Y.}~\bibnamefont {Zhang}},\ }\href
  {\doibase 10.1016/j.nuclphysb.2014.11.009} {\bibfield  {journal} {\bibinfo
  {journal} {Nucl. Phys. B}\ }\textbf {\bibinfo {volume} {890}},\ \bibinfo
  {pages} {17} (\bibinfo {year} {2014})},\ \Eprint
  {http://arxiv.org/abs/1410.0221} {arXiv:1410.0221 [hep-ph]} \BibitemShut
  {NoStop}%
\bibitem [{\citenamefont {Hardy}\ and\ \citenamefont
  {Lasenby}(2017)}]{Hardy:2016kme}%
  \BibitemOpen
  \bibfield  {author} {\bibinfo {author} {\bibfnamefont {E.}~\bibnamefont
  {Hardy}}\ and\ \bibinfo {author} {\bibfnamefont {R.}~\bibnamefont
  {Lasenby}},\ }\href {\doibase 10.1007/JHEP02(2017)033} {\bibfield  {journal}
  {\bibinfo  {journal} {JHEP}\ }\textbf {\bibinfo {volume} {02}},\ \bibinfo
  {pages} {033} (\bibinfo {year} {2017})},\ \Eprint
  {http://arxiv.org/abs/1611.05852} {arXiv:1611.05852 [hep-ph]} \BibitemShut
  {NoStop}%
\bibitem [{\citenamefont {Stetina}\ \emph {et~al.}(2018)\citenamefont
  {Stetina}, \citenamefont {Rrapaj},\ and\ \citenamefont
  {Reddy}}]{Stetina:2017ozh}%
  \BibitemOpen
  \bibfield  {author} {\bibinfo {author} {\bibfnamefont {S.}~\bibnamefont
  {Stetina}}, \bibinfo {author} {\bibfnamefont {E.}~\bibnamefont {Rrapaj}}, \
  and\ \bibinfo {author} {\bibfnamefont {S.}~\bibnamefont {Reddy}},\ }\href
  {\doibase 10.1103/PhysRevC.97.045801} {\bibfield  {journal} {\bibinfo
  {journal} {Phys. Rev. C}\ }\textbf {\bibinfo {volume} {97}},\ \bibinfo
  {pages} {045801} (\bibinfo {year} {2018})},\ \Eprint
  {http://arxiv.org/abs/1712.05447} {arXiv:1712.05447 [astro-ph.HE]}
  \BibitemShut {NoStop}%
\bibitem [{\citenamefont {DeRocco}\ \emph {et~al.}(2019)\citenamefont
  {DeRocco}, \citenamefont {Graham}, \citenamefont {Kasen}, \citenamefont
  {Marques-Tavares},\ and\ \citenamefont {Rajendran}}]{DeRocco:2019njg}%
  \BibitemOpen
  \bibfield  {author} {\bibinfo {author} {\bibfnamefont {W.}~\bibnamefont
  {DeRocco}}, \bibinfo {author} {\bibfnamefont {P.~W.}\ \bibnamefont {Graham}},
  \bibinfo {author} {\bibfnamefont {D.}~\bibnamefont {Kasen}}, \bibinfo
  {author} {\bibfnamefont {G.}~\bibnamefont {Marques-Tavares}}, \ and\ \bibinfo
  {author} {\bibfnamefont {S.}~\bibnamefont {Rajendran}},\ }\href {\doibase
  10.1007/JHEP02(2019)171} {\bibfield  {journal} {\bibinfo  {journal} {JHEP}\
  }\textbf {\bibinfo {volume} {02}},\ \bibinfo {pages} {171} (\bibinfo {year}
  {2019})},\ \Eprint {http://arxiv.org/abs/1901.08596} {arXiv:1901.08596
  [hep-ph]} \BibitemShut {NoStop}%
\bibitem [{\citenamefont {Syvolap}\ and\ \citenamefont
  {Ruchayskiy}(2024)}]{Syvolap:2024hdh}%
  \BibitemOpen
  \bibfield  {author} {\bibinfo {author} {\bibfnamefont {V.}~\bibnamefont
  {Syvolap}}\ and\ \bibinfo {author} {\bibfnamefont {O.}~\bibnamefont
  {Ruchayskiy}},\ }\href@noop {} {\  (\bibinfo {year} {2024})},\ \Eprint
  {http://arxiv.org/abs/2404.19191} {arXiv:2404.19191 [hep-ph]} \BibitemShut
  {NoStop}%
\bibitem [{\citenamefont {Rrapaj}\ and\ \citenamefont
  {Reddy}(2016)}]{Rrapaj:2015wgs}%
  \BibitemOpen
  \bibfield  {author} {\bibinfo {author} {\bibfnamefont {E.}~\bibnamefont
  {Rrapaj}}\ and\ \bibinfo {author} {\bibfnamefont {S.}~\bibnamefont {Reddy}},\
  }\href {\doibase 10.1103/PhysRevC.94.045805} {\bibfield  {journal} {\bibinfo
  {journal} {Phys. Rev. C}\ }\textbf {\bibinfo {volume} {94}},\ \bibinfo
  {pages} {045805} (\bibinfo {year} {2016})},\ \Eprint
  {http://arxiv.org/abs/1511.09136} {arXiv:1511.09136 [nucl-th]} \BibitemShut
  {NoStop}%
\bibitem [{\citenamefont {Li}\ \emph {et~al.}(2011)\citenamefont {Li},
  \citenamefont {Chornock}, \citenamefont {Leaman}, \citenamefont {Filippenko},
  \citenamefont {Poznanski}, \citenamefont {Wang}, \citenamefont
  {Ganeshalingam},\ and\ \citenamefont {Mannucci}}]{Li:2010kd}%
  \BibitemOpen
  \bibfield  {author} {\bibinfo {author} {\bibfnamefont {W.}~\bibnamefont
  {Li}}, \bibinfo {author} {\bibfnamefont {R.}~\bibnamefont {Chornock}},
  \bibinfo {author} {\bibfnamefont {J.}~\bibnamefont {Leaman}}, \bibinfo
  {author} {\bibfnamefont {A.~V.}\ \bibnamefont {Filippenko}}, \bibinfo
  {author} {\bibfnamefont {D.}~\bibnamefont {Poznanski}}, \bibinfo {author}
  {\bibfnamefont {X.}~\bibnamefont {Wang}}, \bibinfo {author} {\bibfnamefont
  {M.}~\bibnamefont {Ganeshalingam}}, \ and\ \bibinfo {author} {\bibfnamefont
  {F.}~\bibnamefont {Mannucci}},\ }\href {\doibase
  10.1111/j.1365-2966.2011.18162.x} {\bibfield  {journal} {\bibinfo  {journal}
  {Mon. Not. Roy. Astron. Soc.}\ }\textbf {\bibinfo {volume} {412}},\ \bibinfo
  {pages} {1473} (\bibinfo {year} {2011})},\ \Eprint
  {http://arxiv.org/abs/1006.4613} {arXiv:1006.4613 [astro-ph.SR]} \BibitemShut
  {NoStop}%
\bibitem [{\citenamefont {{Ginzburg}}\ and\ \citenamefont
  {{Syrovatskii}}(1969)}]{Ginz&Syr}%
  \BibitemOpen
  \bibfield  {author} {\bibinfo {author} {\bibfnamefont {V.~L.}\ \bibnamefont
  {{Ginzburg}}}\ and\ \bibinfo {author} {\bibfnamefont {S.~I.}\ \bibnamefont
  {{Syrovatskii}}},\ }\href@noop {} {\emph {\bibinfo {title} {{The origin of
  cosmic rays}}}}\ (\bibinfo  {publisher} {Gordon \& Breach Publishing Group},\
  \bibinfo {year} {1969})\BibitemShut {NoStop}%
\bibitem [{\citenamefont {{Ginzburg}}\ \emph {et~al.}(1980)\citenamefont
  {{Ginzburg}}, \citenamefont {{Khazan}},\ and\ \citenamefont
  {{Ptuskin}}}]{Ginzburg_H}%
  \BibitemOpen
  \bibfield  {author} {\bibinfo {author} {\bibfnamefont {V.~L.}\ \bibnamefont
  {{Ginzburg}}}, \bibinfo {author} {\bibfnamefont {I.~M.}\ \bibnamefont
  {{Khazan}}}, \ and\ \bibinfo {author} {\bibfnamefont {V.~S.}\ \bibnamefont
  {{Ptuskin}}},\ }\href {\doibase 10.1007/BF00639701} {\bibfield  {journal}
  {\bibinfo  {journal} {Astrophysics and Space Science}\ }\textbf {\bibinfo
  {volume} {68}},\ \bibinfo {pages} {295} (\bibinfo {year} {1980})}\BibitemShut
  {NoStop}%
\bibitem [{\citenamefont {{Strong}}\ and\ \citenamefont
  {{Moskalenko}}(1998)}]{1998ApJ...509..212S}%
  \BibitemOpen
  \bibfield  {author} {\bibinfo {author} {\bibfnamefont {A.~W.}\ \bibnamefont
  {{Strong}}}\ and\ \bibinfo {author} {\bibfnamefont {I.~V.}\ \bibnamefont
  {{Moskalenko}}},\ }\href {\doibase 10.1086/306470} {\bibfield  {journal}
  {\bibinfo  {journal} {The Astrophysical Journal}\ }\textbf {\bibinfo {volume}
  {509}},\ \bibinfo {pages} {212} (\bibinfo {year} {1998})},\ \Eprint
  {http://arxiv.org/abs/astro-ph/9807150} {arXiv:astro-ph/9807150 [astro-ph]}
  \BibitemShut {NoStop}%
\bibitem [{\citenamefont {Strong}\ and\ \citenamefont
  {Moskalenko}(1998)}]{Strong:1998pw}%
  \BibitemOpen
  \bibfield  {author} {\bibinfo {author} {\bibfnamefont {A.~W.}\ \bibnamefont
  {Strong}}\ and\ \bibinfo {author} {\bibfnamefont {I.~V.}\ \bibnamefont
  {Moskalenko}},\ }\href {\doibase 10.1086/306470} {\bibfield  {journal}
  {\bibinfo  {journal} {Astrophys. J.}\ }\textbf {\bibinfo {volume} {509}},\
  \bibinfo {pages} {212} (\bibinfo {year} {1998})},\ \Eprint
  {http://arxiv.org/abs/astro-ph/9807150} {arXiv:astro-ph/9807150} \BibitemShut
  {NoStop}%
\bibitem [{\citenamefont {De~la Torre~Luque}\ \emph
  {et~al.}(2024{\natexlab{d}})\citenamefont {De~la Torre~Luque}, \citenamefont
  {Balaji},\ and\ \citenamefont {Koechler}}]{DelaTorreLuque:2023olp}%
  \BibitemOpen
  \bibfield  {author} {\bibinfo {author} {\bibfnamefont {P.}~\bibnamefont
  {De~la Torre~Luque}}, \bibinfo {author} {\bibfnamefont {S.}~\bibnamefont
  {Balaji}}, \ and\ \bibinfo {author} {\bibfnamefont {J.}~\bibnamefont
  {Koechler}},\ }\href {\doibase 10.3847/1538-4357/ad41e0} {\bibfield
  {journal} {\bibinfo  {journal} {Astrophys. J.}\ }\textbf {\bibinfo {volume}
  {968}},\ \bibinfo {pages} {46} (\bibinfo {year} {2024}{\natexlab{d}})},\
  \Eprint {http://arxiv.org/abs/2311.04979} {arXiv:2311.04979 [hep-ph]}
  \BibitemShut {NoStop}%
\bibitem [{\citenamefont {De~La
  Torre~Luque}(2023)}]{de_la_torre_luque_2023_10076728}%
  \BibitemOpen
  \bibfield  {author} {\bibinfo {author} {\bibfnamefont {P.}~\bibnamefont
  {De~La Torre~Luque}},\ }\href {\doibase 10.5281/zenodo.10076728} {\enquote
  {\bibinfo {title} {"dragon2\_optimized\_dm\&antinuclei"},}\ } (\bibinfo
  {year} {2023})\BibitemShut {NoStop}%
\bibitem [{\citenamefont {Evoli}\ \emph {et~al.}(2017)\citenamefont {Evoli},
  \citenamefont {Gaggero}, \citenamefont {Vittino}, \citenamefont
  {Di~Bernardo}, \citenamefont {Di~Mauro}, \citenamefont {Ligorini},
  \citenamefont {Ullio},\ and\ \citenamefont {Grasso}}]{Evoli:2016xgn}%
  \BibitemOpen
  \bibfield  {author} {\bibinfo {author} {\bibfnamefont {C.}~\bibnamefont
  {Evoli}}, \bibinfo {author} {\bibfnamefont {D.}~\bibnamefont {Gaggero}},
  \bibinfo {author} {\bibfnamefont {A.}~\bibnamefont {Vittino}}, \bibinfo
  {author} {\bibfnamefont {G.}~\bibnamefont {Di~Bernardo}}, \bibinfo {author}
  {\bibfnamefont {M.}~\bibnamefont {Di~Mauro}}, \bibinfo {author}
  {\bibfnamefont {A.}~\bibnamefont {Ligorini}}, \bibinfo {author}
  {\bibfnamefont {P.}~\bibnamefont {Ullio}}, \ and\ \bibinfo {author}
  {\bibfnamefont {D.}~\bibnamefont {Grasso}},\ }\href {\doibase
  10.1088/1475-7516/2017/02/015} {\bibfield  {journal} {\bibinfo  {journal}
  {JCAP}\ }\textbf {\bibinfo {volume} {02}},\ \bibinfo {pages} {015} (\bibinfo
  {year} {2017})},\ \Eprint {http://arxiv.org/abs/1607.07886} {arXiv:1607.07886
  [astro-ph.HE]} \BibitemShut {NoStop}%
\bibitem [{\citenamefont {Evoli}\ \emph {et~al.}(2018)\citenamefont {Evoli},
  \citenamefont {Gaggero}, \citenamefont {Vittino}, \citenamefont {Di~Mauro},
  \citenamefont {Grasso},\ and\ \citenamefont {Mazziotta}}]{Evoli:2017vim}%
  \BibitemOpen
  \bibfield  {author} {\bibinfo {author} {\bibfnamefont {C.}~\bibnamefont
  {Evoli}}, \bibinfo {author} {\bibfnamefont {D.}~\bibnamefont {Gaggero}},
  \bibinfo {author} {\bibfnamefont {A.}~\bibnamefont {Vittino}}, \bibinfo
  {author} {\bibfnamefont {M.}~\bibnamefont {Di~Mauro}}, \bibinfo {author}
  {\bibfnamefont {D.}~\bibnamefont {Grasso}}, \ and\ \bibinfo {author}
  {\bibfnamefont {M.~N.}\ \bibnamefont {Mazziotta}},\ }\href {\doibase
  10.1088/1475-7516/2018/07/006} {\bibfield  {journal} {\bibinfo  {journal}
  {JCAP}\ }\textbf {\bibinfo {volume} {07}},\ \bibinfo {pages} {006} (\bibinfo
  {year} {2018})},\ \Eprint {http://arxiv.org/abs/1711.09616} {arXiv:1711.09616
  [astro-ph.HE]} \BibitemShut {NoStop}%
\bibitem [{\citenamefont {Stone}\ \emph
  {et~al.}(2013{\natexlab{b}})\citenamefont {Stone} \emph
  {et~al.}}]{stone2013voyager}%
  \BibitemOpen
  \bibfield  {author} {\bibinfo {author} {\bibfnamefont {E.}~\bibnamefont
  {Stone}} \emph {et~al.},\ }\href {\doibase 10.1126/science.1236408}
  {\bibfield  {journal} {\bibinfo  {journal} {Science}\ }\textbf {\bibinfo
  {volume} {341}},\ \bibinfo {pages} {150} (\bibinfo {year}
  {2013}{\natexlab{b}})}\BibitemShut {NoStop}%
\bibitem [{\citenamefont {Cummings}\ \emph {et~al.}(2016)\citenamefont
  {Cummings} \emph {et~al.}}]{cummings2016galactic}%
  \BibitemOpen
  \bibfield  {author} {\bibinfo {author} {\bibfnamefont {A.}~\bibnamefont
  {Cummings}} \emph {et~al.},\ }\href {\doibase 10.3847/0004-637X/831/1/18}
  {\bibfield  {journal} {\bibinfo  {journal} {Astrophys. J.}\ }\textbf
  {\bibinfo {volume} {831}},\ \bibinfo {pages} {18} (\bibinfo {year}
  {2016})}\BibitemShut {NoStop}%
\bibitem [{\citenamefont {De~la Torre~Luque}\ \emph
  {et~al.}(2024{\natexlab{e}})\citenamefont {De~la Torre~Luque}, \citenamefont
  {Koechler},\ and\ \citenamefont {Balaji}}]{DelaTorreLuque:2024qms}%
  \BibitemOpen
  \bibfield  {author} {\bibinfo {author} {\bibfnamefont {P.}~\bibnamefont
  {De~la Torre~Luque}}, \bibinfo {author} {\bibfnamefont {J.}~\bibnamefont
  {Koechler}}, \ and\ \bibinfo {author} {\bibfnamefont {S.}~\bibnamefont
  {Balaji}},\ }\href@noop {} {\  (\bibinfo {year} {2024}{\natexlab{e}})},\
  \Eprint {http://arxiv.org/abs/2406.11949} {arXiv:2406.11949 [astro-ph.HE]}
  \BibitemShut {NoStop}%
\bibitem [{\citenamefont {Evoli}\ \emph {et~al.}(2019)\citenamefont {Evoli},
  \citenamefont {Aloisio},\ and\ \citenamefont {Blasi}}]{Evoli:2019wwu}%
  \BibitemOpen
  \bibfield  {author} {\bibinfo {author} {\bibfnamefont {C.}~\bibnamefont
  {Evoli}}, \bibinfo {author} {\bibfnamefont {R.}~\bibnamefont {Aloisio}}, \
  and\ \bibinfo {author} {\bibfnamefont {P.}~\bibnamefont {Blasi}},\ }\href
  {\doibase 10.1103/PhysRevD.99.103023} {\bibfield  {journal} {\bibinfo
  {journal} {Phys. Rev. D}\ }\textbf {\bibinfo {volume} {99}},\ \bibinfo
  {pages} {103023} (\bibinfo {year} {2019})},\ \Eprint
  {http://arxiv.org/abs/1904.10220} {arXiv:1904.10220 [astro-ph.HE]}
  \BibitemShut {NoStop}%
\bibitem [{\citenamefont {Derome}\ \emph {et~al.}(2019)\citenamefont {Derome},
  \citenamefont {Maurin}, \citenamefont {Salati}, \citenamefont {Boudaud},
  \citenamefont {Génolini},\ and\ \citenamefont {Kunzé}}]{Derome_2019}%
  \BibitemOpen
  \bibfield  {author} {\bibinfo {author} {\bibfnamefont {L.}~\bibnamefont
  {Derome}}, \bibinfo {author} {\bibfnamefont {D.}~\bibnamefont {Maurin}},
  \bibinfo {author} {\bibfnamefont {P.}~\bibnamefont {Salati}}, \bibinfo
  {author} {\bibfnamefont {M.}~\bibnamefont {Boudaud}}, \bibinfo {author}
  {\bibfnamefont {Y.}~\bibnamefont {Génolini}}, \ and\ \bibinfo {author}
  {\bibfnamefont {P.}~\bibnamefont {Kunzé}},\ }\href {\doibase
  10.1051/0004-6361/201935717} {\bibfield  {journal} {\bibinfo  {journal}
  {Astronomy \& Astrophysics}\ }\textbf {\bibinfo {volume} {627}},\ \bibinfo
  {pages} {A158} (\bibinfo {year} {2019})}\BibitemShut {NoStop}%
\bibitem [{\citenamefont {Luque}\ \emph {et~al.}(2021)\citenamefont {Luque},
  \citenamefont {Mazziotta}, \citenamefont {Loparco}, \citenamefont {Gargano},\
  and\ \citenamefont {Serini}}]{Luque:2021nxb}%
  \BibitemOpen
  \bibfield  {author} {\bibinfo {author} {\bibfnamefont {P.~D. L.~T.}\
  \bibnamefont {Luque}}, \bibinfo {author} {\bibfnamefont {M.~N.}\ \bibnamefont
  {Mazziotta}}, \bibinfo {author} {\bibfnamefont {F.}~\bibnamefont {Loparco}},
  \bibinfo {author} {\bibfnamefont {F.}~\bibnamefont {Gargano}}, \ and\
  \bibinfo {author} {\bibfnamefont {D.}~\bibnamefont {Serini}},\ }\href
  {\doibase 10.1088/1475-7516/2021/07/010} {\bibfield  {journal} {\bibinfo
  {journal} {JCAP}\ }\textbf {\bibinfo {volume} {07}},\ \bibinfo {pages} {010}
  (\bibinfo {year} {2021})},\ \Eprint {http://arxiv.org/abs/2102.13238}
  {arXiv:2102.13238 [astro-ph.HE]} \BibitemShut {NoStop}%
\bibitem [{\citenamefont {Ferriere}\ \emph {et~al.}(2007)\citenamefont
  {Ferriere}, \citenamefont {Gillard},\ and\ \citenamefont
  {Jean}}]{Ferriere:2007yq}%
  \BibitemOpen
  \bibfield  {author} {\bibinfo {author} {\bibfnamefont {K.}~\bibnamefont
  {Ferriere}}, \bibinfo {author} {\bibfnamefont {W.}~\bibnamefont {Gillard}}, \
  and\ \bibinfo {author} {\bibfnamefont {P.}~\bibnamefont {Jean}},\ }\href
  {\doibase 10.1051/0004-6361:20066992} {\bibfield  {journal} {\bibinfo
  {journal} {Astron. Astrophys.}\ }\textbf {\bibinfo {volume} {467}},\ \bibinfo
  {pages} {611} (\bibinfo {year} {2007})},\ \Eprint
  {http://arxiv.org/abs/astro-ph/0702532} {arXiv:astro-ph/0702532} \BibitemShut
  {NoStop}%
\bibitem [{\citenamefont {Steiman-Cameron}\ \emph {et~al.}(2010)\citenamefont
  {Steiman-Cameron}, \citenamefont {Wolfire},\ and\ \citenamefont
  {Hollenbach}}]{Steiman-Cameron:2010iuq}%
  \BibitemOpen
  \bibfield  {author} {\bibinfo {author} {\bibfnamefont {T.~Y.}\ \bibnamefont
  {Steiman-Cameron}}, \bibinfo {author} {\bibfnamefont {M.}~\bibnamefont
  {Wolfire}}, \ and\ \bibinfo {author} {\bibfnamefont {D.}~\bibnamefont
  {Hollenbach}},\ }\href {\doibase 10.1088/0004-637X/722/2/1460} {\bibfield
  {journal} {\bibinfo  {journal} {Astrophys. J.}\ }\textbf {\bibinfo {volume}
  {722}},\ \bibinfo {pages} {1460} (\bibinfo {year} {2010})}\BibitemShut
  {NoStop}%
\bibitem [{\citenamefont {Lorimer}\ \emph {et~al.}(2006)\citenamefont {Lorimer}
  \emph {et~al.}}]{Lorimer:2006qs}%
  \BibitemOpen
  \bibfield  {author} {\bibinfo {author} {\bibfnamefont {D.~R.}\ \bibnamefont
  {Lorimer}} \emph {et~al.},\ }\href {\doibase
  10.1111/j.1365-2966.2006.10887.x} {\bibfield  {journal} {\bibinfo  {journal}
  {Mon. Not. Roy. Astron. Soc.}\ }\textbf {\bibinfo {volume} {372}},\ \bibinfo
  {pages} {777} (\bibinfo {year} {2006})},\ \Eprint
  {http://arxiv.org/abs/astro-ph/0607640} {arXiv:astro-ph/0607640} \BibitemShut
  {NoStop}%
\bibitem [{\citenamefont {Stone}\ \emph
  {et~al.}(2013{\natexlab{c}})\citenamefont {Stone}, \citenamefont {Cummings},
  \citenamefont {McDonald}, \citenamefont {Heikkila}, \citenamefont {Lal},\
  and\ \citenamefont {Webber}}]{Stone150}%
  \BibitemOpen
  \bibfield  {author} {\bibinfo {author} {\bibfnamefont {E.~C.}\ \bibnamefont
  {Stone}}, \bibinfo {author} {\bibfnamefont {A.~C.}\ \bibnamefont {Cummings}},
  \bibinfo {author} {\bibfnamefont {F.~B.}\ \bibnamefont {McDonald}}, \bibinfo
  {author} {\bibfnamefont {B.~C.}\ \bibnamefont {Heikkila}}, \bibinfo {author}
  {\bibfnamefont {N.}~\bibnamefont {Lal}}, \ and\ \bibinfo {author}
  {\bibfnamefont {W.~R.}\ \bibnamefont {Webber}},\ }\href {\doibase
  10.1126/science.1236408} {\bibfield  {journal} {\bibinfo  {journal}
  {Science}\ }\textbf {\bibinfo {volume} {341}},\ \bibinfo {pages} {150}
  (\bibinfo {year} {2013}{\natexlab{c}})},\ \Eprint
  {http://arxiv.org/abs/https://science.sciencemag.org/content/341/6142/150.full.pdf}
  {https://science.sciencemag.org/content/341/6142/150.full.pdf} \BibitemShut
  {NoStop}%
\bibitem [{\citenamefont {{Strong}}\ \emph
  {et~al.}(1994{\natexlab{a}})\citenamefont {{Strong}}, \citenamefont
  {{Bennett}}, \citenamefont {{Bloemen}}, \citenamefont {{Diehl}},
  \citenamefont {{Hermsen}}, \citenamefont {{Morris}}, \citenamefont
  {{Schoenfelder}}, \citenamefont {{Stacy}}, \citenamefont {{de Vries}},
  \citenamefont {{Varendorff}}, \citenamefont {{Winkler}},\ and\ \citenamefont
  {{Youssefi}}}]{COMPTEL_6010}%
  \BibitemOpen
  \bibfield  {author} {\bibinfo {author} {\bibfnamefont {A.~W.}\ \bibnamefont
  {{Strong}}}, \bibinfo {author} {\bibfnamefont {K.}~\bibnamefont {{Bennett}}},
  \bibinfo {author} {\bibfnamefont {H.}~\bibnamefont {{Bloemen}}}, \bibinfo
  {author} {\bibfnamefont {R.}~\bibnamefont {{Diehl}}}, \bibinfo {author}
  {\bibfnamefont {W.}~\bibnamefont {{Hermsen}}}, \bibinfo {author}
  {\bibfnamefont {D.}~\bibnamefont {{Morris}}}, \bibinfo {author}
  {\bibfnamefont {V.}~\bibnamefont {{Schoenfelder}}}, \bibinfo {author}
  {\bibfnamefont {J.~G.}\ \bibnamefont {{Stacy}}}, \bibinfo {author}
  {\bibfnamefont {C.}~\bibnamefont {{de Vries}}}, \bibinfo {author}
  {\bibfnamefont {M.}~\bibnamefont {{Varendorff}}}, \bibinfo {author}
  {\bibfnamefont {C.}~\bibnamefont {{Winkler}}}, \ and\ \bibinfo {author}
  {\bibfnamefont {G.}~\bibnamefont {{Youssefi}}},\ }\href@noop {} {\bibfield
  {journal} {\bibinfo  {journal} {Astronomy \& Astrophysics}\ }\textbf
  {\bibinfo {volume} {292}},\ \bibinfo {pages} {82} (\bibinfo {year}
  {1994}{\natexlab{a}})}\BibitemShut {NoStop}%
\bibitem [{\citenamefont {{Strong}}\ \emph
  {et~al.}(1994{\natexlab{b}})\citenamefont {{Strong}}, \citenamefont
  {{Bennett}}, \citenamefont {{Bloemen}}, \citenamefont {{Diehl}},
  \citenamefont {{Hermsen}}, \citenamefont {{Morris}}, \citenamefont
  {{Schoenfelder}}, \citenamefont {{Stacy}}, \citenamefont {{de Vries}},
  \citenamefont {{Varendorff}}, \citenamefont {{Winkler}},\ and\ \citenamefont
  {{Youssefi}}}]{COMPTEL1994}%
  \BibitemOpen
  \bibfield  {author} {\bibinfo {author} {\bibfnamefont {A.~W.}\ \bibnamefont
  {{Strong}}}, \bibinfo {author} {\bibfnamefont {K.}~\bibnamefont {{Bennett}}},
  \bibinfo {author} {\bibfnamefont {H.}~\bibnamefont {{Bloemen}}}, \bibinfo
  {author} {\bibfnamefont {R.}~\bibnamefont {{Diehl}}}, \bibinfo {author}
  {\bibfnamefont {W.}~\bibnamefont {{Hermsen}}}, \bibinfo {author}
  {\bibfnamefont {D.}~\bibnamefont {{Morris}}}, \bibinfo {author}
  {\bibfnamefont {V.}~\bibnamefont {{Schoenfelder}}}, \bibinfo {author}
  {\bibfnamefont {J.~G.}\ \bibnamefont {{Stacy}}}, \bibinfo {author}
  {\bibfnamefont {C.}~\bibnamefont {{de Vries}}}, \bibinfo {author}
  {\bibfnamefont {M.}~\bibnamefont {{Varendorff}}}, \bibinfo {author}
  {\bibfnamefont {C.}~\bibnamefont {{Winkler}}}, \ and\ \bibinfo {author}
  {\bibfnamefont {G.}~\bibnamefont {{Youssefi}}},\ }\href@noop {} {\bibfield
  {journal} {\bibinfo  {journal} {A\&A}\ }\textbf {\bibinfo {volume} {292}},\
  \bibinfo {pages} {82} (\bibinfo {year} {1994}{\natexlab{b}})}\BibitemShut
  {NoStop}%
\bibitem [{\citenamefont {Strong}\ \emph
  {et~al.}(2004{\natexlab{b}})\citenamefont {Strong}, \citenamefont
  {Moskalenko},\ and\ \citenamefont {Reimer}}]{EGRET}%
  \BibitemOpen
  \bibfield  {author} {\bibinfo {author} {\bibfnamefont {A.~W.}\ \bibnamefont
  {Strong}}, \bibinfo {author} {\bibfnamefont {I.~V.}\ \bibnamefont
  {Moskalenko}}, \ and\ \bibinfo {author} {\bibfnamefont {O.}~\bibnamefont
  {Reimer}},\ }\href {\doibase 10.1086/423193} {\bibfield  {journal} {\bibinfo
  {journal} {The Astrophysical Journal}\ }\textbf {\bibinfo {volume} {613}},\
  \bibinfo {pages} {962} (\bibinfo {year} {2004}{\natexlab{b}})}\BibitemShut
  {NoStop}%
\bibitem [{\citenamefont {De~la Torre~Luque}\ \emph
  {et~al.}(2024{\natexlab{f}})\citenamefont {De~la Torre~Luque}, \citenamefont
  {Balaji},\ and\ \citenamefont {Silk}}]{DelaTorreLuque:2023cef}%
  \BibitemOpen
  \bibfield  {author} {\bibinfo {author} {\bibfnamefont {P.}~\bibnamefont
  {De~la Torre~Luque}}, \bibinfo {author} {\bibfnamefont {S.}~\bibnamefont
  {Balaji}}, \ and\ \bibinfo {author} {\bibfnamefont {J.}~\bibnamefont
  {Silk}},\ }\href {\doibase 10.3847/2041-8213/ad72f4} {\bibfield  {journal}
  {\bibinfo  {journal} {Astrophys. J. Lett.}\ }\textbf {\bibinfo {volume}
  {973}},\ \bibinfo {pages} {L6} (\bibinfo {year} {2024}{\natexlab{f}})},\
  \Eprint {http://arxiv.org/abs/2312.04907} {arXiv:2312.04907 [hep-ph]}
  \BibitemShut {NoStop}%
\bibitem [{\citenamefont {De~la Torre~Luque}\ \emph
  {et~al.}(2024{\natexlab{g}})\citenamefont {De~la Torre~Luque}, \citenamefont
  {Balaji}, \citenamefont {Fairbairn}, \citenamefont {Sala},\ and\
  \citenamefont {Silk}}]{DelaTorreLuque:2024wfz}%
  \BibitemOpen
  \bibfield  {author} {\bibinfo {author} {\bibfnamefont {P.}~\bibnamefont
  {De~la Torre~Luque}}, \bibinfo {author} {\bibfnamefont {S.}~\bibnamefont
  {Balaji}}, \bibinfo {author} {\bibfnamefont {M.}~\bibnamefont {Fairbairn}},
  \bibinfo {author} {\bibfnamefont {F.}~\bibnamefont {Sala}}, \ and\ \bibinfo
  {author} {\bibfnamefont {J.}~\bibnamefont {Silk}},\ }\href@noop {} {\
  (\bibinfo {year} {2024}{\natexlab{g}})},\ \Eprint
  {http://arxiv.org/abs/2410.16379} {arXiv:2410.16379 [astro-ph.HE]}
  \BibitemShut {NoStop}%
\bibitem [{\citenamefont {Cordes}\ and\ \citenamefont
  {Lazio}(2003)}]{Cordes:2003ik}%
  \BibitemOpen
  \bibfield  {author} {\bibinfo {author} {\bibfnamefont {J.~M.}\ \bibnamefont
  {Cordes}}\ and\ \bibinfo {author} {\bibfnamefont {T.~J.~W.}\ \bibnamefont
  {Lazio}},\ }\href@noop {} {\  (\bibinfo {year} {2003})},\ \Eprint
  {http://arxiv.org/abs/astro-ph/0301598} {arXiv:astro-ph/0301598} \BibitemShut
  {NoStop}%
\bibitem [{\citenamefont {Cordes}\ and\ \citenamefont
  {Lazio}(2002)}]{Cordes:2002wz}%
  \BibitemOpen
  \bibfield  {author} {\bibinfo {author} {\bibfnamefont {J.~M.}\ \bibnamefont
  {Cordes}}\ and\ \bibinfo {author} {\bibfnamefont {T.~J.~W.}\ \bibnamefont
  {Lazio}},\ }\href@noop {} {\  (\bibinfo {year} {2002})},\ \Eprint
  {http://arxiv.org/abs/astro-ph/0207156} {arXiv:astro-ph/0207156} \BibitemShut
  {NoStop}%
\bibitem [{\citenamefont {Beacom}\ and\ \citenamefont
  {Yuksel}(2006)}]{Beacom:2005qv}%
  \BibitemOpen
  \bibfield  {author} {\bibinfo {author} {\bibfnamefont {J.~F.}\ \bibnamefont
  {Beacom}}\ and\ \bibinfo {author} {\bibfnamefont {H.}~\bibnamefont
  {Yuksel}},\ }\href {\doibase 10.1103/PhysRevLett.97.071102} {\bibfield
  {journal} {\bibinfo  {journal} {Phys. Rev. Lett.}\ }\textbf {\bibinfo
  {volume} {97}},\ \bibinfo {pages} {071102} (\bibinfo {year} {2006})},\
  \Eprint {http://arxiv.org/abs/astro-ph/0512411} {arXiv:astro-ph/0512411}
  \BibitemShut {NoStop}%
\bibitem [{\citenamefont {Dirac}(1930)}]{Dirac}%
  \BibitemOpen
  \bibfield  {author} {\bibinfo {author} {\bibfnamefont {P.~A.~M.}\
  \bibnamefont {Dirac}},\ }\href {\doibase 10.1017/S0305004100016091}
  {\bibfield  {journal} {\bibinfo  {journal} {Mathematical Proceedings of the
  Cambridge Philosophical Society}\ }\textbf {\bibinfo {volume} {26}},\
  \bibinfo {pages} {361–375} (\bibinfo {year} {1930})}\BibitemShut {NoStop}%
\bibitem [{\citenamefont {{Kappadath}}(1998)}]{Comptel}%
  \BibitemOpen
  \bibfield  {author} {\bibinfo {author} {\bibfnamefont {S.~C.}\ \bibnamefont
  {{Kappadath}}},\ }\emph {\bibinfo {title} {{Measurement of ...}}},\
  \href@noop {} {Ph.D. thesis},\ \bibinfo  {school} {University of New
  Hampshire} (\bibinfo {year} {1998})\BibitemShut {NoStop}%
\bibitem [{\citenamefont {de~la Torre~Luque}\ \emph {et~al.}(2022)\citenamefont
  {de~la Torre~Luque}, \citenamefont {Mazziotta}, \citenamefont {Ferrari},
  \citenamefont {Loparco}, \citenamefont {Sala},\ and\ \citenamefont
  {Serini}}]{delaTorreLuque:2022vhm}%
  \BibitemOpen
  \bibfield  {author} {\bibinfo {author} {\bibfnamefont {P.}~\bibnamefont
  {de~la Torre~Luque}}, \bibinfo {author} {\bibfnamefont {M.~N.}\ \bibnamefont
  {Mazziotta}}, \bibinfo {author} {\bibfnamefont {A.}~\bibnamefont {Ferrari}},
  \bibinfo {author} {\bibfnamefont {F.}~\bibnamefont {Loparco}}, \bibinfo
  {author} {\bibfnamefont {P.}~\bibnamefont {Sala}}, \ and\ \bibinfo {author}
  {\bibfnamefont {D.}~\bibnamefont {Serini}},\ }\href {\doibase
  10.1088/1475-7516/2022/07/008} {\bibfield  {journal} {\bibinfo  {journal}
  {JCAP}\ }\textbf {\bibinfo {volume} {07}},\ \bibinfo {pages} {008} (\bibinfo
  {year} {2022})},\ \Eprint {http://arxiv.org/abs/2202.03559} {arXiv:2202.03559
  [astro-ph.HE]} \BibitemShut {NoStop}%
\bibitem [{\citenamefont {De~la Torre~Luque}\ \emph {et~al.}(2023)\citenamefont
  {De~la Torre~Luque}, \citenamefont {Loparco},\ and\ \citenamefont
  {Mazziotta}}]{DelaTorreLuque:2023zyd}%
  \BibitemOpen
  \bibfield  {author} {\bibinfo {author} {\bibfnamefont {P.}~\bibnamefont
  {De~la Torre~Luque}}, \bibinfo {author} {\bibfnamefont {F.}~\bibnamefont
  {Loparco}}, \ and\ \bibinfo {author} {\bibfnamefont {M.~N.}\ \bibnamefont
  {Mazziotta}},\ }\href {\doibase 10.1088/1475-7516/2023/10/011} {\bibfield
  {journal} {\bibinfo  {journal} {JCAP}\ }\textbf {\bibinfo {volume} {10}},\
  \bibinfo {pages} {011} (\bibinfo {year} {2023})},\ \Eprint
  {http://arxiv.org/abs/2305.02958} {arXiv:2305.02958 [astro-ph.HE]}
  \BibitemShut {NoStop}%
\bibitem [{\citenamefont {{Heinbach}}\ and\ \citenamefont
  {{Simon}}(1995)}]{1995ApJ...441..209H}%
  \BibitemOpen
  \bibfield  {author} {\bibinfo {author} {\bibfnamefont {U.}~\bibnamefont
  {{Heinbach}}}\ and\ \bibinfo {author} {\bibfnamefont {M.}~\bibnamefont
  {{Simon}}},\ }\href {\doibase 10.1086/175350} {\bibfield  {journal} {\bibinfo
   {journal} {The Astrophysical Journal}\ }\textbf {\bibinfo {volume} {441}},\
  \bibinfo {pages} {209} (\bibinfo {year} {1995})}\BibitemShut {NoStop}%
\bibitem [{\citenamefont {Drury}\ and\ \citenamefont {Strong}(2015)}]{Drury_1}%
  \BibitemOpen
  \bibfield  {author} {\bibinfo {author} {\bibfnamefont {L.~O.}\ \bibnamefont
  {Drury}}\ and\ \bibinfo {author} {\bibfnamefont {A.~W.}\ \bibnamefont
  {Strong}},\ }\href@noop {} {\enquote {\bibinfo {title} {Cosmic-ray diffusive
  reacceleration: a critical look},}\ } (\bibinfo {year} {2015}),\ \Eprint
  {http://arxiv.org/abs/1508.02675} {arXiv:1508.02675 [astro-ph.HE]}
  \BibitemShut {NoStop}%
\bibitem [{\citenamefont {Seo}\ and\ \citenamefont
  {Ptuskin}(1994)}]{seo1994stochastic}%
  \BibitemOpen
  \bibfield  {author} {\bibinfo {author} {\bibfnamefont {E.-S.}\ \bibnamefont
  {Seo}}\ and\ \bibinfo {author} {\bibfnamefont {V.~S.}\ \bibnamefont
  {Ptuskin}},\ }\href@noop {} {\bibfield  {journal} {\bibinfo  {journal} {ApJ}\
  }\textbf {\bibinfo {volume} {431}},\ \bibinfo {pages} {705} (\bibinfo {year}
  {1994})}\BibitemShut {NoStop}%
\bibitem [{\citenamefont {Osborne}\ and\ \citenamefont
  {Ptuskin}(1987)}]{osborne1987cosmic}%
  \BibitemOpen
  \bibfield  {author} {\bibinfo {author} {\bibfnamefont {J.}~\bibnamefont
  {Osborne}}\ and\ \bibinfo {author} {\bibfnamefont {V.}~\bibnamefont
  {Ptuskin}},\ }in\ \href@noop {} {\emph {\bibinfo {booktitle} {International
  Cosmic Ray Conference}}},\ Vol.~\bibinfo {volume} {2}\ (\bibinfo {year}
  {1987})\ p.\ \bibinfo {pages} {218}\BibitemShut {NoStop}%
\bibitem [{\citenamefont {{Weinrich}}\ \emph {et~al.}(2020)\citenamefont
  {{Weinrich}}, \citenamefont {{Boudaud}}, \citenamefont {{Derome}},
  \citenamefont {{G{\'e}nolini}}, \citenamefont {{Lavalle}}, \citenamefont
  {{Maurin}}, \citenamefont {{Salati}}, \citenamefont {{Serpico}},\ and\
  \citenamefont {{Weymann-Despres}}}]{Weinrich_halo}%
  \BibitemOpen
  \bibfield  {author} {\bibinfo {author} {\bibfnamefont {N.}~\bibnamefont
  {{Weinrich}}}, \bibinfo {author} {\bibfnamefont {M.}~\bibnamefont
  {{Boudaud}}}, \bibinfo {author} {\bibfnamefont {L.}~\bibnamefont {{Derome}}},
  \bibinfo {author} {\bibfnamefont {Y.}~\bibnamefont {{G{\'e}nolini}}},
  \bibinfo {author} {\bibfnamefont {J.}~\bibnamefont {{Lavalle}}}, \bibinfo
  {author} {\bibfnamefont {D.}~\bibnamefont {{Maurin}}}, \bibinfo {author}
  {\bibfnamefont {P.}~\bibnamefont {{Salati}}}, \bibinfo {author}
  {\bibfnamefont {P.}~\bibnamefont {{Serpico}}}, \ and\ \bibinfo {author}
  {\bibfnamefont {G.}~\bibnamefont {{Weymann-Despres}}},\ }\href {\doibase
  10.1051/0004-6361/202038064} {\bibfield  {journal} {\bibinfo  {journal} {The
  Astrophysical Journal}\ }\textbf {\bibinfo {volume} {639}},\ \bibinfo {eid}
  {A74} (\bibinfo {year} {2020})},\ \Eprint {http://arxiv.org/abs/2004.00441}
  {arXiv:2004.00441 [astro-ph.HE]} \BibitemShut {NoStop}%
\bibitem [{\citenamefont {{Drury, Luke O\'{}C.}}\ and\ \citenamefont {{Strong,
  Andrew W.}}(2017)}]{Drudy_VA_Energetics}%
  \BibitemOpen
  \bibfield  {author} {\bibinfo {author} {\bibnamefont {{Drury, Luke
  O\'{}C.}}}\ and\ \bibinfo {author} {\bibnamefont {{Strong, Andrew W.}}},\
  }\href {\doibase 10.1051/0004-6361/201629526} {\bibfield  {journal} {\bibinfo
   {journal} {A\&A}\ }\textbf {\bibinfo {volume} {597}},\ \bibinfo {pages}
  {A117} (\bibinfo {year} {2017})}\BibitemShut {NoStop}%
\bibitem [{\citenamefont {Foreman-Mackey}\ \emph {et~al.}(2013)\citenamefont
  {Foreman-Mackey}, \citenamefont {Hogg}, \citenamefont {Lang},\ and\
  \citenamefont {Goodman}}]{emcee}%
  \BibitemOpen
  \bibfield  {author} {\bibinfo {author} {\bibfnamefont {D.}~\bibnamefont
  {Foreman-Mackey}}, \bibinfo {author} {\bibfnamefont {D.~W.}\ \bibnamefont
  {Hogg}}, \bibinfo {author} {\bibfnamefont {D.}~\bibnamefont {Lang}}, \ and\
  \bibinfo {author} {\bibfnamefont {J.}~\bibnamefont {Goodman}},\ }\href
  {\doibase 10.1086/670067} {\bibfield  {journal} {\bibinfo  {journal} {Publ.
  Astron. Soc. Pac.}\ }\textbf {\bibinfo {volume} {125}},\ \bibinfo {pages}
  {306} (\bibinfo {year} {2013})},\ \Eprint {http://arxiv.org/abs/1202.3665}
  {arXiv:1202.3665 [astro-ph.IM]} \BibitemShut {NoStop}%
\bibitem [{\citenamefont {De~la Torre~Luque}\ \emph
  {et~al.}(2024{\natexlab{h}})\citenamefont {De~la Torre~Luque}, \citenamefont
  {Winkler},\ and\ \citenamefont {Linden}}]{DelaTorreLuque:2024ozf}%
  \BibitemOpen
  \bibfield  {author} {\bibinfo {author} {\bibfnamefont {P.}~\bibnamefont
  {De~la Torre~Luque}}, \bibinfo {author} {\bibfnamefont {M.~W.}\ \bibnamefont
  {Winkler}}, \ and\ \bibinfo {author} {\bibfnamefont {T.}~\bibnamefont
  {Linden}},\ }\href {\doibase 10.1088/1475-7516/2024/05/104} {\bibfield
  {journal} {\bibinfo  {journal} {JCAP}\ }\textbf {\bibinfo {volume} {05}},\
  \bibinfo {pages} {104} (\bibinfo {year} {2024}{\natexlab{h}})},\ \Eprint
  {http://arxiv.org/abs/2401.10329} {arXiv:2401.10329 [astro-ph.HE]}
  \BibitemShut {NoStop}%
\bibitem [{\citenamefont {Luque}(2021)}]{Luque:2021ddh}%
  \BibitemOpen
  \bibfield  {author} {\bibinfo {author} {\bibfnamefont {P.~D. L.~T.}\
  \bibnamefont {Luque}},\ }\href {\doibase 10.1088/1475-7516/2021/11/018}
  {\bibfield  {journal} {\bibinfo  {journal} {JCAP}\ }\textbf {\bibinfo
  {volume} {11}},\ \bibinfo {pages} {018} (\bibinfo {year} {2021})},\ \Eprint
  {http://arxiv.org/abs/2107.06863} {arXiv:2107.06863 [astro-ph.HE]}
  \BibitemShut {NoStop}%
\bibitem [{\citenamefont {De~La Torre~Luque}\ \emph {et~al.}(2024)\citenamefont
  {De~La Torre~Luque}, \citenamefont {Smirnov},\ and\ \citenamefont
  {Linden}}]{DeLaTorreLuque:2023fyg}%
  \BibitemOpen
  \bibfield  {author} {\bibinfo {author} {\bibfnamefont {P.}~\bibnamefont
  {De~La Torre~Luque}}, \bibinfo {author} {\bibfnamefont {J.}~\bibnamefont
  {Smirnov}}, \ and\ \bibinfo {author} {\bibfnamefont {T.}~\bibnamefont
  {Linden}},\ }\href {\doibase 10.1103/PhysRevD.109.L041301} {\bibfield
  {journal} {\bibinfo  {journal} {Phys. Rev. D}\ }\textbf {\bibinfo {volume}
  {109}},\ \bibinfo {pages} {L041301} (\bibinfo {year} {2024})},\ \Eprint
  {http://arxiv.org/abs/2309.03281} {arXiv:2309.03281 [hep-ph]} \BibitemShut
  {NoStop}%
\bibitem [{\citenamefont {Sung}\ \emph {et~al.}(2019)\citenamefont {Sung},
  \citenamefont {Tu},\ and\ \citenamefont {Wu}}]{Sung:2019xie}%
  \BibitemOpen
  \bibfield  {author} {\bibinfo {author} {\bibfnamefont {A.}~\bibnamefont
  {Sung}}, \bibinfo {author} {\bibfnamefont {H.}~\bibnamefont {Tu}}, \ and\
  \bibinfo {author} {\bibfnamefont {M.-R.}\ \bibnamefont {Wu}},\ }\href
  {\doibase 10.1103/PhysRevD.99.121305} {\bibfield  {journal} {\bibinfo
  {journal} {Phys. Rev. D}\ }\textbf {\bibinfo {volume} {99}},\ \bibinfo
  {pages} {121305} (\bibinfo {year} {2019})},\ \Eprint
  {http://arxiv.org/abs/1903.07923} {arXiv:1903.07923 [hep-ph]} \BibitemShut
  {NoStop}%
\bibitem [{\citenamefont {Redondo}\ and\ \citenamefont
  {Postma}(2009)}]{Redondo:2008ec}%
  \BibitemOpen
  \bibfield  {author} {\bibinfo {author} {\bibfnamefont {J.}~\bibnamefont
  {Redondo}}\ and\ \bibinfo {author} {\bibfnamefont {M.}~\bibnamefont
  {Postma}},\ }\href {\doibase 10.1088/1475-7516/2009/02/005} {\bibfield
  {journal} {\bibinfo  {journal} {JCAP}\ }\textbf {\bibinfo {volume} {02}},\
  \bibinfo {pages} {005} (\bibinfo {year} {2009})},\ \Eprint
  {http://arxiv.org/abs/0811.0326} {arXiv:0811.0326 [hep-ph]} \BibitemShut
  {NoStop}%
\bibitem [{\citenamefont {Fradette}\ \emph {et~al.}(2014)\citenamefont
  {Fradette}, \citenamefont {Pospelov}, \citenamefont {Pradler},\ and\
  \citenamefont {Ritz}}]{Fradette:2014sza}%
  \BibitemOpen
  \bibfield  {author} {\bibinfo {author} {\bibfnamefont {A.}~\bibnamefont
  {Fradette}}, \bibinfo {author} {\bibfnamefont {M.}~\bibnamefont {Pospelov}},
  \bibinfo {author} {\bibfnamefont {J.}~\bibnamefont {Pradler}}, \ and\
  \bibinfo {author} {\bibfnamefont {A.}~\bibnamefont {Ritz}},\ }\href {\doibase
  10.1103/PhysRevD.90.035022} {\bibfield  {journal} {\bibinfo  {journal} {Phys.
  Rev. D}\ }\textbf {\bibinfo {volume} {90}},\ \bibinfo {pages} {035022}
  (\bibinfo {year} {2014})},\ \Eprint {http://arxiv.org/abs/1407.0993}
  {arXiv:1407.0993 [hep-ph]} \BibitemShut {NoStop}%
\bibitem [{\citenamefont {Li}\ \emph {et~al.}(2020)\citenamefont {Li},
  \citenamefont {Fuller},\ and\ \citenamefont {Grohs}}]{Li:2020roy}%
  \BibitemOpen
  \bibfield  {author} {\bibinfo {author} {\bibfnamefont {J.-T.}\ \bibnamefont
  {Li}}, \bibinfo {author} {\bibfnamefont {G.~M.}\ \bibnamefont {Fuller}}, \
  and\ \bibinfo {author} {\bibfnamefont {E.}~\bibnamefont {Grohs}},\ }\href
  {\doibase 10.1088/1475-7516/2020/12/049} {\bibfield  {journal} {\bibinfo
  {journal} {JCAP}\ }\textbf {\bibinfo {volume} {12}},\ \bibinfo {pages} {049}
  (\bibinfo {year} {2020})},\ \Eprint {http://arxiv.org/abs/2009.14325}
  {arXiv:2009.14325 [astro-ph.CO]} \BibitemShut {NoStop}%
\bibitem [{\citenamefont {Sabti}\ \emph {et~al.}(2020)\citenamefont {Sabti},
  \citenamefont {Magalich},\ and\ \citenamefont {Filimonova}}]{Sabti:2020yrt}%
  \BibitemOpen
  \bibfield  {author} {\bibinfo {author} {\bibfnamefont {N.}~\bibnamefont
  {Sabti}}, \bibinfo {author} {\bibfnamefont {A.}~\bibnamefont {Magalich}}, \
  and\ \bibinfo {author} {\bibfnamefont {A.}~\bibnamefont {Filimonova}},\
  }\href {\doibase 10.1088/1475-7516/2020/11/056} {\bibfield  {journal}
  {\bibinfo  {journal} {JCAP}\ }\textbf {\bibinfo {volume} {11}},\ \bibinfo
  {pages} {056} (\bibinfo {year} {2020})},\ \Eprint
  {http://arxiv.org/abs/2006.07387} {arXiv:2006.07387 [hep-ph]} \BibitemShut
  {NoStop}%
\bibitem [{\citenamefont {Boyarsky}\ \emph {et~al.}(2021)\citenamefont
  {Boyarsky}, \citenamefont {Ovchynnikov}, \citenamefont {Ruchayskiy},\ and\
  \citenamefont {Syvolap}}]{Boyarsky:2020dzc}%
  \BibitemOpen
  \bibfield  {author} {\bibinfo {author} {\bibfnamefont {A.}~\bibnamefont
  {Boyarsky}}, \bibinfo {author} {\bibfnamefont {M.}~\bibnamefont
  {Ovchynnikov}}, \bibinfo {author} {\bibfnamefont {O.}~\bibnamefont
  {Ruchayskiy}}, \ and\ \bibinfo {author} {\bibfnamefont {V.}~\bibnamefont
  {Syvolap}},\ }\href {\doibase 10.1103/PhysRevD.104.023517} {\bibfield
  {journal} {\bibinfo  {journal} {Phys. Rev. D}\ }\textbf {\bibinfo {volume}
  {104}},\ \bibinfo {pages} {023517} (\bibinfo {year} {2021})},\ \Eprint
  {http://arxiv.org/abs/2008.00749} {arXiv:2008.00749 [hep-ph]} \BibitemShut
  {NoStop}%
\bibitem [{\citenamefont {Mastrototaro}\ \emph {et~al.}(2021)\citenamefont
  {Mastrototaro}, \citenamefont {Serpico}, \citenamefont {Mirizzi},\ and\
  \citenamefont {Saviano}}]{Mastrototaro:2021wzl}%
  \BibitemOpen
  \bibfield  {author} {\bibinfo {author} {\bibfnamefont {L.}~\bibnamefont
  {Mastrototaro}}, \bibinfo {author} {\bibfnamefont {P.~D.}\ \bibnamefont
  {Serpico}}, \bibinfo {author} {\bibfnamefont {A.}~\bibnamefont {Mirizzi}}, \
  and\ \bibinfo {author} {\bibfnamefont {N.}~\bibnamefont {Saviano}},\ }\href
  {\doibase 10.1103/PhysRevD.104.016026} {\bibfield  {journal} {\bibinfo
  {journal} {Phys. Rev. D}\ }\textbf {\bibinfo {volume} {104}},\ \bibinfo
  {pages} {016026} (\bibinfo {year} {2021})},\ \Eprint
  {http://arxiv.org/abs/2104.11752} {arXiv:2104.11752 [hep-ph]} \BibitemShut
  {NoStop}%
\bibitem [{\citenamefont {Diamond}\ and\ \citenamefont
  {Marques-Tavares}(2022)}]{Diamond:2021ekg}%
  \BibitemOpen
  \bibfield  {author} {\bibinfo {author} {\bibfnamefont {M.~D.}\ \bibnamefont
  {Diamond}}\ and\ \bibinfo {author} {\bibfnamefont {G.}~\bibnamefont
  {Marques-Tavares}},\ }\href {\doibase 10.1103/PhysRevLett.128.211101}
  {\bibfield  {journal} {\bibinfo  {journal} {Phys. Rev. Lett.}\ }\textbf
  {\bibinfo {volume} {128}},\ \bibinfo {pages} {211101} (\bibinfo {year}
  {2022})},\ \Eprint {http://arxiv.org/abs/2106.03879} {arXiv:2106.03879
  [hep-ph]} \BibitemShut {NoStop}%
\bibitem [{\citenamefont {Ferreira}\ \emph {et~al.}(2022)\citenamefont
  {Ferreira}, \citenamefont {Marsh},\ and\ \citenamefont
  {M\"uller}}]{Ferreira:2022xlw}%
  \BibitemOpen
  \bibfield  {author} {\bibinfo {author} {\bibfnamefont {R.~Z.}\ \bibnamefont
  {Ferreira}}, \bibinfo {author} {\bibfnamefont {M.~C.~D.}\ \bibnamefont
  {Marsh}}, \ and\ \bibinfo {author} {\bibfnamefont {E.}~\bibnamefont
  {M\"uller}},\ }\href {\doibase 10.1088/1475-7516/2022/11/057} {\bibfield
  {journal} {\bibinfo  {journal} {JCAP}\ }\textbf {\bibinfo {volume} {11}},\
  \bibinfo {pages} {057} (\bibinfo {year} {2022})},\ \Eprint
  {http://arxiv.org/abs/2205.07896} {arXiv:2205.07896 [hep-ph]} \BibitemShut
  {NoStop}%
\end{thebibliography}%

\end{document}